\documentclass[a4paper,11pt]{article}
\usepackage{jheppub} 
\usepackage{lineno}
\usepackage[compat=1.1.0]{tikz-feynman}
\usepackage{titlesec}
\usepackage{appendix}
\usepackage{subfig}
\usepackage[dvipsnames]{xcolor}
\usepackage{multirow}

\def\OMIT#1{{}}

\usepackage[export]{adjustbox}
\usepackage[breakable]{tcolorbox}
\usepackage{array}
\newcommand{\PreserveBackslash}[1]{\let\temp=\\#1\let\\=\temp}
\newcolumntype{C}[1]{>{\PreserveBackslash\centering}p{#1}}

\makeatletter
\renewcommand\paragraph{\@startsection{paragraph}{4}%
  {\z@}%
  {1.5ex \@plus 1ex \@minus .2ex}%
  {1ex \@plus .2ex}%
  {\normalfont\normalsize\bfseries}%
}
\makeatother
\DeclareMathOperator*{\SumInt}{%
\mathchoice%
  {\ooalign{$\displaystyle\sum$\cr\hidewidth$\displaystyle\int$\hidewidth\cr}}
  {\ooalign{\raisebox{.14\height}{\scalebox{.7}{$\textstyle\sum$}}\cr\hidewidth$\textstyle\int$\hidewidth\cr}}
  {\ooalign{\raisebox{.2\height}{\scalebox{.6}{$\scriptstyle\sum$}}\cr$\scriptstyle\int$\cr}}
  {\ooalign{\raisebox{.2\height}{\scalebox{.6}{$\scriptstyle\sum$}}\cr$\scriptstyle\int$\cr}}
}

\title{Hybrid Partial Dressing: Correct Effective Potentials at All Temperatures}

\author[a]{Rapha\"el Berthiaume,}
\author[a]{David Curtin,}
\author[a]{Michael Luke,}
\author[a]{Andrija Rasovic}
\author[b]{and Jyotirmoy Roy}
\affiliation[a]{Department of Physics, University of Toronto,\\60 St. George Street, Toronto, ON M5S 1A7, Canada}
\affiliation[b]{Department of Physics, Duke University, \\ Durham, NC 27708, USA}

\emailAdd{raphael.berthiaume@mail.utoronto.ca, d.curtin@utoronto.ca, michael.luke@utoronto.ca, a.rasovic@mail.utoronto.ca, jyotirmoy.roy@duke.edu}

\abstract{
We develop Hybrid Partial Dressing (HPD), a simple diagrammatic resummation scheme for finite-temperature effective potentials in scalar field theory which is valid at all temperatures. HPD improves on earlier resummation schemes such as Daisy Resummation and Partial Dressing by being two-loop exact at all temperatures and free of the overlapping momentum problem. Renormalization group improvement is straightforward to implement. We illustrate the scheme in multi-scalar theories with a spontaneously broken $\mathbb{Z}_2$ symmetry, and compare it against existing approaches, showing in particular that it reproduces the two-loop Dimensional Reduction (DR) potential exactly in the high-temperature regime where DR applies, while remaining valid at all temperatures. This makes HPD suitable for the general study of strong first order phase transitions beyond the Standard Model, as we demonstrate through gravitational wave predictions from HPD and its RG-improved version, RGHPD.
}

\begin{document}
\maketitle
\flushbottom


\section{Introduction}
\label{sec:intro}   
Phase transitions in the early universe provide a natural setting for addressing some of the most fundamental questions in particle physics and cosmology. One particularly well-motivated framework is electroweak baryogenesis, which can explain the observed matter–antimatter asymmetry if Beyond Standard Model (BSM) effects make the  electroweak phase transition strongly first order~\cite{Trodden:1998qg,Cline:2006ts,Morrissey:2012db,White:2016nbo,Garbrecht:2018mrp,Curtin:2016urg, Dine:1992wr, Schicho:2022wty, Kajantie:1995dw, Athron:2023xlk, Arnold:1992rz, Gavela:1994dt, Cline:1996mga, Patel:2012pi,Chala:2018ari, Curtin:2022ovx, Laine:2017hdk, Laine:2016hma, Bahl:2024ykv, Espinosa:1992gq, Espinosa:1992kf, Gould:2021dzl, Gould:2021oba, Gould:2022ran}. Testing this possibility is a major goal for future colliders, which have the potential to confirm or exclude the conditions required for this baryogenesis mechanism.
Importantly, the relevance of strong first-order transitions is not restricted to the electroweak scale. Similar dynamics can arise in hidden sectors \cite{Schwaller:2015tja,Baldes:2018emh,Breitbach:2018ddu,Croon:2018erz,Hall:2019ank,Baldes:2017rcu,Geller:2018mwu,Croon:2019rqu,Hall:2019rld,Chao:2020adk,Dent:2022bcd}, during symmetry breaking in grand unified theories \cite{Hashino:2018zsi,Huang:2017laj,Croon:2018kqn,Brdar:2019fur,Huang:2020bbe}, in conformal extensions of the Standard Model \cite{Prokopec:2018tnq,Kierkla:2022odc}, and in a variety of other motivated scenarios \cite{Caldwell:2022qsj}. 
It is therefore important to obtain reliable predictions for the detectable signatures of cosmological first-order phase transitions. 
The most important of these signatures are stochastic gravitational waves produced by the dynamics of bubble nucleation and collision during the phase transition~\cite{Ramsey-Musolf:2019lsf,Profumo:2007wc,Delaunay:2007wb,Huang:2016cjm,Chala:2018ari,Croon:2020cgk,Grojean:2006bp,Alves:2018jsw,Alves:2020bpi,Vaskonen:2016yiu,Dorsch:2016nrg,Chao:2017vrq,Wang:2019pet,Demidov:2017lzf,Ahriche:2018rao,Huang:2017rzf,Mohamadnejad:2019vzg,Baldes:2018nel,Huang:2018aja,Ellis:2019flb,Alves:2018oct,Alves:2019igs,Cline:2021iff,Chao:2021xqv,Liu:2021mhn,Zhang:2021alu,Cai:2022bcf},
which are potentially detectable by upcoming gravitational-wave observatories such as LISA \cite{Caprini:2019egz}, the Einstein Telescope \cite{Punturo:2010zz}, Cosmic Explorer \cite{Yagi:2011wg}, and related experiments \cite{AEDGE:2019nxb,Hild:2010id,Sesana:2019vho,Theia:2017xtk}.

Predicting the gravitational waves from a first order phase transition starts with a reliable finite-temperature quantum field theory (FTQFT) computation of the temperature-dependent effective potential $V_\mathrm{eff}(\phi, T)$ for the scalar(s) or order parameter(s) which undergo the transition.
In the imaginary time formalism of FTQFT, the time dimension becomes compactified with a radius $\sim 1/T$, leading to 4D fields being represented as an infinite tower of 3D Matsubara modes.
For a scalar with mass $m$ and quartic coupling $\lambda$ which acquires a nonzero vacuum expectation value (VEV) at zero temperature, hard thermal loops of non-zero-modes generate a thermal mass $\Pi \sim \lambda T^2$ which restores the symmetry at high temperatures.
A phase transition therefore occurs near the temperature where the thermal mass and the tachyonic scalar mass at the origin cancel almost exactly. 
At this critical temperature, the naive loop expansion parameter $\alpha \sim \lambda T^2/m^2$ approaches unity (see Ref.~\cite{Schicho:2022wty} for a recent discussion). 
Consequently, the perturbative series becomes unreliable in the presence of long-wavelength modes (i.e. when the temperature is much higher than the zero-mode scalar mass), leading to the well-known infrared divergence problem \cite{Quiros:1999jp,Laine:2016hma} of FTQFT.

To obtain a controlled expansion, the contributions from hard thermal loops must be resummed, after which the effective loop expansion parameter is 
$\beta \sim \lambda T/m$, which roughly counts additional zero-mode loops. At the phase transition, $\beta \sim \sqrt{\lambda}$, which restores perturbativity with the caveat that achieving the equivalent of `one-loop precision' relative to the tree-level contribution at zero temperature requires an FTQFT calculation that includes all contributions up to order $\beta^2$. This involves not only accounting for the tower of non-zero-mode contributions via the abovementioned resummation, but must also include higher-order zero-mode diagrams beyond one-loop to capture all $\mathcal{O}(\beta^2)$ contributions. 
Given the extraordinary sensitivity of gravitational wave observables to the properties of the phase transition (e.g. $\Omega_{GW} h^2 \sim T^{-18}$), it is therefore crucial to have a reliable and general way of obtaining $V_\mathrm{eff}(\phi,T)$ to second order in $\beta$ for BSM theories, especially if gravitational wave signatures are to be correlated with other experimental probes like Higgs coupling measurements~\cite{Cepeda:2019klc,ATLAS:2019mfr,DiMicco:2019ngk,FCC:2025lpp,deBlas:2019rxi,Ramsey-Musolf:2019lsf, DiVita:2017eyz,DiVita:2017vrr,Curtin:2014jma,Profumo:2014opa,Huang:2016cjm,Kotwal:2016tex,Katz:2014bha,Morrissey:2012db,Huang:2017jws,Chen:2017qcz,Carena:2022yvx,Anisha:2025zbc}.

In phenomenological studies of the phase transition, the most commonly employed method for resumming the non-zero-modes is Daisy Resummation~\cite{Parwani:1991gq,Arnold:1992fb,Arnold:1992rz}, which reorganizes the perturbative series such that long-wavelength modes are effectively screened by hard thermal loops.
Unfortunately, all commonly used incarnations of this scheme, including Parwani resummation~\cite{Parwani:1991gq}, Arnold-Espinosa resummation~\cite{Arnold:1992rz}, and Full Dressing~\cite{Espinosa:1992gq,Espinosa:1992kf}, are only correct at first order in $\beta$ and display large scale variation in the predicted peak gravitational-wave amplitude \cite{Gould:2021oba,Croon:2020cgk} even when the effective thermal potentials are  RG-improved. 
These commonly used calculational methods are therefore inherently unsuitable for correctly predicting gravitational wave signatures of first order phase transitions.

The state of the art for systematically computing $V_\mathrm{eff}(\phi,T)$ is Dimensional Reduction (DR). In the high temperature limit, in which the
particle masses in the plasma are parametrically smaller than the temperature, the non-zero Matsubara modes may be integrated out of the theory, matching to a three-dimensional effective field theory (3D EFT).
This matching automatically resums all the non-zero-mode contributions, while the 3D EFT describes the long-distance physics of the light zero mode. 
Furthermore, renormalization group improvement of DR is straightforwardly achieved by evolving the parameters of the effective theory. 
When this matching is performed at two-loop order (NLO DR),
the resulting effective potential is RG-improved and has $\beta^2$ precision. As a result, NLO DR can supply  percent-level predictions for many gravitational wave signals~\cite{Croon:2020cgk}. Moreover, recent studies demonstrate that perturbative computations within weakly coupled 3D EFTs agree well with non-perturbative lattice simulations, suggesting that infrared contributions at higher loops are not numerically significant \cite{Ekstedt:2022zro,Gould:2022ran}.

Dimensional Reduction is theoretically rigorous and systematically improvable, with well-defined order parameters $\sim \beta, m/T$. However, DR is only valid in the limit $\pi T\gg m$, with known corrections proportional to powers $m/ \pi T$~\cite{Chala:2024xll, Chala:2025oul,  Bernardo:2025vkz} but no way to resum them to all orders. DR, in its standard form, can therefore not be used to study strong or supercooled phase transitions where sizable couplings induce large field-dependent masses that drive the system outside the high-$T$ regime during the phase transition, invalidating the EFT construction. Its use is only possible in (near-)conformal theories,
where the barrier is generated thermally close to the origin and the tunnelling field remains in the high-$T$ regime; DR can then be applied to the nucleation dynamics, with the broken minimum determined from the zero-$T$ potential and the two regimes matched through RG running~\cite{Kierkla:2023von,Christiansen:2025xhv,Puchades-Ibanez:2026ksh,Kierkla:2026bnm}. NLO DR calculations are also technically complex, requiring  the computation of $\mathcal{O}(10^2)$ two-loop diagrams at finite temperature. Although recent automation tools have streamlined parts of this process \cite{Ekstedt:2022bff}, users must still carefully determine whether multiple dynamical fields should be retained or integrated out, and whether to employ the soft or ultrasoft potential in different regions of parameter space.

The systematic study of general first order phase transitions in general BSM theories requires an efficient, and ideally simple, method of computing $V_\mathrm{eff}(\phi,T)$ at $\beta^2$ order that does not rely on the high-temperature approximation (making $\beta$ the only relevant power counting parameter).
In this work we develop Hybrid Partial Dressing (HPD), a generalization and correction of standard daisy and super-daisy resummation, for the calculation of finite-temperature effective potentials. Unlike DR, HPD is not an EFT, but instead a diagrammatic technique which consistently resums all graphs up to order $\beta^2$ to all orders in $\alpha$, while not relying on the high-$T$ expansion. Moreover, $V_\mathrm{eff}^\mathrm{HPD}$ reduces to the two-loop zero temperature potential in the $T\rightarrow 0$ limit, while reproducing the NLO DR potential in the high-temperature limit $T \gg m$.
As a result of this formal consistency, HPD is easily RG-improvable in the same manner as the zero-temperature potential, to define RGHPD. It is also readily generalizable to more complicated theories (which we will present in the upcoming publication~\cite{HPDfuturecite}) and higher order, while being technically and numerically straight-forward to implement. 

HPD has its origins in standard Partial Dressing (PD), proposed thirty years ago by Boyd, Brahm and Hsu~\cite{Boyd:1993tz}. PD involves using the one-loop potential to define a gap equation of the form $M^2 = f(M^2)$ for the resummed thermal mass $M(\phi)$, substituting the solution of that gap equation into the one-loop tadpole $d V_\mathrm{eff}/d\phi$, and integrating to obtain the final effective potential. This correctly resums daisies and superdaisies, improving on generalizations of simple Daisy resummation like Full Dressing~\cite{Espinosa:1992gq,Espinosa:1992kf}, which overcount diagrams at $\beta^2$ order. However, PD required cumbersome sunset correction factors, and was never formally developed beyond single-scalar theories or beyond the high-temperature limit.

In this work we lift both of these restrictions.\footnote{This work is motivated by earlier attempts to generalize PD, including studies on Optimized Partial Dressing (OPD)~\cite{Curtin:2016urg}, in particular the recent analytical
result~\cite{Curtin:2022ovx} showing that an RG-improved PD-type calculation will have the same parametrically small $\mathcal{O}(\lambda^3)$ scale variation as the NLO DR effective potential. However, our work reveals various shortcomings in OPD that make it incorrect at $\mathcal{O}(\beta^2)$, and (RG)HPD completely supersedes OPD and PD for all future studies.} We begin in Section~\ref{sec:101} with a brief review of finite-temperature effective potential calculations, the convergence issues of FTQFT, and the strategies commonly used to resum the non-zero modes (Parwani/Arnold-Espinosa/Full Dressing and Dimensional Reduction). In Section~\ref{s.partialdressing} we review PD and carefully generalize it beyond the single-scalar case, correcting various mistakes in previous derivations in the literature, and show that the hybrid formulation makes both this generalization and the extension beyond the high-temperature regime dramatically simpler. All of these derivations are performed at general temperature, without relying on the high-temperature limit. Section~\ref{sec:rgi} then shows that $V_\mathrm{eff}^\mathrm{HPD}$ can be RG improved in the usual fashion, without introducing the overcounting mistakes that normally accompany thermal mass resummation, by simply substituting the running parameters to define $V_\mathrm{eff}^\mathrm{RGHPD}$.

To quantify what this buys us, we define a consistent all-temperature loop expansion parameter $\tilde\beta$, which is $\sim \lambda$ at low temperatures and $\sim \sqrt{\lambda}$ at the phase transition (in the high-temperature limit), and show that (RG)HPD captures all terms at order $\tilde\beta^2$ with small $\mathcal{O}(\lambda^3)$ scale variation. The renormalization scale in $V_\mathrm{eff}^\mathrm{RGHPD}$ must be chosen to track the most important dynamics in each regime, $\mu_R^2 \sim (\pi T)^2 + m^2$; with this choice we prove that RGHPD exactly reproduces the NLO DR results in the high-temperature limit and also approaches the RG-improved zero-temperature two-loop effective potential as $\phi/T \to \infty$. Section~\ref{sec:executivesummary} summarizes these results and illustrates (RG)HPD diagrammatically, and Section~\ref{sec. NumPres} provides a concise numerical recipe for evaluating $V^\mathrm{RGHPD}_\mathrm{eff}$.

Section~\ref{sec:numeric} puts the method to work in a $\phi_1^4 \times \phi_2^4$
two-scalar theory, comparing RGHPD against HPD, Parwani resummation and NLO DR for the
effective potential and the associated phase transition and gravitational wave observables. We find equivalence with NLO DR in the high-$T$ limit, and well-controlled predictions beyond the high-temperature regime where DR starts to fail. The theoretical uncertainty of (RG)HPD from neglected higher-order contributions, as captured by the scale variation, matches NLO DR in the high-temperature limit and is parametrically much smaller than in Parwani resummation; once the high-temperature approximation begins to fail, (RG)HPD uncertainties stay manageably small while NLO DR errors blow up. Interestingly, in the intermediate regime where NLO DR starts to supply unreliable answers, its theoretical uncertainties underestimate the actual errors, emphasizing the need for (RG)HPD to supply accurate and reliable predictions there. We conclude in Section~\ref{s.conclusions}, and collect various technical details in the Appendices.

\section{Review of Finite-Temperature Effective Potential Calculations} \label{sec:101}

We begin by defining our benchmark toy model and
reviewing standard procedures for computing effective potentials within FTQFT. 
We discuss partial dressing separately in Section~\ref{s.partialdressing}.

A comment on notation: We will use $V_\mathrm{eff}^\mathrm{X}$ to denote the effective potential, with X = `zero' for the zero-temperature one-loop fixed-order potential (Section~\ref{s.Veffzero}); X = `RGzero' for the same potential RG-improved with one-loop running couplings (Section~\ref{s.RGIzeroT}); X = `Par' for the one-loop finite-temperature effective potential computed using Parwani thermal resummation (Section~\ref{s.fulldressing}); X = `DR' for the high-temperature effective potential computed using NLO DR (Section~\ref{s.dimensionalreduction}), which by its nature includes the effect of running the couplings to the temperature scale; X = `PD' and `HPD' for the finite-temperature effective potential computed using standard Partial Dressing (Section~\ref{s.singlephiPD} and~\ref{s.twofieldpartialdressing})
and Hybrid Partial Dressing (Sections~\ref{s.singlephiHPD} and~\ref{s.twophiHPD}) thermal resummation; and finally X = `RGHPD' for the RG-improved finite-temperature Hybrid Partial Dressing effective potential introduced in Section~\ref{s.RGIfiniteT}. The one-loop effective potentials are composed of a tree-level contribution $V_\mathrm{tree}$ and the loop-level contribution $V_1 = V_{1,\mathrm{zero}} + V_{1,\mathrm{th}}$, where $V_\mathrm{1, zero}$ is the one-loop zero-temperature effective potential contribution (often called the Coleman-Weinberg potential), and $V_\mathrm{1,\mathrm{th}}$ is explicitly only the thermal contribution at one-loop level, not including the zero-temperature piece. Similar subscripts distinguish zero-temperature and thermal contributions of various specific loop diagrams throughout this work.

\subsection{Scalar Theory Model Setup}

In this paper, we will focus on purely scalar theories where only one scalar acquires a VEV at a time. Generalization of our new results to models involving fermions, gauge bosons, and more general field trajectories is currently being prepared for a follow-up analysis. 
Analytical proofs are first presented, for simplicity, in a single real scalar theory with tree-level potential
\begin{gather}
    \label{e.lagrangiansinglephi}
    V_\textrm{tree} = \frac{1}{2} \nu^2\phi^2 + \frac{\lambda}{4!} \phi^4 \ .
\end{gather}
However, this model is unsuitable for quantitative studies of strong first-order phase transitions: in any perturbative scheme, the expansion parameter becomes large as the phase transition becomes strong, since the only interaction available to radiatively drive generation of the thermal energy barrier is a large scalar self-coupling.

The minimal model that can generate a first order phase transition in the perturbative regime is a theory of two real scalars where each obeys a $\mathbb{Z}_2$ symmetry $\phi_i \to - \phi_i$, which gets broken by $\langle \phi_1 \rangle \neq 0$. The  tree-level potential of the two-field model is given by
\begin{gather}
    \label{e.lagrangian}
    V_\mathrm{tree} =
    \frac{1}{2} \nu_1^2 \phi_1^2 + \frac{1}{2} \nu_2^2 \phi_2^2 + \frac{\lambda_1}{4!} \phi_1^4 + \frac{\lambda_2 }{4!} \phi_2^4 + \frac{\lambda_{12}}{4}\phi_1^2\phi_2^2 \ ,
\end{gather}
where we have a phase transition when $\nu_1^2 <0$ and we assume $\nu_2^2 >0$.

At each point, when relevant, we explain how the analytical derivations generalize to the two-field case. For the numerical studies presented in this paper, we will focus on two benchmarks points given in Table~\ref{tab:bm}. These points are chosen as such that both give a first order phase transition. In case of BP1, phase transition occurs in the regime where high-$T$ approximation is valid, while for BP2, the high-temperature approximation starts to break down as $M_2(\phi_c) \sim \pi T$. 
Both of these benchmark points feature $\lambda_1 \ll \lambda_2, \lambda_{12}$, and at high temperatures, the effective quartic is dominated by the radiative corrections.

Note that for our simple toy model, it is challenging to generate a first-order phase transition that entirely exits the high-temperature regime, preventing us from explicitly demonstrating the reliability of (RG)HPD in a regime where NLO DR fails completely. However, BP2 will demonstrate that theoretical uncertainties for (RG)HPD remain under control, and in particular errors due to neglected three-loop thermal diagrams remain negligible, while for NLO DR, both scale variation and errors from higher dimensional operators (corresponding to the effect of the high-temperature assumption) start to grow dramatically. Furthermore, in this intermediate regime, the theoretical uncertainties of NLO DR actually underestimate the systematic shift of the NLO DR prediction from the true value.

\begin{table}
\begin{center}
\begin{tabular}{||c c c||} 
 \hline
 Parameters & BP1 & BP2 \\ [0.5ex] 
 \hline
$ m_1^\text{phys}$ & 42.5 GeV & 180 GeV  \\ 
 \hline
$ m_2^\text{phys}$& 400 GeV & 1200 GeV \\
 \hline
 VEV $\langle \phi_1 \rangle$ & 1100 GeV & 1000 GeV  \\
 \hline
 $\lambda_2$ & 1.2 & 0.2   \\
 \hline
 $\lambda_{12}$& 0.26 & 2.85  \\ 
 \hline \hline
 $\lambda_{1}$ & 0.0045 &  0.098 \\  
 \hline 
\end{tabular}
\end{center}
\caption{
Definition of the two benchmark points of the two-scalar model~\eqref{e.lagrangian} used in our study, in terms of physical zero-temperature observables: masses, VEV, and effective zero-momentum quartic couplings of the second scalar, all matched to Lagrangian parameters from the two-loop 
RG-improved zero-temperature effective potential.  Both benchmarks feature a strong first order phase transition, with the transition at BP1 being entirely within the high-temperature regime, while the transition at BP2 is in an intermediate regime with $M_2(\phi_c) \sim \pi T$.  Note we also show the zero-momentum quartic of $\phi_1$ (which acquires a VEV), which is derived from the first five matching conditions, to demonstrate $\lambda_1 \ll \lambda_2, \lambda_{12}$ required for a first order phase transition in this model.
\label{tab:bm}
}
\end{table}

\subsection{Effective Potential at $T=0$}
\label{s.Veffzero}
We cover this textbook material briefly for completeness and to establish notation. 
We begin with a single real scalar theory.
The effective potential at zero temperature is given in terms of 1PI $n$-point functions through the relation:
\begin{gather}
    \label{e.VeffzeroT}
    V_\text{eff}^\mathrm{zero} = - \sum_{n=0}^\infty \frac{1}{n!}\phi^n_c \bar{\Gamma}^{(n)}(0) = V_\mathrm{tree}(\phi_c) + V_{1,\mathrm{zero}}(\phi_c)+\cdots \ .
\end{gather}
If we want to compute one-loop effective potential, it is equivalent of computing the series of diagrams:
\begin{equation}
V_{1,\mathrm{zero}}=\quad
\includegraphics[valign=c, scale=1]{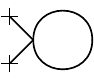} \quad + \quad 
\includegraphics[valign=c,  scale=1]{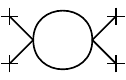} 
\quad + \quad \cdots \quad ,
\label{eq:cw-pot}
\end{equation}
where the $\times$ indicate $\phi_c$ insertion. Using this, together with the symmetry factors for $n$-th diagram in  Eq. \ref{eq:cw-pot}, one can write the effective potential at one-loop.
\begin{gather}
    V_{1,\mathrm{zero}} (\phi_c) = i \sum_{n=1}^{\infty}\int \frac{d^4 p}{(2\pi)^4} \frac{1}{2n} \left[\frac{\lambda \phi_c^2/2}{p^2 - \nu^2 + i \epsilon}\right]^n \ .
    \end{gather}
After Wick's rotation ($p_E = (-i p^0, \vec{p})$) and defining the field-dependent mass as $m^2(\phi_c) = \nu^2 + \frac{1}{2} \lambda \phi_c^2$, we end up with a simple expression for one-loop effective potential we get:
\begin{gather}
    V_{1,\mathrm{zero}}(\phi_c) = \frac{1}{2} \int \frac{d^4 p}{(2 \pi)^4} \log \left[p^2 + m^2 (\phi_c)\right] \ .
    \end{gather}
We work within renormalized perturbation theory in the $\overline{\mathrm{MS}}$ scheme, which regularizes this UV-divergent expression. The one-loop effective potential (dubbed Coleman-Weinberg potential) takes form:
\begin{gather}
\label{e.Vcw}
    V_{1,\mathrm{zero}} = \frac{m^4 (\phi_c)}{64 \pi^2} \left[\log \left(\frac{m^2(\phi_c)}{\mu_0^2}\right) -\frac{3}{2}\right] \ ,
\end{gather}
where $\mu_0$ is the renormalization scale. The generalization to the two-field model is straight forward:
\begin{gather}
\label{e.Vcwtwophi}
    V_{1,\mathrm{zero}} = \sum_i \frac{m_i^4 (\phi_{1c},\phi_{2c})}{64 \pi^2} \left[\log \left(\frac{m_i^2(\phi_{1c},\phi_{2c})}{\mu_0^2}\right) -\frac{3}{2}\right] \ ,
\end{gather}
with $m_{1,2}^2(\phi_{1c},\phi_{2c}) = \nu_{1,2}^2 + \frac{1}{2}\lambda_{1,2} \phi^2_{1c,2c} + \frac{1}{2}\lambda_{12} \phi_{2c,1c}^2$.

Results beyond one-loop are readily available in the literature. We summarize the two-loop effective potential for the one- and two-scalar theory, which we use in HPD, in Appendix~\ref{a.potential}.


\subsection{Effective Potential at Finite Temperature}

\label{s.VefffiniteT}

To calculate the effective potential at finite temperature, we will work in the imaginary time formalism (for review, see~\cite{Quiros:1999jp}). When introducing a temperature, the Kubo-Martin-Schwinger (KMS) relations~\cite{Kubo:1957mj,Martin:1959jp} force the Green's functions to become periodic:
\begin{gather}
    G(\tau-\beta_T) = G(\tau) \quad \text{for} \quad 0\leq \tau \leq \beta_T \ ,
\end{gather}
where $\beta_T=1/T$. In other words, the spacetime manifold becomes $\mathbb{R}^4 \rightarrow \mathbb{R}^3 \times S^1$. This implies that Green's function is periodic in the temporal direction with period $\beta_T$. Then, the Fourier transform of the Green's function,
\begin{gather}
    \tilde{G}(\omega_n,\vec{p}) = \int_{\alpha_T-\beta_T}^{\alpha_T} d\tau e^{i\omega_n \tau}\int d^3x e^{-i \vec{p} \cdot \vec{x}}G(\tau, \vec{x}) \label{eq:gpw} \ ,
\end{gather}
should be independent of $\alpha_T$. 
For that to be satisfied, $e^{i\omega_n\beta_T} = 1 \implies \omega_n = 2\pi n T$. $\omega_n$ are called Matsubara frequencies. This leads us to finite temperature propagator:
\begin{gather}
    \tilde{G}(\omega_n,\vec{p}) = \frac{1}{\vec{p}^2 + m^2 + \omega_n^2} \ ,
\end{gather}
where $m^2 = m^2 (\phi_c)$.
This modifies our propagator Feynman rules, as well as transforming the temporal integral into a sum over Matsubara frequencies $\omega_n$. The theory effectively contains an infinite tower of particles, each with mass $m_n^2 = m^2 + (2n \pi T)^2$. 

 Using the finite-temperature propagator and replacing the integral $d p^0$ with the sum for the single-field model, the vacuum diagrams in Eq.~\ref{eq:cw-pot} give
\begin{gather}
    V_{1}(\phi_c) = \frac{T}{2} \sum_{n=-\infty}^\infty \int \frac{d^3 p}{(2\pi)^3} \log(\omega_n^2 + \omega^2) \ ,
\end{gather}
where $\omega ^2 = p^2 + m^2$. Doing the sum explicitly (and ignoring $\phi_c$ and $\omega$ independent terms), we get a finite result:
\begin{gather}
    V_{1}(\phi_c) = \int \frac{d^3 p }{(2\pi)^3} \left[\frac{\omega}{2} + T \log (1- e^{-\omega/T})\right]
    \label{eq:V1split-ZeroandT} \ .
\end{gather}
The first term just becomes the zero temperature Coleman-Weinberg potential $V_{1,\mathrm{zero}}$, whilst the second one is given in terms of bosonic thermal functions:
\begin{gather}
    \label{e.VTH}
    V_{1,\mathrm{th}} = 
    T\int \frac{d^3 p}{(2\pi)^3} \log(1-e^{-\omega/T}) = \frac{T^4}{2\pi^2} J_B \left( \frac{m^2(\phi_c)}{T^2} \right) \ ,
\end{gather}
where $J_B$ is the bosonic thermal function,
\begin{gather}
    J_B(y^2) = \int_0^\infty dx\text{ } x^2  \log \left[1-e^{- \sqrt{x^2+ y^2}}\right]
    = - \sum_{n=1}^{\infty} \frac{y^2}{n^2} K_2 (n y) \ ,
    \label{eq: Tf}
\end{gather}
which vanish as $T\to0$. The generalization for the two-field case is again straightforward, simply replacing $m^2(\phi_c) \to m_{1,2}^2(\phi_{1c},\phi_{2c})$ above.

Although $J_B$ does not have a closed form solution in terms of elementary functions, one can write the simple analytical high-$T$ and high-$T$ series expansions:
\begin{alignat}{2}
     J_B(y^2) &\simeq - \frac{\pi^2}{45} + \frac{\pi^2}{12} y^2 - \frac{\pi}{6} (y^2)^{3/2} - \frac{1}{32}y^4 \log \frac{y^2}{a_b}  &&\quad \text{for }  y^2 \ll 1  \\
    J_B(y^2) &\simeq - \sum_{n=1}^{N_\textrm{max}} \frac{y^2}{n^2} K_2 (n y) &&\quad \text{for } y^2 \gg 1, \quad N_\textrm{max} \geq 1 \label{eq:Tf} \ . 
\end{alignat}
where $a_b = 16 \pi^2 \exp(3/2 - 2 \gamma_E)$.
These expansions are in fact so good (agreement at sub-percent level over whole range of $m/T$) that one never needs to evaluate the full numerical integral.  
We explicitly compare the bosonic thermal function and its derivatives to the high-$T$ and high-$T$ series expansion with different $N_\mathrm{max}$ in  Appendix \ref{thermalint_lowhighTlimits}. 
This informs how we treat these thermal functions in this paper:
\begin{itemize}
    \item We use the high-$T$ expansion for certain analytical proofs (for effects that only dominate at high temperature) due to its simplicity.
    \item 
    To avoid relying on the high temperature expansion in our numerical calculations, we compute both the thermal potential and its derivatives using the ``low-temperature'' series Eq.~\ref{eq:Tf} with $N_\mathrm{max} = 50$, which gives numerically excellent agreement with the full thermal integrals even for $y \to 0$.\footnote{Previous iterations of the (optimized) partial dressing procedures \cite{Curtin:2016urg,Curtin:2022ovx} defined a piece-wise analytical approximation by joining the high- and high-$T$ approximations to $J_B$. However, this construction can introduce unphysical kinks in the intermediate region ($y^2 \simeq 1$), complicating the numerical solution of the gap equation.} (This is to be contrasted with the high-$T$ expansion, which even with many terms never works for $m \gg T$.)
    \item For the two-loop finite-temperature sunset, we perform the full numerical integral to obtain results valid at all temperatures, see Appendix~\ref{a.sunsetlowT}. This can be encoded in a simple lookup table for efficient computation. 
\end{itemize}

\subsection{IR Problem in FTQFT}
\label{s.IRproblemFTQFT}
    
The breakdown of perturbation theory at finite temperature due to infrared (IR) divergences is a central issue in finite-temperature QFT. Unlike the zero-temperature case, where soft divergences are usually associated with the emission of real massless quanta, at finite temperature they manifest themselves already in self-energy corrections, altering the very notion of particle masses and dispersion relations in a thermal bath.
 
At high temperatures, the Bose–Einstein distribution enhances the occupation number of low-momentum modes,
\begin{equation}
n_B(E) = \frac{1}{e^{E/T} - 1} \sim \frac{T}{E}, \quad (E \ll T),
\end{equation}
so that the contributions of diagrams with soft propagators are strongly amplified. In scalar theories such as $\phi^4$, higher-loop diagrams (like daisy and super-daisy topologies, which we define below) become equally important as the one-loop correction. The presence of factors $\sim T/m$ indicates that the infrared sector dominates the dynamics near the critical point where $m^2 \to 0$.

To make this more concrete, let us consider the simple, $\phi^4 $ theory in the high-temperature limit $T \gg m$ to demonstrate the IR problem. One-loop self energy correction goes as: 
\begin{equation}
\delta m_T^2=\quad \raisebox{2.3pt}{\includegraphics[scale=1]{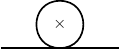}}
\sim \lambda T^2 .
\label{e.lambdaT2massCorrection}
\end{equation}
We use ``$\times$" and $\bullet$ to identify non-zero mode and zero-mode Matsubara loops, respectively.
We explicitly evaluate the above thermal loop integral, and other diagrams used in our analysis, in Appendix~\ref{thermalint_loopints}, but the most important scaling to keep in mind is 
\begin{equation}
    I^\bullet_j(m) = 
    \includegraphics[valign=c, scale=1]{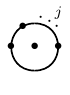}
    \sim T m^{3-2j} 
    \ \ , \ \ 
    I^\times_j(m) = 
    \includegraphics[valign=c, scale=1]{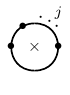}
    \sim T^{4-2j} \ ,
\end{equation}
where $j$ is just the number of propagators in each loop and we only keep the lowest-order terms in the high-temperature expansion.
The leading corrections  to Eq.~\eqref{e.lambdaT2massCorrection} come from `petals' of the same non-zero mode loop attached to a zero-mode loop to make a `daisy' diagram:
\begin{equation}
\delta m_T^2=\quad
\raisebox{2pt}{\includegraphics[scale=1]{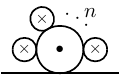}}
\sim \lambda^{n+1} \frac{T^{2n+1}}{m^{2n-1}} 
= \lambda T m \left( \lambda \frac{T^2}{m^2} \right)^n .
\end{equation}
This series of diagrams defines the expansion parameter 
\begin{equation}
    \label{e.alphasinglephi}
    \alpha \equiv \left. 
    \frac{-1}{4} \includegraphics[valign=c,scale=1] {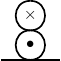} \ \right/ \frac{-1}{2} \includegraphics[valign=c,scale=1] {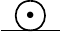} 
    = \frac{\lambda}{48} \frac{T^2}{m^2}  \ .
\end{equation}
where we are careful to include numerical factors in the definition of $\alpha$ for the single-field $\phi^4$ theory.  
It is also useful to ask how big the above daisy diagrams actually are compared to the leading thermal mass correction Eqn.~\eqref{e.lambdaT2massCorrection}:
\begin{equation}
    \label{e.daisydiagram}
    \left. \includegraphics[scale=1]{ExtraDiagram/Quartic_ProgScalar1_ZeroScalar1_DaisyCHAIN.pdf}
    \right/ 
    \includegraphics[scale=1]{ProveHPDtwoScalarDiagrams/Gap_equations/Quartic_ProgScalar1_NonzeroScalar1.pdf}
    \sim \alpha^n \frac{m}{T}  =  \alpha^{n-1} \lambda \frac{T}{m}
\end{equation}
This motivates the definition of a second expansion parameter by taking the above ratio for $n = 1$,
\begin{equation}
    \label{e.betasinglephi}
    \beta \equiv \left. 
    \frac{-1}{4} \includegraphics[valign=c,scale=1] {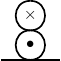}  \ \right/ \frac{-1}{2} \includegraphics[valign=c,scale=1] {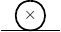}   = \frac{\lambda}{16 \pi} \frac{T}{m} 
    \ .
\end{equation}
Adopting the convention of labeling the order of each diagram relative to dominant hard thermal loop
Eqn.~\eqref{e.lambdaT2massCorrection},
the daisies in Eqn.~\eqref{e.daisydiagram} are therefore $\mathcal{O}(\beta \alpha^{n-1})$. 
As we will see, for a large class of diagrams $\alpha$ counts hard thermal loops (i.e. non-zero-mode loops), while $\beta$ counts zero-mode loops. Finally, we introduce 
\begin{equation}
    \gamma \equiv \frac{\phi^2}{T^2} \ .
\end{equation}
This is not an expansion parameter, but it helps us organize our diagrams by how important they are close to the origin of field space.
We show all contributions to the thermal mass of the single scalar model, up to order $\beta^2$, in Figure~\ref{f.masscontributions}.
Daisy diagrams of order $\mathcal{O}(\beta)$ are in the second and third row, while super-daisy diagrams specifically refer to their $\mathcal{O}(\beta^2)$ generalization in the the fourth row, distinct from the sunset contributions in the fifth row which are also $\mathcal{O}(\beta^2)$.

\newcommand{\diag}[2][0pt]{\raisebox{#1}{\includegraphics[scale=0.9325]{Figure1_diagram/#2.pdf}}}
\newcommand{\diagC}[2][0pt]{\raisebox{#1}{\includegraphics[valign=c,scale=0.9325]{Figure1_diagram/#2.pdf}}}
\newcommand{\ca}[1]{{\color{red}#1}}
\newcommand{\cb}[1]{{\color{orange}#1}}
\newcommand{\cg}[1]{{\color{blue}#1}}

\begin{figure}[h]
\centering
\begin{tabular}{| c | c | c c c c c c c c| }
\hline 
$\mathcal{O}(\cb{1})$ & \multicolumn{1}{c}{} & & & & & & & & \\
 & \multicolumn{1}{c}{} & & \diag[2.3pt]{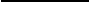} & $+$ & \diag[2.3pt]{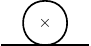} & & & & \\
\hline \hline
$\mathcal{O}(\cb{\beta})$ & $\mathcal{O}(\cg{\gamma^0})$ & &
  $\ca{\alpha^{-1}}$ & &
  $\ca{\alpha^{0}}$ & & & &
  $\ca{\alpha^{n-1}}$ \\

     & & $+$ & 
  \diag[2.3pt]{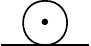} & $+$ & \diag[2.3pt]{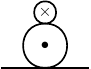} &
  $+$ & $\cdots$ & $+$ & \diag[2.3pt]{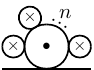} \\
\cline{2-10}
\multirow{2}{*}{i.e. \parbox{2cm}{\centering 1-loop \\   + \\ $\alpha$-resum }} & $\mathcal{O}(\cg{\gamma^1})$ & &
  $\ca{\alpha^{-1}}$ & &
  $\ca{\alpha^{0}}$ & & & &
  $\ca{\alpha^{n-1}}$ \\
& & & & & & & & & \\[-12pt]
& & $+$ &
  \diagC{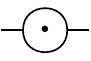} & $+$ & \diagC[0.4pt]{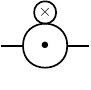} & $+$ & $\cdots$ & $+$ & \diagC[0.1pt]{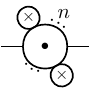} \\
\hline \hline
$\mathcal{O}(\cb{\beta^2})$ & $\mathcal{O}(\cg{\gamma^0})$ & &
  $\ca{\alpha^{-1}}$ & & $\ca{\alpha^{0}}$ & & & &
  $\ca{\alpha^{n-1}}$ \\
& & & & & & & & & \\[-12pt]
& & $+$ &
  \diag[2.3pt]{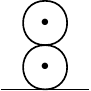} & $+$ & \diag[2.3pt]{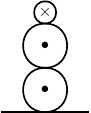} & $+$
  & $\cdots$  & $+$ & \diag[0.025pt]{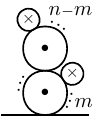} \\
& & $+$ &
  \diagC[0.05pt]{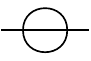} & $+$ & \diagC[0.05pt]{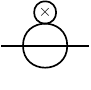} & $+$
  & $\cdots$  & $+$ & \diagC{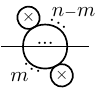} \\
& & \textcolor{gray}{$+$} &
  \diag[2.3pt]{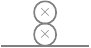} & \textcolor{gray}{$+$} & \multicolumn{5}{c|}{\textcolor{gray}{extra terms higher order in $\beta$}}   \\
\cline{2-10}
& $\mathcal{O}(\cg{\gamma^1})$ & &
  $\ca{\alpha^{0}}$ & & $\ca{\alpha^1}$ & & & &
  $\ca{\alpha^{n}}$ \\
& & & & & & & & & \\[-12pt]
& & $+$ &
  \diagC[9pt]{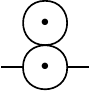} & $+$ & \diagC[14.5pt]{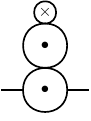} &
  $+$ & $\cdots$ & $+$ & \diagC[10.6pt]{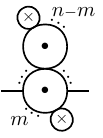} \\
& & $+$ &
  \diag[2.3pt]{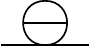} & $+$ & \diag[2.3pt]{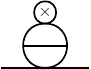} & $+$ &
  $\cdots$ & $+$ & \diag[-1.2pt]{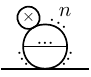} \\
\cline{3-10}
& & & $\ca{\alpha^{-1}} $ & &  & & & & \\
& & \textcolor{gray}{$+$} &
  \diagC{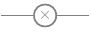} & \textcolor{gray}{$+$} & \multicolumn{5}{c|}{\textcolor{gray}{extra terms higher order in $\beta$}}   \\
\cline{2-10}
\multirow{2}{*}{i.e. \parbox{2cm}{\centering 2-loop \\ + \\ $\alpha$-resum }} & $\mathcal{O}(\cg{\gamma^2})$ & &
  $\ca{\alpha^1}$ & & $\ca{\alpha^2}$ & & & &
  $\ca{\alpha^{n+1}}$ \\
& & $+$ &
  \diagC{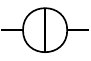} & $+$ & \diagC[3.1pt]{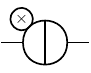}  & $+$ &
  $\cdots$ & $+$ & \diagC[2pt]{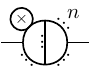} \\
\hline
\end{tabular}
\caption{
All diagrams contributing to the thermal mass in the single-scalar theory up to $\mathcal{O}(\beta^2)$, organized by their order in $\beta$ (rows) and $\alpha$ (columns). Loops with $\bullet$/$\times$ denote zero- and non-zero modes, while unmarked loops include both. During a phase transition, the hard thermal loop is of same size as the tree-level mass, so both are shown as being $\mathcal{O}(1)$ in the first row. The order of other diagrams is understood as being relative to that dominant hard thermal loop. This illustrates how, after resumming $\alpha$ to all orders, $\beta$ is the loop expansion parameter of our finite-temperature effective potential calculation.
}
\label{f.masscontributions}
\end{figure}

The IR problem of FTQFT can now be made explicit.
During a phase transition, the thermal mass $\Pi = \frac{\lambda}{24} T^2 + \mathcal{O}(T m)$ has to nearly cancel the tachyonic zero-temperature mass at the origin. Assuming for clarity that $|\Pi/m^2| = 1$ exactly, this implies that $\alpha = 1/2 \sim 1$ near any phase transition. 
Naively, this would not be a problem if $\beta$ was always so small that you could neglect the $\mathcal{O}(\beta)$ daisy contributions to the potential. However, as we explain below, reliable phase transition predictions require higher order precision in $\beta$.

Resummation of hard thermal loops would account for all diagrams of order $\alpha^n$, $n \geq 0$, at a given order in $\beta$. In Figure~\ref{f.masscontributions}, this would effectively merge all diagrams in a single row into one resummed contribution. 
$\beta$ is therefore the 
actual loop expansion parameter of our finite temperature calculation.

Making the same assumptions as above,
$\beta = \sqrt{\frac{3 \lambda}{32 \pi^2}}$ during the phase transition. 
Convergence of the finite-temperature loop calculation is therefore guaranteed as long as
$\beta < 1$, corresponding to the quartic coupling satisfying the standard 4D loop expansion criterion $\lambda \lesssim 16 \pi^2$. However, the square root changes the power counting compared to a standard loop calculation: a two-loop $\mathcal{O}(\beta^2)$ finite-temperature calculation is required to capture corrections to the tree-level potential that are the same size as one-loop corrections at zero temperature.\footnote{Furthermore, the relative error term of a $\beta^2$-order  calculation is order $\beta^3 \sim (\lambda/16 \pi^2)^{3/2}$, which is actually still larger than the order $(\lambda/16 \pi^2)^2$ error term of a one-loop zero-temperature calculation.} 
Therefore, achieving the equivalent of one-loop zero-temperature precision at finite temperature requires a two-loop effective potential calculation that includes, at minimum, all $\mathcal{O}(\beta^2)$ contributions.\footnote{It is well-known that a strong first-order phase transition in a pure single-scalar theory requires $\beta \sim 1$, requiring non-perturbative methods to understand their phase transition dynamics. However, for multi-scalar theories with sizeable quartics between two scalars, analogous to phenomenologically viable scalar extensions of the SM Higgs sector, the phase transition can be strong while keeping all $\beta$-parameters small, motivating our two-scalar benchmark toy model.}

At this point, it is also worth reminding ourselves of an additional subtlety regarding the precise numerical interpretation of $\beta$ for phase transition calculations. A small $\beta$ is indeed required for convergence of the loop calculation of the effective potential. However, the precision of phase transition observables is not directly captured by the (often very small) numerical size of $\beta$, since by its nature, the phase transition relies on a precise cancellation between tree-level and loop-level contributions to the effective mass at the critical temperature, which greatly amplifies numerical errors. This is only captured by  the variation of phase transition observables under scale variation or the addition of various other error terms to the potential, as we demonstrate in Section~\ref{sec:numeric}, and this informs the requirement of including all $\mathcal{O}(\beta^2)$ contributions in $V_\mathrm{eff}$.

Including all diagrams of order $\alpha^n$ for a given order in $\beta$ requires reorganizing perturbation theory to resum all the  dangerous 
IR-sensitive contributions of non-zero Matsubara modes. 
This can be achieved in two ways.
The first method is to simply redefine the 4D scalar mass to include the dominant thermal correction (and possibly other terms).
The simplest incarnation of this approach is Daisy resummation,
see Section~\ref{s.fulldressing},
with an effective scalar propagator
\begin{equation}
    D(p) = \frac{1}{p^2 - m^2 - \Pi(T)} \ .
\end{equation}
where $\Pi(T) \sim \lambda T^2$ corresponds to the leading-order thermal mass. However, one-loop Daisy resummation and its relatives only capture contributions up to $\mathcal{O}(\beta)$. 
Going to higher order motivates the second approach, Dimensional Reduction, see Section~\ref{s.dimensionalreduction}. At high $T$, non-zero Matsubara modes acquire masses $\sim 2\pi T$ and decouple from long-distance dynamics. The low-energy sector is described by a three-dimensional Euclidean theory of the zero modes.
Matching this low-energy effective theory to the full theory naturally resums the non-zero-mode contributions by including them in the effective zero-mode mass and other couplings.
Performing both the matching and the zero-mode effective potential calculation at two-loop order will then include all contributions of order $\beta^2$.
This framework makes explicit how IR divergences are absorbed into renormalized, temperature-dependent parameters.

Daisy resummation is technically simple and hence by far the most common method for computing the effective thermal potential  in BSM phenomenological studies of the phase transition, but its $\mathcal{O}(\beta)$ precision makes it a rough estimate at best. 
Dimensional reduction, on the other hand, is well-controlled, precise and accurate, but fundamentally limited to the high-temperature regime only. The required two-loop matching calculations are also technically complex.

This motivates our RG-improved Hybrid Partial Dressing (RGHPD) procedure in Sections~\ref{s.partialdressing} and~\ref{sec:rgi}, which combines something close to the simplicity of Parwani resummation with the $
\beta^2$ accuracy of NLO Dimensional Reduction, all while avoiding reliance on the high-temperature expansion. However, before constructing RGHPD, it is useful to review standard Daisy resummation and Dimensional Reduction in more detail.

\subsection{Parwani Resummation/Arnold-Espinosa Resummation/Full Dressing}
\label{s.fulldressing}

The most common way of dealing with IR divergences in the phenomenological studies is  Parwani Resummation \cite{Parwani:1991gq}. It consists of replacing the field dependent mass $m^2(\phi)$  (or $m_i^2(\phi_1,\phi_2)$) with the thermal self-energy on the level of the potential:
\begin{gather}
\nonumber
    m^2(\phi) \rightarrow M^2(\phi) = m^2(\phi) + \Pi(T)\\
      \label{e.fulldressing}
    V_\mathrm{eff}^\mathrm{Par} = V_\mathrm{tree} + V_1(M^2(\phi))
\end{gather}
Most commonly, only the leading piece is kept in $\Pi(T) \equiv \lambda T^2 /24$. This is equivalent of only keeping the correction coming from non-zero Matsubara modes running in the loop:
\begin{align}
    M^2(\phi) =m^2(\phi) + \frac{\lambda T^2}{24}
    = 
\includegraphics[valign=c, scale=1]{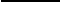}
+
\raisebox{2.3pt}{\includegraphics[scale=1]{ProveHPDtwoScalarDiagrams/Gap_equations/Quartic_ProgScalar1_NonzeroScalar1.pdf}} \ .
\end{align}
It is easy to see that performing this substitution in the one-loop effective potential captures all $\mathcal{O}(\beta)$ contributions in Figure~\ref{f.masscontributions}. 
However, this resummation breaks down outside of the high-temperature limit, and does not include $\mathcal{O}(\beta^2)$ contributions from superdaisies and sunsets.

Daisy resummation generalizes immediately to the two-scalar theory, simply replacing each scalar's mass in the effective potential by $m_i^2(\phi_1) + \Pi_i$. The neglected $\mathcal{O}(\beta^2)$ superdaisy and sunset contributions can be used to define an error term:
\begin{align}
\label{e.deltaVTParwani}
 \delta V^\textrm{Par}_T &= 
 -\frac{1}{8} \left( \includegraphics[valign=c,scale=1]{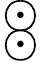} +
                    2  \includegraphics[valign=c,scale=1]{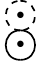} + \includegraphics[valign=c,scale=1]{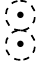} \right)
 -\frac{1}{12} \includegraphics[valign=c,scale=1]{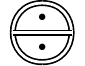}
 -\frac{1}{4} \includegraphics[valign=c,scale=1]{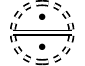}
 \\
 &=\frac{T^2 }{128 \pi ^2} \left( \lambda_{1} m_1^2 + 2 \lambda_{12} m_1 m_2 + \lambda_2 m_2^2  \right) + V_\text{sun}(M_1,M_2) \ ,
 \end{align}
 where two-scalar sunset diagram is given by:
 \begin{align}
     V_\text{sun}(M_1,M_2) &= (\lambda_1 \phi)^2 \frac{T^2}{384 \pi^2}  \left\{ \log\left(\frac{M_1^2}{\mu T}\right) + 1.65 \right\} \nonumber\\
     &+ (\lambda_{12} \phi)^2 \frac{T^2}{128 \pi^2}  \left\{\log\left(\frac{(M_1 + 2M_2)^2}{9\mu T}\right) + 1.65 \right\} \ ,
 \end{align}
see Appendix~\ref{thermalint_loopints}.
Since Parwani is by far the most common resummation scheme in BSM phenomenology calculations, we will compare its prediction for $V_\mathrm{eff}$ to Dimensional Reduction and partial dressing in the numerical studies of Section~\ref{sec:numeric}, using the above error terms to quantify the theoretical uncertainty of the Parwani prediction.

There are several attempts in the literature to improve the simple Parwani Scheme. In \cite{Arnold:1992fb,Espinosa:1992gq} it was demonstrated that replacing $m^2(\phi) \rightarrow M^2(\phi)$ everywhere in the potential will generate field independent pieces proportional to $T^4$ in the potential. They proposed a modified scheme, dubbed Arnold-Espinosa resummation where you only do the resummation in the piece of the thermal potential coming from zero Matsubara mode. This is achieved by adding a ring potential.
\begin{align}
    V_\mathrm{eff}^\mathrm{AE} &= V_\mathrm{tree} + V_{1,\mathrm{zero}} (m^2 (\phi)) + V_{1,\mathrm{th}} (m^2(\phi)) + V_\mathrm{ring}\\
    V_\mathrm{ring} &= -\frac{\lambda T}{12 \pi } \left( M^3(\phi) - m^3(\phi)\right) \ .
\end{align}
The Ring potential amounts to resumming the IR divergent contributions to the zero Matsubara mode. It is equivalent to performing Parwani resummation, under assumptions that only thermal mass of the zero Matsubara mode is relevant, but it does not address the shortcoming that the thermal mass correction is evaluated at lowest order in the high-temperature expansion, and is also only accurate to order $\beta$.

Parwani or Arnold-Espinosa resummation is technically the simplest to perform, and is hence the most popular for BSM phenomenology studies (see, for example~\cite{Basler:2016obg, Ritter:2021hgu, Carena:2019une, Baldes:2018nel, Shuve:2017jgj, Wainwright:2011qy}). This is sufficient for crude estimates of the effective potential properties, but insufficient for precise quantiative predictions, since the inclusion of only the most dominant thermal mass correction results in error terms of order $\mathcal{O}(\beta^2)$, which in practice are significant~\cite{Gould:2021oba} 

In some cases, the full thermal potential instead of the high-$T$ approximation is used. However, even though it is more accurate to do so near $m \sim T$, in those cases the assumption used for adding $V_\text{ring}$ is explicitly violated (since the thermal masses of the non-zero Matsubara modes matters).
This can be addressed by instead considering the full thermal self-energy, which can be computed from the gap equation. Diagramatically, one can write the thermal self-energy at one-loop order: 
\begin{align} \begin{array}{ccccccccccc}
    \includegraphics[valign=c,scale=0.95]{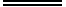}
    & = & 
    \includegraphics[valign=c,scale=0.95]{ProveHPDtwoScalarDiagrams/Gap_equations/Prog_scalar1.pdf}
    & + &
    \raisebox{2.3pt}{\includegraphics[scale=0.95]{ProveHPDtwoScalarDiagrams/Gap_equations/Quartic_ProgScalar1_NonzeroScalar1.pdf}}
    & + &
    \raisebox{2.3pt}{\includegraphics[scale=0.95]{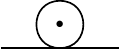}}
    & + &
    \raisebox{2.1pt}{\includegraphics[valign=c,scale=0.95]{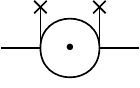}}
    & + &
    \raisebox{2.1pt}{\includegraphics[valign=c,scale=0.95]{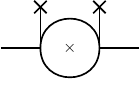}} \ ,
\end{array} \end{align}
where double lines represent propagators with resummed mass $M^2$, and again ``$\times$" (``$\bullet$") are non-zero (zero) Matsubara modes. Algebraically, we can write:
\begin{align}
    M^2(\phi) &= \frac{\partial^2 V}{\partial \phi^2} \nonumber \\
              &= m^2 (\phi) + \frac{\lambda}{2} \left( \frac{ T^2}{12}- \frac{ m(\phi) T}{4 \pi} -\frac{ m^2(\phi)}{16\pi^2} L_R \right)
     - \frac{(\lambda \phi)^2}{2} \left( \frac{T}{8 \pi m(\phi)} + \frac{ L_R }{16\pi^2} \right) \ ,
\end{align}
where we define $L_R \equiv \log\left(\frac{\mu_0^2}{16 \pi^2 e^{-2\gamma_E}T^2}\right)$. If one replaces $m(\phi) \to M(\phi)$ in all but the tree-level term on the RHS,
 
this resums not only all daisies but also super-daisies into the effective mass~\cite{Boyd:1992xn}:
\begin{align} \begin{array}{ccccccc}
    \includegraphics[valign=c,scale=1]{ProveHPDtwoScalarDiagrams/Gap_equations/DressProg_scalar1.pdf}
    & = & 
    \includegraphics[valign=c,scale=1]{ProveHPDtwoScalarDiagrams/Gap_equations/Prog_scalar1.pdf}
    & + &
    \raisebox{1.2pt}{\includegraphics[scale=1]{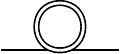}}
    & + &
    \raisebox{1.2pt}{\includegraphics[valign=c,scale=1]{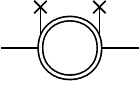}} \\
    M^2 & = & m^2  & + & \frac{\lambda}{2} \left(\frac{T^2}{12}  - \frac{ M T}{4\pi} - \frac{ M^2}{16 \pi^2} L_R  \right) & - & \frac{(\lambda \phi)^2}{2} \left(\frac{ T}{8 \pi M} + \frac{L_R}{16\pi^2} \right) \ , 
\label{eq:gapeqn}
\end{array} \end{align}
Now, we can replace $m^2$ in the potential with the resummed mass $M^2$ as in Eqn.~\eqref{e.fulldressing}~\cite{Espinosa:1992gq}, $V_\mathrm{eff}^\mathrm{FD} = V_\mathrm{tree} + V_1(M^2(\phi))$ and solve this gap equation analytically or numerically. This procedure is referred to as Full Dressing.
The procedure for the two-field model is exactly analogous, with two gap equations defined as $M_i(\phi_1, \phi_2) = \frac{\partial^2 V}{\partial \phi_i^2}$ (for one scalar VEV, the surviving $\mathbb{Z}_2$ symmetry prevents mixing) and solved simultaneously.

Full dressing attempt to capture all daisy and super-daisy contributions to the effective potential, but actually miscounts certain superdaisy contributions~\cite{Boyd:1993tz}. It also misses sunsets. 
Full dressing therefore has an $\mathcal{O}(\beta^2)$ error term:
\begin{equation}
\label{e.deltaVTFullDressing}
     \delta V^\textrm{FD}_T  = 
    -\frac{\{3\}}{8} \includegraphics[valign=c,scale=1]{ExtraDiagram/figure-8_vacuum_phi1.pdf}
    -\frac{1}{12} \includegraphics[valign=c,scale=1]{ProveHPDtwoScalarDiagrams/Tadpoles/VACUUMsunset_RESUM_Scalar111.pdf}
    = \{3\} \frac{\lambda T^2 m^2}{128 \pi ^2}  +V_\textrm{sun} \ ,
\end{equation}
where $\{3\}$ indicates the overcounting mistake.
Therefore, despite its increased complexity, full dressing does not parametrically increase the precision of the Parwani calculation.

\subsection{Dimensional Reduction}
\label{s.dimensionalreduction}
A different approach to resolving the IR problems of finite-temperature QFT is to utilize hierarchies that occur at high-temperature. Namely, when we are in high-temperature regime, there is a hierarchy $2\pi T \gg m$. This means that there is a clear separation between zero and non-zero Matsubara modes. Therefore, one can integrate out all the non-zero Matsubara modes (since $m_n^2 \gg m_0^2$), and as such obtain the 3D effective field theory (EFT) with only zero mode as a dynamical field. This procedure is called Dimensional Reduction \cite{Kajantie:1995kf,Laine:2016hma,Moore:2000jw,Gould:2021dzl,Ghiglieri:2020dpq,Chala:2025oul,Kierkla:2023von}.

Let us again first work in the single scalar theory.
If we remember the KMS relations, we know that once the temperature is introduced, the field theory now lives on $\mathbb{R}^3 \times S^1$. Working in imaginary time formalism, we can utilize the Matsubara mode decomposition (which is practically Fourier decomposition in field space).
\begin{gather}
    \phi = \sqrt{T}\sum_{n=-\infty}^{\infty} \phi_n e^{i \omega_n \tau} \quad , \quad \omega_n = 2 \pi n T \ .
\end{gather}
Even though both perturbative 4D FTQFT approaches and DR use imaginary time formalism, the key difference between them is that DR integrates out the non-zero Matsubara modes, leaving an effective 3D theory with only the zero modes included as dynamical fields.

To demonstrate how DR works, we perform the decomposition. The Lagrangian becomes:
\begin{gather}
    \mathcal{L} = T \left(\sum_{n=-\infty}^{\infty}\frac{1}{2} (\partial_i \phi_n)^2 -\frac{1}{2}\nu^2 \phi_n^2 - \frac12 (2\pi n T)^2 \phi_n^2 - \frac{\lambda T}{4!} \phi_n^4\right) \ .
 \end{gather}
 Now define $m_n^2 = \nu^2 + (2\pi n T)^2$ and we have:
 \begin{gather}
     \mathcal{L} = \frac{T}{2} \sum_n\left(  (\partial_i \phi_n)^2 - m_n^2 \phi_n^2\right) - \frac{\lambda T^2}{4!}\sum_{\substack{n_1,n_2, \\ n_3,n_4}} \phi_{n_1}\phi_{n_2}\phi_{n_3}\phi_{n_4} \delta( \raisebox{0.2ex}{$\scriptstyle n_1+n_2+n_3+n_4$}) \equiv T  \mathcal{L}_3 \ .
 \end{gather}
 $\mathcal{L}_3$ describes at $\phi^4$ theory in 3D. However, since $m_n^2 \gg m_0^2$, to get the EFT, we need to integrate out all the $\phi_n$ for $n\neq 0$ via matching relations of a full 4D theory to a 3D EFT.  After integrating out, we obtain:
 \begin{gather}
     \mathcal{L}^\text{EFT}_3 = \frac{1}{2} (\partial_i \phi_3)^2 -\frac{1}{2} m_3^2 \phi_3^2 - \frac{\lambda _3}{4! }\phi_3^4 \label{e.dr} \ ,
 \end{gather}
 where $m_3, \lambda_3$ we obtain by matching the 4D theory with 3D theory at scale $2\pi T$ (characteristic mass scale of non-zero Matsubara modes). The matching proceeds by equating $n$-point functions of the 4D theory to the ones in 3D EFT, up to chosen loop order. However, due to the fact that we are integrating out infinite degrees of freedom, the number of diagrams one need to match quickly becomes very large. Luckily, there have been developments to make the matching procedure easier. For instance, DRalgo \cite{Ekstedt:2022bff} provides a relatively simple Mathematica package that allows easier computation of matching relations in dimensionally reduced theory. Doing this for $\phi^4$ one obtains relations at NLO. 
 \begin{align}
     \phi_3 &= \phi_0/\sqrt{T}\\
     m_3^2 &= \bar{\mu}^2 + \frac{\bar{\lambda} T^2}{24} + \frac{\lambda_3^2}{6 (4\pi)^2} \left(\log\left(\frac{\mu_3}{3T}\right)-c\right)\\
     \lambda_3 &= \bar{\lambda} T\\
     c &= - \log \left(\frac{3 e ^{\gamma_E/2} A ^6}{4 \pi}\right) \ ,
 \end{align}
 where overline means renormalized quantities in 4D theory, $A$ is Glaschier constant, and $\mu_3$ is a scale at which we evaluate the physical quantities.

At NLO (2-loop matching and 2-loop effective potential in the EFT), DR procedure captures a finite subset of IR-sensitive diagrams. At this order, DR captures all daisy and all leading super-daisy diagrams with at most two zero mode loops. To estimate the effect of neglected $\mathcal{O}(\beta^3)$ terms in our numerical comparisons, we study how much the DR effective potential changes when adding terms of size parametrically equal to the basketball diagram:
 \begin{align}      
 \label{e.deltaVbasketballDR}
    \delta V^\textrm{DR}_{\includegraphics[valign=c,scale=0.5]{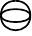}}
    \approx \frac{T^3 \phi^2}{6144 \pi^3} \left[ \frac{\lambda_1^3}{ m_{1,3d}}+\frac{3 \lambda_1 \lambda_{12}^2}{m_{1,3d}+2m_{2,3d}} + \frac{3 \lambda_{12}^3}{2m_{1,3d}+m_{2,3d}}+\frac{\lambda_{12} \lambda_2^2}{m_{2,3d}}\right] \ .
\end{align}
See Sec.~\ref{sec. TET} for the complete calculation.
We also have to check the convergence of the high-temperature expansion. Following~\cite{Chala:2025oul,Bernardo:2025vkz}, we check the sensitivity of the effective potential to the addition of a dimension-6 operator, which could arise from matching at higher order and is suppressed in the high-$T$ limit but can significantly affect the potential (and therefore both critical and nucleation temperatures, as well as other observables) as one exits the high-$T$ regime:
\begin{gather}
    \label{e.deltaVdim6DR}
    \delta V^\textrm{DR}_\textrm{dim-6} = T \left(\frac{\left(\lambda_1^3+\lambda_{12}^3\right) \zeta(3)}{6144 \pi^4 T^3} \phi^6\right) \ .
\end{gather}
The above error term is indeed dominant as $T \to M$ (e.g. in our BP2), but interestingly, we find that at higher temperatures (like for our BP1), a larger error is generated by the explicit difference between the high-temperature limit of the two-loop sunset contribution, see Eqns.~\eqref{e.singleScalarSun_FULL} and~\eqref{e.sunset2field}  below, and the full finite-temperature sunset, see Appendix~\ref{a.sunsetlowT}. We  denote this second error term as
\begin{equation}
    \delta V^\mathrm{DR}_\mathrm{sun} \equiv V_\mathrm{sun}^\mathrm{full} - V_\mathrm{sun}^\mathrm{zero \times th} \ .
    \label{e.deltaVsunDR}
\end{equation}
Of course, $\delta V_\mathrm{dim-6}^\mathrm{DR}$ and $\delta V_\mathrm{sun}^\mathrm{DR}$ are just different estimates of all the missing terms from imposing the high-temperature approximation. The former is an estimate of all contributions to the $\phi^6$ term, the latter resums all $\phi^{n\geq 6}$ terms for the sunset graph only. We use both in our numerical investigations.

Note that the DR procedure, being essentially just a standard EFT calculation, already includes the effects of running couplings. Specifically, the zero-temperature couplings are evolved up to the matching scale $\mu_m = \mu_R = 4\pi T/e^{\gamma_E}$
, where the theory is then matched to the 3D EFT. The 3D EFT then has its own new 3D RGEs, which are used to evolve the couplings down to the scale $\mu_3 = b \pi T/2$, for some $\mathcal{O}(1)$ constant $b$. The dominant scale variation of the NLO DR prediction for phase transition observables can then be captured by varying $b$, while the dependence on the exact matching scale $\mu_m$ is subdominant by comparison.

\subsection{Functional methods}

We also briefly mention functional methods for computing effective potential at finite temperature. In particular we focus on Cornwall, Jackiw and Toubulis method (CJT), also dubbed 2PI \cite{Cornwall,Amelino-Camelia:1992qfe,Klose:2026pux,Banik:2026ljf}. 
Unlike any method that utilizes the 1PI effective action or potential, CJT works at the level of the 2PI effective action:
\begin{gather}
    \Gamma[\phi] = S_{cl}[\phi] + \Gamma^{(1)} [\phi] + \Gamma^{(2)}[\phi] -\textrm{Tr}\left(\frac{\delta \Gamma^{(2)}(\phi,G)}{\delta G}G\right) \label{e.2pi} \ ,
\end{gather}
where $\Gamma^{(1)}$ is the improved 1-loop vacuum graph, $\Gamma^{(2)}$  consists of 2PI graphs with improved propagators and unimproved vertices. $G$ is improved propagator that solves the gap equation
\begin{gather}
    G^{-1} = D^{-1}+ 2\frac{\delta \Gamma^{(2)}[\phi,G]}{\delta G} \ .
\end{gather}
If one expands this diagrammatically, one can show that this generates the same subset of diagrams that partial dressing does \cite{Boyd:1993tz}, except for the fact that CJT correctly accounts for overlapping momenta in the sunset and lollipop contributions. However, unlike any of the above schemes, CJT is not limited to zero external momentum, since it resums full propagators, not just the mass. This makes it suitable for the study of non-equillibrium dynamics at finite temperature.

While this paper was in preparation, two related 2PI calculations appeared, which also compute finite-temperature effective potentials without relying on the high-temperature approximation. Klose presented a one-loop abelian-Higgs calculation in a real-time 2PI scheme~\cite{Klose:2026pux}, which is gauge invariant, does not intrinsically rely on the high-temperature approximation, and reduces to one-loop DR at high temperature. Banik and Kainulainen presented a 2PI calculation in the Hartree approximation~\cite{Banik:2026ljf} for a two-scalar model closely analogous to ours, in which the thermal masses solve a coupled gap equation rather than being extracted from a high-temperature expansion, so that the potential retains the correct Boltzmann suppression at low temperature; much of that work is devoted to renormalizing the 2PI action in the presence of scalar mixing and connecting the resulting parameters to physical masses and four-point functions, and they further show the Hartree potential to be exactly scale independent and the wave-function renormalization factors to have a negligible effect on the nucleation temperature. Both calculations are, however, leading order in the couplings, resumming daisy and super-daisy topologies but omitting the two-loop sunset that, as we argue throughout this work, is necessary for realistic precision. Much like Ref.~\cite{Navarrete:2025yxy} discussed in Section~\ref{sec:RGHPDdiagrammaticproof}, a two-loop version of either calculation would be equivalent to HPD, simply because there is just one true thermally resummed two-loop result, which any correct calculation would arrive at. However, the technical implementation and generalization  to more complicated theories will be much simpler, both analytically and practically, in the HPD scheme, as we will show in our upcoming analysis on HPD in theories beyond the pure scalar case~\cite{HPDfuturecite}.

\section{Partial Dressing (PD) and Hybrid Partial Dressing (HPD)}
\label{s.partialdressing}

In \cite{Boyd:1993tz}, it was shown that full dressing overcounts some higher-loop superdaisy diagrams, and the correct procedure is to solve Eq. \ref{eq:gapeqn}, and then replace $m^2 \rightarrow M^2$ in the tadpole (derivative) of the potential, $\frac{\partial V}{\partial\phi}$. This procedure is known as Partial Dressing (PD), and was further studied in
\cite{Curtin:2016urg, Curtin:2022ovx, Bittar:2025lcr, Bahl:2024ykv}.
While the naive partial dressing procedure fixes the overcounting problem, \cite{Boyd:1993tz} also showed that some correction terms in the gap equation and the tadpole itself were needed to obtain an effective potential that is fully correct at order $\beta^2$.  These terms are fairly simple in the high-temperature limit, but becomes much more complicated for arbitrary temperatures.

In this section we will review the derivation of \cite{Boyd:1993tz}  to introduce partial dressing and the hybrid method for a single $\phi^4$ theory. We then carefully generalize both procedures to our toy model of two scalars, demonstrating how the overlapping momentum problem makes the original PD procedure impractical, while Hybrid Partial Dressing (HPD) generalizes trivially, correctly resumming daisies and super-daisies to all orders while accounting for the remaining $\mathcal{O}(\beta^2)$ sunset  contributions with analytical potential correction terms. 
Furthermore, by never relying on the high-temperature limit in our derivations, we make explicit that the HPD potential is valid at all temperatures, and reduces to the standard two-loop effective potential at zero temperature.

\subsection{Review of Partial Dressing for single $\phi^4$ theory}
\label{s.partialdressingreview}

In this section, we review the  partial dressing derivations from~\cite{Boyd:1993tz} for a single $\phi^4$ theory. 
We start by outlining how the partial dressing effective potential is constructed, then demonstrate the exact difference between full and partial dressing and the need for correction factors. We then review how to construct the equivalent potential using Hybrid Partial Dressing.

\subsubsection{Partial Dressing}
\label{s.singlephiPD}
First recall the form of the unresummed one-loop finite-temperature effective potential:
\begin{align}
    V_\mathrm{eff} &=V_\mathrm{tree} + V_{1,\mathrm{zero}} + V_{1,\mathrm{th}}\\
    V_{1,\mathrm{zero}} &= \frac{m^4(\phi)}{64\pi^2}\left(\log\frac{m^2(\phi)}{\mu^2}-\frac{3}{2}\right)\\
    V_{1,\mathrm{th}} &= \frac{T^4}{2\pi^2} J_B\left(\frac{m^2}{T^2}\right) \ .
\end{align}
$V_{1,\mathrm{zero}}$ is the zero-temperature one loop effective potential Eqn.~\eqref{e.Vcw}, $V_{1,\mathrm{th}}$ is the one-loop thermal contribution Eqn.~\eqref{e.VTH}. 

The gap equation that we must solve for the partial dressing effective potential is identical to Eq. \ref{eq:gapeqn}, except it includes a sunset correction factor of $\zeta$ (whose origin we will explain soon). Using $V_1 = V_{1,\mathrm{zero}} + V_{1,\mathrm{th}}$, this gives:
\begin{eqnarray} 
\nonumber 
M^2 &=& \left.\frac{d^2V_\mathrm{eff}}{d\phi^2}\right|_{m^2\rightarrow M^2}
\\
&=& \frac{d^2 V_\mathrm{tree}}{d \phi^2} 
+ \left[
\frac{\partial^2 m^2}{\partial \phi^2} \left( \left.\frac{\partial 
V_1
}{\partial m^2} \right|_{m^2 \to M^2} \right)
+
\zeta 
\left(\frac{\partial m^2}{\partial \phi}\right)^2 \left( \left.  \frac{\partial^2 
V_1
}{\partial (m^2)^2} \right|_{m^2 \to M^2} \right)
\right]
\label{e.MsqgapeqnbeyondhighT}
\\
&=& m^2(\phi) + \lambda \left( \frac{T^2}{24} - \frac{M T}{8\pi} - \frac{M^2}{32\pi^2}L_R \right) - \zeta (\lambda \phi)^2 \left( \frac{T}{16 \pi M}  + \frac{L_R}{32\pi^2} \right)\label{e.Msqgapeqn} \ .
\end{eqnarray}
The last line explicitly makes use of the high-$T$ expansion for the purpose of our analytical derivations, but numerically the procedure is well-defined by the second line.
Once the gap equation is solved (either algebraically or numerically), one performs a replacement $m^2 \rightarrow M^2$ in the derivative (tadpole) of the potential, $\frac{\partial V}{\partial \phi}$.
\begin{equation}
    \label{e.VfullPDfromV'one}
    V_\mathrm{eff}(\phi, T) = V_\mathrm{tree}(\phi) + \int d\phi \left[ V_{1,\mathrm{zero}}'(\phi, m^2 \to M^2) + V'_{1,\mathrm{th}}(\phi,T; m^2 \to M^2)\right] \ .
\end{equation}
This procedure captures all daisy and super-daisy diagrams, without overcounting. To demonstrate this, we will explicitly compare it with full-dressing procedure, and demonstrate that partial dressing correctly captures the most important subset of diagrams, while full-dressing does not.

The definitions of the $\alpha$ and $\beta$ expansion parameters in Eqns.~\eqref{e.alphasinglephi} and~\eqref{e.betasinglephi} apply verbatim to the potential tadpole calculation, since the effective mass contributions in the diagrammatic ratios are replaced by the corresponding potential tadpole contributions, resulting in the same ratio of loop integrals.
The first loop diagram contributing to the effective potential at non-zero $\gamma$ order is the sunset vacuum graph, which is $\mathcal{O}(\gamma \beta^2)$. Higher order diagrams in $\gamma$ are even higher order in $\beta$. This can be seen for the effective mass in Figure~\ref{f.masscontributions}, but it applies to the tadpoles analogously. Therefore, for an effective potential calculation with $\beta^2$ precision, we only ever have to work to first order in $\gamma$ at all loop orders in our analytical derivations.

In the high-$T$ expansion, the  PD gap equation Eqn.~\eqref{e.Msqgapeqn}  can be solved algebraically, though the full expression is long and not very instructive. To order $\mathcal{O}(\lambda ^3 ,\gamma^0)$, the solution is 
\begin{equation} 
M= m-\frac{T \lambda}{16 \pi}+\frac{T^2 \lambda}{48 m}+\frac{T^2 \lambda^2}{512 m \pi^2}-\frac{T^4 \lambda^2}{4608 m^3}-\frac{T^4 \lambda^3}{24576 m^3 \pi^2}+\frac{T^6 \lambda^3}{221184 m^5} \ .   \label{e.gapsoln}
\end{equation}

However full dressing overcounts some super-daisy diagrams compared to partial dressing. To illustrate this, we compare tadpole diagrams directly.
Rather than applying the full dressing procedure via $m^2\to M^2$ in the potential, we implement full dressing at the tadpole level by dressing both the propagator and the three-point vertex. Therefore, the key difference between the full and partial dressing is the fact that full dressing also dresses an effective three point vertex ($c_3 = \lambda \phi$). Once we do the replacement of $m^2 \rightarrow M^2$ at the level of the potential, the effective 3-point coupling $C_3$ for full dressing becomes:
\begin{gather}
    c_3 =  \frac{dm^2}{d\phi} \rightarrow C_3 =  \frac{d M^2}{d\phi}\\
    C_3 =c_3\left[1+\left(-\frac{3 \lambda T}{16 \pi m}+\frac{ \lambda^2 T^3}{256 \pi m^3}-\frac{ \lambda^3 T^5}{8192 \pi m^5}+\frac{3 \lambda^3 T^3}{8192 \pi^3 m^3}\right)+\mathcal{O}\left(\lambda^4, \gamma^1\right)\right] \ .
\end{gather}
The terms in the parentheses is where the difference between full dressing and partial dressing occur. To compare them directly, we will work with derivative of the high-temperature potential:
\begin{gather}
    V'_\text{pd} \equiv 
    \left[ V_{1,\mathrm{zero}}' + V'_{1,\mathrm{th}}\right]_{m^2 \to M^2}
    \stackrel{\text{High-T}}{=} 
    \lambda \phi \left(\frac{T^2}{24}- \frac{TM}{8\pi}- \frac{M^2}{32\pi^2}L_R\right) \ .
\end{gather}
Diagrammatically, in partial dressing, the first two terms correspond to:
\begin{align}
    V_\text{pd}' &=
    -\frac{1}{2} \includegraphics[valign=c,width=1.5cm]{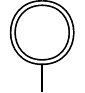}
    \supset   -\frac{1}{2} \left(\includegraphics[valign=c,width=1.5cm]{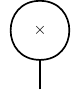}
      +\includegraphics[valign=c,width=1.5cm]{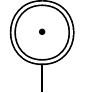} \right)
    = c_3 \frac{T^2}{24}
+ c_3 \frac{-T M}{8 \pi} \\
& = (\lambda \phi) \frac{T^2}{24}
+ (\lambda \phi) \left[\frac{-T m}{8 \pi}
-\frac{\lambda T^3}{384 \pi m}
+\frac{ \lambda T^2}{128 \pi^2}
+\frac{\lambda^2 T^5}{36864 \pi m^3}
-\frac{ \lambda^2 T^3}{4096 \pi^3 m} \right. \label{e.pd} \\ 
& \hspace{4.8cm}\left.
-\frac{\lambda^3 T^7}{1769472 \pi m^5}
+\frac{ \lambda^3 T^5}{196608 \pi^3 m^3}
+\mathcal{O}\left(\lambda^4, \gamma^1\right)\right] 
\nonumber \ .
\end{align}
In full dressing, the tadpole of the corresponding loop potential looks slightly different:
\begin{align}
    V_\text{fd}' &=
    -\frac{1}{2} \includegraphics[valign=c,scale=1]{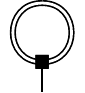}
    \supset -\frac{1}{2} \left( \includegraphics[valign=c,scale=1]{ProveHPDtwoScalarDiagrams/Tadpoles/TAD_NonzeroScalar1.pdf}
+
\includegraphics[valign=c,scale=1]{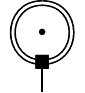} \right)
= c_3 \frac{T^2}{24} + C_3 \frac{-T M}{8 \pi} \\
&= (\lambda \phi) \frac{T^2}{24}
+ (\lambda \phi ) \left[\frac{-T m}{8 \pi}-\frac{\lambda T^3}{384 \pi m}+\frac{ \lambda T^2}{32 \pi^2}+\frac{\lambda^2 T^5}{36864 \pi m^3}-\frac{ 7\lambda^2 T^3}{4096 \pi^3 m} \right. \label{e.fd} \\ &\hspace{4.8cm} \left. -\frac{\lambda^3 T^7}{1769472 \pi m^5}+\frac{7 \lambda^3 T^5}{196608 \pi^3 m^3}+\mathcal{O}\left(\lambda^4, \gamma^1\right)\right] \nonumber \ .
\end{align}
The difference will be induced by the difference between dressed and undressed 3-point vertex (the difference in terms of Eq.~\ref{e.fd} and Eq.~\ref{e.pd}):
\begin{align}
    \Delta &= -\frac{T M}{8 \pi} (C_3 -c_3) =
    (\lambda \phi) \left[ 
    \frac{3 \lambda T^2}{128 \pi^2}- \frac{3\lambda^2 T^3}{2048\pi^3 m} + \frac{\lambda^3 T^5}{32768\pi^3m^3} + \mathcal{O}(\lambda^4, \gamma^1)
    \right] \ .
    \label{e.errorfd}
\end{align}
We have shown explicitly what the difference between the two approaches is, but that does not tell us which one is correct. In order to show which resummation method is correct, one needs to compute the diagrams explicitly. The differences come from diagrams of $\mathcal{O}(\beta^2)$ and $\mathcal{O}(\beta^3)$. First difference comes from diagram with 2 zero and 0 non-zero Matsubara loops (ie. $\mathcal{O}(\alpha^{-1} \beta^2 \gamma^0) )$: 
\begin{align}
    -\frac{1}{4} \includegraphics[valign=c,scale=1]{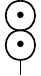}
&= -\frac{1}{4} (\lambda \phi) (\lambda) \underbrace{\left[T \int \frac{d^3 p}{(2\pi)^3} \frac{1}{p^2 + m^2}\right]}_{I^\bullet_1(m^2)} \underbrace{\left[T \int \frac{d^3 p}{(2\pi)^3} \frac{1}{(p^2 + m^2)^2}\right]}_{I^\bullet_2(m^2)} = \lambda \phi \frac{\lambda T^2}{128 \pi^2} \label{e.figure8} \ ,
\end{align}
where we used thermal integrals found in Appendix \ref{thermalint_loopints}. We can see that the partial dressing correctly captures this term (and full dressing does not). Similarly, two other graphs $\mathcal{O}(\alpha^{-1} \beta^3 \gamma^0)$ and $\mathcal{O}(\alpha^0 \beta^3 \gamma^0)$:
\begin{align}
&-\frac18\left(
    \includegraphics[valign=c,scale=1]{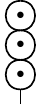} +
    \includegraphics[valign=c,scale=1]{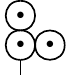}
    \right)
= -\frac{\lambda^3 \phi}{8} \left(I_2^{\bullet2}(m^2) I_1^\bullet(m^2) + I_3^\bullet(m^2) I_1^{\bullet2}(m^2)\right)
= \frac{-(\lambda \phi )\lambda T^3}{4096 \pi^3 m} ,\\ \nonumber
&-\frac{1}{16} \left(\includegraphics[valign=c,scale=1]{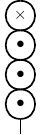} 
+2  \includegraphics[valign=c,scale=1]{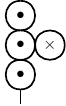} 
+ 2\includegraphics[valign=c,scale=1]{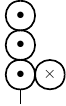} 
+2 \includegraphics[valign=c,scale=1]{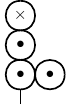} 
+ 2 \, \includegraphics[valign=c,scale=1]{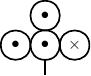} 
+ \includegraphics[valign=c,scale=1]{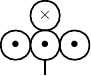}\right) \\
&=-\frac{\lambda^3}{16} (\lambda \phi) I^\times_1(m^2)\left[I_2^{\bullet3}(m^2) + 6 I_1^\bullet(m^2)I_2^\bullet(m^2)I_3^\bullet(m^2)+ 3I_4^\bullet(m^2)I_1^{\bullet2}(m^2)\right] \nonumber \\ 
&=\lambda \phi \frac{\lambda^3 T^5}{196608 \pi^3 m^3} \ .
\end{align}
Once again, we see that partial dressing correctly captures these diagrams, and full-dressing miscounts them. This shows that full dressing makes an error of order $\mathcal{O}(\beta^2)$.

However, naive partial dressing must still be augmented to obtain the correct potential to order $\beta^2$. Consider the sunset correction factor in the modified gap equation Eq.~\ref{e.Msqgapeqn}. 
Without it, the gap equation as written in Eq.~\ref{eq:gapeqn} captures parts of the setting sun two-loop graph, which is of the same $\mathcal{O}(\beta^2)$ as super-daisies, but with a coefficient that is wrong by a factor of 2/3:
\begin{align}
    V_\text{pd}' &\supset  -\frac{1}{2} \includegraphics[valign=c,scale=1]{ProveHPDtwoScalarDiagrams/Tadpoles/TAD_RESUM_LoopScalar1.pdf} 
\supset -\frac{1}{4} \;\includegraphics[valign=c,scale=1]{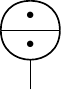}
= \lambda \phi \frac{\lambda^2 \phi^2 T}{384 \pi^2 m^2} \ .
\label{e.sun}
\end{align}
This is the reason for the sunset correction factor $\zeta$ in Eqn.~\eqref{e.Msqgapeqn}, and setting it to 
\begin{equation}
    \label{e.zetasinglephi}
    \zeta = \frac{2}{3} ,
\end{equation} 
ensures that the partial dressing effective potential in Eqn.~\eqref{e.VfullPDfromV'one} correctly includes the sunset contributions to the tadpole.

Even with this correction, Eqn.~\eqref{e.Msqgapeqn} misses lollipop contributions to the tadpole. We must add them as a separate tadpole correction term.\footnote{Sometimes referred to as `lollipop correction terms' since they account for missing lollipop graphs, but we will consistently use the more descriptive `tadpole correction term' to refer to additional terms manually added to the tadpole in partial dressing and its variations, in order to explicitly distinguish them from correction factors that manually rescale terms in the gap equation.} In other words, we simply have to manually add to the one-loop tadpole some missing two-loop lollipop-type contributions of order $\mathcal{O}(\alpha^{-1}\beta^2\gamma^0)$, which are not generated by the partial dressing procedure: 
\begin{align}
    V_\textrm{lol}' & =   -\frac{1}{6} \includegraphics[valign=c,scale=1] {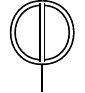} = \frac{\lambda^2 \phi}{16 \pi^2} \left( \log\left(\frac{M_1^2}{\mu_0 T}\right) +1.65 \right) I_{1,\mathrm{th}}(M_1) \label{e.lollipop} \ ,
\end{align}
where  $I_{1,\mathrm{th}}(M) = \frac{T^2}{12} - \frac{TM}{4 \pi} - \frac{M^2}{16 \pi^2} \log\left(\frac{M^2}{a_b T^2}\right)$ is the one-loop thermal loop integral (without zero-temperature contributions).

This ultimately raises the precision of the PD calculation to order $\beta^2$, meaning the error is order $\beta^3$. The final partial dressing potential is therefore
\begin{equation}
    \label{e.VfullPDfromV'}
    V_\mathrm{eff}^\mathrm{PD}(\phi, T) = V_\mathrm{tree}(\phi) + \int d\phi \left[ V_{1,\mathrm{zero}}' + V'_{1,\mathrm{th}}+ V'_\mathrm{lol}\right]_{m^2 \to M^2} \ ,
\end{equation}
where $M^2$ is obtained by solving the gap equation~\eqref{e.Msqgapeqn} with the sunset correction factor from Eqn.~\eqref{e.zetasinglephi}.

\subsubsection{Hybrid Partial Dressing}
\label{s.singlephiHPD}

There are two types of tadpole graphs that partial dressing either miscounts or misses, illustrated in Eqns.~\eqref{e.sun} and~\eqref{e.lollipop} respectively. Both of these tadpole graphs originate from the same sunset vacuum graph.
This demonstrates that the gap equation correctly resums all $\mathcal{O}(\gamma^0)$ diagrams but makes mistakes at higher orders in $\gamma$.
The authors of~\cite{Boyd:1993tz} mention that one can generalize partial dressing to a hybrid approach that uses the  gap equation to resum those diagrams it can capture correctly, i.e. connected vacuum bubbles without dependence on external momenta, while adding the missing potential contributions up to some desired order in $\beta$ manually. 
We denote this procedure Hybrid Partial Dressing, and we will show that it generalizes much more readily to more complicated theories and to arbitrary temperatures.

Instead of deriving and inserting a sunset correction factor in the gap equation, and adding a tadpole correction term to account for the pieces that are still missing, the hybrid method solves both problems at once in the following way:
\begin{enumerate}
    \item Remove the $\phi^2$ terms from the gap equation so it only resums daisies and super-daisies, which it does correctly to all orders in $\beta$:
\begin{equation}
M^2= m^2(\phi) + \frac{\lambda}{2} \left( \frac{ T^2}{12} - \frac{M T}{4\pi} - \frac{ M^2}{16\pi^2}L_R \right) \label{e.MsqgapeqnHPD} \ .
\end{equation} 
This can be generalized to an analytical expression which can be evaluated beyond the high-$T$ regime by replacing one of the $\phi$-derivatives by a partial derivative in the HPD-analogue of Eqn.~\eqref{e.MsqgapeqnbeyondhighT}:
\begin{eqnarray}  
M^2 &=& \left.\frac{\partial}{\partial \phi} \frac{dV}{d\phi}\right|_{m^2\rightarrow M^2}
= \frac{d^2 V_\mathrm{tree}}{d \phi^2} 
+ 
\frac{\partial^2 m^2}{\partial \phi^2} \left( \left.\frac{\partial 
V_1
}{\partial m^2} \right|_{m^2 \to M^2} \right) \ .
\label{e.HPDMsqgapeqnbeyondhighT}
\end{eqnarray}
\item Add the missing sunset vacuum graph to the potential, with the resummed mass inserted:
\begin{equation}
    \label{e.singleScalarSundiagram}
    V_\textrm{sun}^\mathrm{full} = -\frac{1}{12} \includegraphics[valign=c,scale=1]{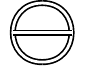} \nonumber \ .
\end{equation}
The complete sunset can be thought of as comprising three components, see Appendix~\ref{a.potential} and~\ref{a.sunsetlowT}: (zero-$T$ loop)$^2$ + (zero-$T$ loop) $\times$ (thermal loop) + (thermal loop)$^2$. The first term $V_\mathrm{sun}^\mathrm{zero}$ is simply the zero-temperature sunset, see Eqn.~\eqref{e.sunsetOneScalatZeroT} in Appendix~\ref{a.potential}. The second and third terms are given by the expressions and lookup tables presented in~\ref{a.sunsetlowT}, without any reliance on the high-temperature approximation. 
\end{enumerate}
The final HPD potential is then given by:
\begin{align}
    \label{e.VfullHPDphi1}
    V_\mathrm{eff}^\mathrm{HPD}(\phi, T) = 
    V_\mathrm{tree}(\phi) &+ \int d\phi \left[ V_{1,\mathrm{zero}}'(\phi, m^2 \to M^2) + V'_{1,\mathrm{th}}(\phi,T, m^2 \to M^2)\right] \nonumber \\
     &+V_\textrm{sun}^\mathrm{full}(\phi, T, m^2 \rightarrow M^2) 
\end{align}
where $M^2$ is the solution to the HPD gap equation Eqn.~\eqref{e.HPDMsqgapeqnbeyondhighT}.

The hybrid and original partial dressing approaches are practically and analytically equivalent in the single scalar case in the high-temperature limit,  with an error term of order $\mathcal{O}(\beta^3)$.
However, as the above construction illustrates, it is straightforward to generalize the HPD potential beyond the high-temperature regime, while the PD potential would need a generalized calculation for the sunset correction factor and tadpole correction term.
Furthermore, as we will show, generalization to theories with multiple scalars is also much more straightforward in HPD.

We use the full zero- and finite-temperature sunset graphs in all our numerical calculations. 
However, it is instructive to study the relative importance of these different terms in different regimes. 
The zero-temperature sunset is not stricly required to achieve $\beta^2$ precision at high temperatures, but including it will ensure that the HPD effective potential converges to the standard two-loop effective potential as $T \to 0$ or $\phi\rightarrow \infty$, as long as the finite-temperature expressions used elsewhere are also defined for general temperature. Furthermore, for strong phase transitions with large couplings, the dominant theoretical uncertainty can result from zero-temperature scale variation, which is reduced by using a full two-loop effective potential.
In the high-temperature regime, one could neglect the (thermal loop)$^2$ piece, which is $\mathcal{O}(\beta^3)$, and only keep the lowest-order $\mathcal{O}(\beta^2)$ piece of the (zero-$T$ loop) $\times$ (thermal loop) piece,
\begin{equation}
    V_\textrm{sun}^\mathrm{zero \times th}  = \frac{(\lambda \phi)^2T^2}{384 \pi^2} \left( \log\left(\frac{M^2}{\mu_0 T}\right) +1.65 \right)\label{e.singleScalarSun_FULL} \ ,
\end{equation}
This is effectively the sunset contribution included in NLO DR. However, we find that even for many benchmarks in the high-temperature regime, like our BP1, the difference between the above approximation and the full sunset actually supplies the dominant thermal error compared to a generic estimate of higher-order operators. This underlines the importance of using the full sunset without any high-temperature approximation in our HPD calculations.

\subsection{Generalizing Partial Dressing to two scalars without mixing}
\label{s.PDmultiplescalar}

We now carefully generalize partial dressing to our benchmark model Eqn.~\eqref{e.lagrangian} of two scalars with a $\mathbb{Z}_2$ $\phi_i \to - \phi_i$ symmetry, with only $\phi_1$ acquiring a VEV.
Our results can be trivially generalized to many scalars as long as only one obtains a VEV.

While partial dressing has been used for multiple scalar theories in the literature before, to our knowledge our derivation is the first one to carefully account for all the diagrams that must be included for $\mathcal{O}(\beta^2)$ accuracy, and explicitly reduces to the two-loop potential in the $T\to 0$ limit.
In standard partial dressing, we find that new mass-dependent sunset correction factors are required even in the high-temperature limit, due to the gap equation resummation not correctly accounting for overlapping momenta in sunsets. The arbitrary-temperature generalizations of these correction factors would  be even more complicated.
This indicates that generalization of standard Partial Dressing to more realistic theories would be highly nontrivial.

In contrast, Hybrid Partial Dressing generalizes almost trivially, and we show by explicit computation that this generalization is correct to $\mathcal{O}(\beta^2)$. The overlapping momenta problem is avoided simply because sunsets are excluded from the gap equation, and instead added at the potential level with the resummed mass inserted.
Hybrid Partial Dressing also has two other important advantages. First, it is numerically much simpler and more stable to solve than partial dressing, involving only a simple quadratic gap equation and a single analytical potential correction term, which are easily solved numerically beyond the high-temperature limit by iteration. 
Second, Hybrid Partial Dressing can be systematically extended to higher accuracy by simply augmenting the potential correction term to include needed three-loop tadpoles, see discussion in Section~\ref{sec:executivesummary}.

We have verified that extending Hybrid Partial Dressing to more complicated theories is procedurally straightforward, and we will present the generalization of HPD to renormalizable theories with arbitrary scalar sectors, fermions, and gauge bosons in an upcoming publication~\cite{HPDfuturecite}.

\subsubsection{Expansion Parameters}

For multiple scalars, the single $\alpha$ and $\beta$ parameters generalize to
\begin{align}
\label{e.alphaijdefinitionNEW}
\alpha_{ij} &= 
\left. \frac{-1}{4}\includegraphics[valign=c,scale=1] {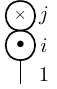}
 \right/ \frac{-1}{2}
\includegraphics[valign=c,scale=1] {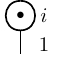}
= \frac{ \frac14 \lambda_{ij} I_2^\bullet(m_i) I_1^\times(m_j)}{\frac12 I_1^\bullet(m_i)}
= \frac{\lambda_{ij}}{48} \frac{T^2}{m_i^2 } \ ,\\
\label{e.betaijdefinitionNEW}
\beta_{ij} &= 
\left. \frac{-1}{4}\includegraphics[valign=c,scale=1] {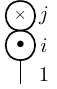}
 \right/ \frac{-1}{2}
\includegraphics[valign=c,scale=1] {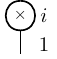}
= \frac{ \frac14 \lambda_{ij} I_2^\bullet(m_i) I_1^\times(m_j)}{\frac12 I_1^\times(m_i)}
=\frac{\lambda_{ij}}{16\pi} \frac{T}{m_i} \ ,
\end{align}
where the field labels $i,j \in \{1,2\}$ do not commute (so $\beta_{12} \neq \beta_{21}$),
$m_{1,2} = \nu_{1,2}^2 + \frac{1}{2}\lambda_{1,2} \phi_{1,2} + \frac{1}{2}\lambda_{12} \phi_{2,1}^2$,
$\lambda_{11} = \lambda_{1}$, $\lambda_{22} = \lambda_{2}$, and $\lambda_{21} = \lambda_{12}$. The above $\alpha_{ij}$ and $\beta_{ij}$ are bookkeeping devices to count, respectively, non-zero mode and zero mode loops. For notational convenience we often refer to a contribution of order $\mathcal{O}(\beta_{i_{1}j_{1}}\ldots \beta_{i_{n}j_{n}})$ as being of order $\mathcal{O}(\beta^n)$. 
However, our generalization of $\beta$ to two scalars, $\beta_{ij}$, introduces a complication when interpreting it as an expansion parameter, since the denominator is not necessarily the leading tadpole contribution. Therefore, we define the expansion parameter $\Tilde{\beta}_{ij}$ from Eq.~\eqref{e.betaijdefinitionNEW} by replacing the denominator with the sum of all one-loop tadpoles. Since our analysis extends beyond the high-temperature limit, it is also useful to generalize this diagrammatic ratio to all temperatures:
\begin{equation}
    \label{e.betaijdefinitionALLT}
    \tilde \beta_{ij} = \left. \frac{-1}{4}\includegraphics[valign=c,scale=1]{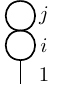}
    \right/ \frac{-1}{2}
    \sum_k \includegraphics[valign=c,scale=1]{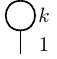} 
    \
    = \frac{\frac14 \lambda_{1i}  \lambda_{ij} I_1(m_j) I_2(m_i)}{\frac12 \lambda_{1}  I_1(m_1) +  \frac12 \lambda_{12}  I_1(m_2)} \ .
    \end{equation}
The calculation is under control as long as all $\Tilde{\beta}_{ij}$ are significantly smaller than 1. At high temperatures when $\beta_{ij} $ is valid, we have $\tilde  \beta_{ij} \le \beta_{ij}$:
\begin{equation}
    \tilde \beta_{ij} \stackrel{\text{high-T}}{=} \left. \frac{-1}{4}\includegraphics[valign=c,scale=1]{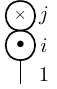}
    \right/ \frac{-1}{2}
    \sum_k \includegraphics[valign=c,scale=1]{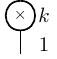} \
    = 
    \frac12 \frac{\lambda_{1i} \lambda_{ij}}{\lambda_1 + \lambda_{12}} I_2^\bullet(m_i)
    = \left( \frac{\lambda_{1i}}{\lambda_1 + \lambda_{12}}\right) \beta_{ij} 
    \ ,
    \label{e.TILDEbetaijdefinitionHIGHT}
    \end{equation} 
where $\beta_{ij}$ is defined in Eq.~\eqref{e.betaijdefinitionNEW}, $I_1(m_i) \stackrel{\text{high-T}}{\to} I^\times_1(m_i)$ and $I_2(m_i) \stackrel{\text{high-T}}{\to} I^\bullet_2(m_i)$.

Consider a first order phase transition induced by a strong coupling between two scalars $\lambda_{12}$. Since the $M_1$ thermal mass is dominated by $\phi_2$ loops, we can again estimate $\tilde \beta_{ij}$ during the phase transition by setting $\Pi = \frac{1}{24}\lambda_{12} T^2 = |m_1^2|$ and $\lambda_{12} \gg \lambda_1$
    \begin{equation} 
        \tilde \beta_{ij} \stackrel{\text{high-T}}{=} \sqrt{\frac{3 \lambda_{ij}}{32\pi^2} \  \frac{\lambda_{1i}^2 \lambda_{ij}}{\lambda_{12}^3} } \ \frac{m_1}{m_i} \ . 
        \label{e.TILDEbetaijdefinitionHIGHT+PT}
    \end{equation}
The condition $\tilde \beta_{ij} < 1$ continues to hold whenever the familiar convergence criterion $\lambda_{ij} < 16\pi^2$ is satisfied.
In fact, all $\tilde \beta_{ij}$ parameters are smaller than $\sim \sqrt{\lambda_{ij}/16 \pi}$: 
$\tilde \beta_{1j}$ 
is suppressed by the small coupling ratio, while 
$\tilde \beta_{2j}$
is suppressed by the small mass ratio. (In the latter case, the coupling ratio may be $\gtrsim 1$ if $\lambda_{22}$ is large, but this would never overwhelm the parametric suppression of the mass ratio.)

Let us also examine the low-temperature limiting behaviour of $\tilde \beta_{ij}$. 
Up to $\log$ factors, 
\begin{equation}
    \tilde \beta_{ij} \stackrel{\text{zero-$T$}}{\sim} \frac{\lambda_{ij}}{32 \pi^2} \ \frac{\lambda_{1i} m_j^2}{\lambda_1 m_1^2 + \lambda_{12} m_2^2} \\
    \approx
    \frac{\lambda_{ij}}{32 \pi^2} \left(\frac{\lambda_{1i}}{ \lambda_{12}}\right) \left(\frac{ m_j}{  m_2}\right)^2 \ ,
\end{equation}
where the last approximation uses $m_2 \gg m_1$ and $\lambda_{12} \gg \lambda_1$. 
Just as in the high-temperature case, $\tilde \beta_{ij} < 1$ is guaranteed if $\lambda_{ij} < 16 \pi^2$ (the mass and coupling ratios again only suppress $\tilde \beta_{ij}$). However, we now notice that $\tilde \beta_{ij}$ scales with the quartic couplings instead of their square root. This corresponds to the fact that at zero temperature, a two-loop calculation has $\mathcal{O}(\lambda_{ij}^2)$ precision relative to the tree-level result. 

$\tilde \beta_{ij}$ is therefore the consistent loop expansion parameter at all temperatures, and an $\mathcal{O}(\tilde \beta^3)$ error term of HPD will reflect the fact that it includes the full two-loop potential at high and low temperatures.
However, as we discussed in Section~\ref{s.IRproblemFTQFT}, the actual numerical error of phase transition observables is much larger than a naive $\mathcal{O}(\tilde \beta^3)$ estimate, owing to their reliance on tree vs loop cancellations.

\subsubsection{Partial Dressing}
\label{s.twofieldpartialdressing}

The tree-level mass matrix for the two-scalar theory in Eqn.~\eqref{e.lagrangian} is 
\begin{gather}
\mathbf{m}^2 = \begin{pmatrix}
        \frac{\partial ^2 V_\textrm{tree}}{\partial \phi_1^2} &    \frac{\partial ^2 V_\textrm{tree}}{\partial \phi_1\partial \phi_2} \\
        \frac{\partial ^2 V_\textrm{tree}}{\partial \phi_1\partial \phi_2} & \frac{\partial ^2 V_\textrm{tree}}{\partial \phi_2^2} &    
    \end{pmatrix} = \begin{pmatrix}
        \nu_1^2 + \frac{\lambda_1 \phi_1^2}{2} + \frac{\lambda_{12} \phi_2^2}{2} & \lambda_{12}\phi_1\phi_2\\
        \lambda_{12}\phi_1\phi_2 &  \nu_2^2 + \frac{\lambda_2 \phi_2^2}{2} + \frac{\lambda_{12} \phi_1^2}{2}
    \end{pmatrix} \ ,
\end{gather}
with mass eigenvalues $m^2_K = m^2_K(\phi_1, \phi_2)$ for general values of $\phi_1, \phi_2$. For our case of interest, where $\phi_2 = 0$, $m^2_K = \mathbf{m}^2_{KK}$. However, the off-diagonal elements play an important role in the gap equation.

The single-field gap equation 
Eqn.~\eqref{e.MsqgapeqnbeyondhighT} can be generalized to multiple fields as
\begin{gather}
    \label{e.PDtwoscalargapeqn}
    M_i^2 = \left.\frac{d^2 V_\textrm{tree}}{d \phi_i^2}\right|_{(\phi_1,\phi_2)\rightarrow (\phi,0)} + \left.\frac{d^2}{d \phi_i^2} \sum_K
    V_1\left(m_K(\phi_1, \phi_2)\right)
    \right|_{(\phi_1,\phi_2)\rightarrow (\phi,0), \, m^2_i \to M^2_i} \ ,
\end{gather}
where $V_1 = V_{1,\mathrm{zero}} + V_{1,\mathrm{th}}$. Crucially, we must take care to use mass eigenvalues $m_{K}(\phi_1,\phi_2)$ for arbitrary field values in the loop potentials above, and only set $\phi_2 \to 0$ \emph{after} evaluating the above derivatives, at which point we can replace $m_{K = i}^2$ by $M_i^2$. 
We then arrive at the following two gap equations for the resummed mass of $\phi_1$ and $\phi_2$:
\begin{align}
\nonumber 
    M_1^2 &= m_1^2 (\phi) + \lambda_1 V_1'(M_1) + \lambda_{12} V_1'(M_2) 
    + \zeta_1 (\lambda_1\phi)^2 V_1''(M_1) + 
    \zeta_2 
    (\lambda_{12} \phi)^2 V_1''(M_2) \ , \\
    M_2^2 &= m_2^2 (\phi) + \lambda_{12} V_1'(M_1) + \lambda_{2} V_1'(M_2)+ 
    2 \zeta_3
    (\lambda_{12} \phi)^2 \frac{ V_1'(M_1) - V_1'(M_2) }{M_1^2-M_2^2} \ ,
\end{align}
where we have again inserted sunset correction factors $\zeta_{1,2,3}$ in front of all the $\phi^2$ terms, in anticipation of the needed corrections. 
In the high-temperature approximation, the gap equations simplifies to
\begin{align}
    M_1^2 = m_1^2(\phi) &+ \frac{\lambda_1}{2}   I_1(M_1) + \frac{\lambda_{12}}{2} I_1(M_2) 
                          - \zeta_1 \frac{(\lambda_1 \phi)^2}{2} I_2(M_1)
                           - \zeta_2 \frac{(\lambda_{12} \phi)^2}{2} I_2(M_2) \label{e.gapeqPD2scalar}
\\
    M_2^2 = m_2^2(\phi) & + \frac{\lambda_{12}}{2} I_1(M_1) + \frac{\lambda_{2}}{2} I_1(M_2) - \zeta_3 (\lambda_{12} \phi)^2  I_2(M_1,M_2) \nonumber \ ,
\end{align}
where $I_1(M) = \left(\frac{T^2}{12} - \frac{T M}{4\pi} -\frac{M^2}{16\pi^2}L_R \right)$, $I_2(M_i,M_j) = \left(\frac{T}{4 \pi (M_i+M_j)} -\frac{L_R}{16\pi^2}\right)$.
In diagrammatic form, adopting the convention of denoting $\phi_1$ (which acquires a VEV) with a solid line and $\phi_2$ with a dashed line, this can be written as
\begin{align} 
    \includegraphics[valign=c,scale=1]{ProveHPDtwoScalarDiagrams/Gap_equations/DressProg_scalar1.pdf}
    &=
    \includegraphics[valign=c,scale=1]{ProveHPDtwoScalarDiagrams/Gap_equations/Prog_scalar1.pdf}
    +
    \raisebox{1.5pt}{\includegraphics[scale=1]{ProveHPDtwoScalarDiagrams/Gap_equations/Quartic_Progscalar1_ResumLoopscalar1.pdf}}
    +
    \raisebox{1.5pt}{\includegraphics[scale=1]{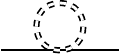}}
    +
    \zeta_1 \raisebox{1.5pt}{\includegraphics[valign=c,scale=1]{ProveHPDtwoScalarDiagrams/Gap_equations/Cubic_ProgScalar1_ResumLoopScalar1.pdf}}
    +
    \zeta_2 \raisebox{1.5pt}{\includegraphics[valign=c,scale=1]{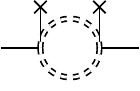}}
    \nonumber \ ,\\ 
    \includegraphics[valign=c,scale=1]{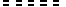}
    &=
    \includegraphics[valign=c,scale=1]{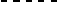}
    +
    \raisebox{1.5pt}{\includegraphics[scale=1]{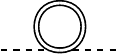}}
    +
    \raisebox{1.5pt}{\includegraphics[scale=1]{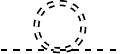}}
    +
    \zeta_3 \raisebox{1.5pt}{\includegraphics[valign=c,scale=1]{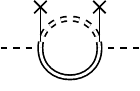}} \ .
\end{align}
We emphasize that the term multiplied by $\zeta_3$ would be missing if we used the diagonal terms $\mathbf{m}^2_{KK}$ instead of the full mass eigenstates $m^2_K(\phi_1, \phi_2)$ for general $\phi_1, \phi_2$  in Eqn.~\eqref{e.PDtwoscalargapeqn}.

In this non-mixing scenario, $\zeta_{2,3}$ are mass-dependent and not the standard $2/3$. The 4-propagator integral defined in Eq.~\ref{eq:H4} correctly captures overlapping momenta for the sunset tadpoles. The sunset correction factor is the ratio between the correct result $\mathcal{H}_4$ and the naive result that consists of splitting the integral into 2 two-propagator integrals $I_2 I_2$
\begin{equation}
    \zeta(m_1,m_2,m_3) = \frac{\mathcal{H}_4(m_1,m_2,m_3)}{I_2(m_1,m_1)I_2(m_2,m_3)} \ .
\end{equation}
In the high-temperature limit, our 3 sunset correction factors are
\begin{align}
    \zeta_1  &= \frac{\mathcal{H}_4(m_1,m_1,m_1)}{I^\bullet_2(m_1,m_1)I^\bullet_2(m_1,m_1)} = \frac{2}{3} \ ,\label{e.zeta1}\\
    \zeta_2 (m_1, m_2) &= \frac{\mathcal{H}_4(m_1,m_2,m_2)}{I^\bullet_2(m_1,m_1)I^\bullet_2(m_2,m_2)} = \frac{2m_2}{m_1+2m_2}  \ , \label{e.zeta2}\\
    \zeta_3 (m_1, m_2) &=\frac{\mathcal{H}_4(m_2,m_1,m_2)}{I^\bullet_2(m_2,m_2)I^\bullet_2(m_1,m_2)} =\frac{m_1+m_2}{m_1+2m_2}   \label{e.zeta3} \ .
\end{align}
More complicated expressions would be needed at arbitrary temperature.

With these correction factors in the gap equation, we will correctly capture the sunset tadpoles. Promoting the sunset correction factors to depend on the resummed masses $\zeta_i(m_1\to M_1 , m_2\to M_2)$ while solving the gap equations captures the non-zero mode daisy chain which is also of $\mathcal{O}(\beta^2)$. 
Finally, just like in the single-field case, we must also add the missing lollipop contributions to the tadpole: 
\begin{align}
    \label{e.PD2philol}
    V'_{\text{LOL}} &= 
    -\frac{1}{6} \includegraphics[valign=c,width=1.5cm]{ProveHPDtwoScalarDiagrams/Tadpoles/Lollipop_RESUM_Scalar111.pdf}  -
    \frac{1}{2} \includegraphics[valign=c,width=1.5cm]{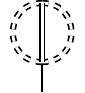}  \\
    &  = \frac{\lambda_1^2 \phi}{16 \pi^2}  \left( \log\left(\frac{M_1^2}{\Bar{\mu}T}\right) + 1.65 \right) I_{1,\mathrm{th}}(M_1)  \\
    &+ \frac{\lambda_{12}^2 \phi}{16 \pi^2}  \left( \log\left(\frac{(M_1 + 2M_2)^2}{9\Bar{\mu}T}\right) + 1.65 \right)
\left( I_{1,\mathrm{th}}(M_1) + 2 I_{1,\mathrm{th}}(M_2) \right) \nonumber \ .
\end{align}
In the high-temperature approximation, we can again solve the gap equation, Eq.~\eqref{e.gapeqPD2scalar} analytically. Inserting the solution into $V'_1$ and adding the lollipops in Eqn.~\eqref{e.PD2philol}
gives the correct result to order $\beta^2$:
\begin{equation}
     \begin{aligned}
     {V^\mathrm{PD}_{\text{eff}}}' =\, V_\mathrm{tree}' +  & (\lambda_1 \phi) \left[\frac{T^2}{24} -\frac{TM_1}{8\pi} - \frac{M_1^2 L_R}{32\pi^2} \right] + (\lambda_{12} \phi) \left[\frac{T^2}{24} -\frac{TM_2}{8\pi} - \frac{M_2^2 L_R}{32\pi^2} \right] \\
     &+ 
     \label{e.VprimePDtwoscalar}
     \lambda_1^2 \phi \frac{T^2}{192 \pi^2}  \left\{ \log\left(\frac{M_1^2}{\Bar{\mu}T}\right) + 1.65 \right\}  \\
     &+
   \lambda_{12}^2 \phi \frac{T^2}{64 \pi^2}  \left\{\log\left(\frac{(M_1 + 2M_2)^2}{9\Bar{\mu}T}\right) + 1.65 \right\} \ .
     \end{aligned}
\end{equation}
This is verified by comparing to the full diagrammatic calculation in e.g. Appendix~\ref{a.beta2proof}.
A similar proof to the one-scalar case also demonstrates that full dressing commits overcounting mistakes in the two-scalar case, again limiting its accuracy to $\mathcal{O}(\beta)$.

We close this section by commenting on previous incorrect implementations of partial dressing for multi-field theories in the literature. In particular, Ref~\cite{Curtin:2022ovx} corresponds to the above but with $\zeta_3 \to 0$ (incorrectly taking $ m^2_K(\phi_1, \phi_2) \to \mathbf{m}^2_{KK}$ before taking derivatives in the gap equation) and $\zeta_{1,2} = 1$ (neglecting sunset correction factors). 
This does not negate the important analytical result of that study, which showed that an RG-improved one-loop partial dressing calculation has  parametrically the same $\mathcal{O}(\beta^3)$ scale variation as a corresponding 2-loop Dimensional Reduction calculation, which partially motivated this work, but nevertheless it means that the calculational framework for Optimized Partial Dressing is missing several important terms which limit its accuracy to $\mathcal{O}(\beta)$. For future calculations, HPD and its upcoming generalizations to more realistic theories~\cite{HPDfuturecite} entirely supersedes OPD. 
Ref.~\cite{Bahl:2024ykv} focuses on the two-field scenario where both fields get a VEV, introducing mixing between the scalars.
They also discuss the non-mixing case of multiple scalars where only one field acquires a VEV for which partial dressing would work as we derived above. In the non-mixing case, the authors advocate for setting all sunset correction factors to $2/3$, in analogy to the single-field case. However, as we show above, this would introduce error terms of order $\mathcal{O}(\beta^2)$, similar to OPD.

\subsubsection{Hybrid Partial Dressing}
\label{s.twophiHPD}

Unlike for the partial dressing case, we can generalize the Hybrid Partial Dressing procedure for the two-field case trivially, by exact analogy to the single-field case.
\begin{enumerate}
    \item Remove $\phi^2$ terms from the gap equations, to only resum daisies and super-daisies:
\begin{align}
    M_{1}^{2} &=  m_{1}^2+ \frac{\lambda_{1}}{2} \left(\frac{T^2}{12} - \frac{T M_1}{4 \pi} -\frac{M_1^2}{16\pi^2}L_R \right) + \frac{\lambda_{12}}{2} \left(\frac{T^2}{12} - \frac{T M_2}{4 \pi} -\frac{M_2^2}{16\pi^2}L_R\right) \ , \\
    M_{2}^{2} &=  m_{2}^2 + \frac{\lambda_{12}}{2} \left(\frac{T^2}{12} - \frac{T M_1}{4 \pi} -\frac{M_1^2}{16\pi^2}L_R \right) + \frac{\lambda_{2}}{2} \left(\frac{T^2}{12} - \frac{T M_2}{4 \pi} -\frac{M_2^2}{16\pi^2}L_R\right) \ .
\end{align}
Once again this can be generalized to an analytical expression that can be evaluated beyond the high-$T$ regime by replacing one of the $\phi$-derivatives by a partial derivative in Eqn.~\eqref{e.PDtwoscalargapeqn}:
\begin{gather}
    \label{e.HPDtwoscalargapeqn}
    M_i^2 = \left.\frac{\partial}{\partial \phi_i}\frac{d V_\textrm{tree}}{d \phi_i}\right|_{(\phi_1,\phi_2)\rightarrow (\phi,0)} + \left.\frac{\partial}{\partial \phi_i}\frac{d}{d \phi_i} \sum_K
    V_1\left(m_K(\phi_1, \phi_2)\right)
    \right|_{ \scriptsize
    \begin{matrix}
        (\phi_1,\phi_2)\rightarrow (\phi,0) \\  m^2_i \to M^2_i 
    \end{matrix} \ ,
    }
\end{gather}
Derivatives of the mass eigenvalues $m_{K}(\phi_1, \phi_2)$ must be evaluated for general nonzero values of both fields. (This is technically unnecessary for the pure scalar theories we study in this paper, due to the truncation of the gap equation; we will verify in~\cite{HPDfuturecite} whether this is unnecessary for general theories.)

\item Add the missing sunset vacuum graphs to the potential, with the resummed mass inserted:
\begin{equation}
    \nonumber 
    V_\text{sun}^\mathrm{full} =  
    -\frac{1}{12} \includegraphics[valign=c,width=1.5cm]{ProveHPDtwoScalarDiagrams/Tadpoles/VACUUMsunset_RESUM_Scalar111_FULL.pdf} 
    - \frac{1}{4} \includegraphics[valign=c,width=1.5cm]{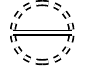} \ .
\end{equation}
The sunset discussion  discussion Section~\ref{s.singlephiHPD} applies verbatim, with the two-scalar zero-temperature sunset given by Eqn.~\eqref{e.sunsetTwoScalatZeroT} in Appendix~\ref{a.potential}, and the thermal pieces computed without any high-temperature approximation in Appendix~\ref{a.sunsetlowT}. 

\end{enumerate}
The final HPD potential is then given by:
\begin{align}
   \nonumber 
    V_\mathrm{eff}^\mathrm{HPD}(\phi_1, &\phi_2 = 0, T) = \ V_\mathrm{tree}(\phi_1, \phi_2 = 0) 
    \\ \nonumber 
    & + \int d\phi_1 \left[ V_{1,\mathrm{zero}}'(\phi_1, \phi_2 = 0, m_i^2 \to M_i^2) + V'_{1,\mathrm{th}}(\phi_1, \phi_2 = 0,T, m_i^2 \to M_i^2)\right] \nonumber 
    \\  
    & +V_\textrm{sun}^\mathrm{full}(\phi_1, \phi_2 = 0, T, m_i^2 \rightarrow M_i^2)  
    \label{e.VfullHPDphi2}
\end{align}
where $M_i^2$ are the solutions to the HPD gap equations Eqn.~\eqref{e.HPDtwoscalargapeqn}, and the full sunset is given by Eqn.~\eqref{e.sunsetTwoScalatZeroT} and  Appendix~\ref{a.sunsetlowT}. 

In the high-temperature expansion, this can again be solved analytically. 
The high-temperature limit of the two-scalar sunset is:
\begin{align}
\label{e.sunset2field} 
    V_\textrm{sun}^\mathrm{zero \times th} &=  
     \frac{(\lambda_1 \phi_1)^2 T^2}{384 \pi^2}  \left( \log\left(\frac{M_1^2}{\Bar{\mu}T}\right) + 1.65 \right) 
\\ &
+ \frac{(\lambda_{12} \phi_1  )^2 T^2}{128 \pi^2}  \left( \log\left(\frac{(M_1 + 2M_2)^2}{9\Bar{\mu}T}\right) + 1.65 \right) \nonumber \ .
\end{align}
Taking the derivative of $V_\mathrm{sun}$ to produce a tadpole for comparison to the PD result then gives
\begin{equation}
     \begin{aligned}
     {V^\mathrm{HPD}_\mathrm{eff}}' = V_\mathrm{tree}' +  & (\lambda_1 \phi_1) \left[\frac{T^2}{24} -\frac{TM_1}{8\pi} - \frac{M_1^2 L_R}{32\pi^2}  \right] + 
     (\lambda_{12} \phi_1) \left[\frac{T^2}{24}  -\frac{TM_2}{8\pi} - \frac{M_2^2 L_R}{32\pi^2} \right] \\
     & \label{e.VprimeHPDtwoscalar}
     +\frac{d}{d \phi_1}\left[
     (\lambda_1 \phi)^2 \frac{T^2}{384 \pi^2}  \left\{ \log\left(\frac{M_1^2}{\Bar{\mu}T}\right) + 1.65 \right\} \right] \\
     &+\frac{d}{d \phi_1}\left[
    (\lambda_{12} \phi)^2 \frac{T^2}{128 \pi^2}  \left\{\log\left(\frac{(M_1 + 2M_2)^2}{9\Bar{\mu}T}\right) + 1.65 \right\}
     \right] \ ,
     \end{aligned}
\end{equation}
which agrees with the partial dressing result Eqn.~\ref{e.VprimePDtwoscalar}. (Note that in evaluating the derivatives on the 2nd and 3rd lines, the $\phi$-dependence of $M_i^2$ will add $\partial M_i^2/\partial \phi_1$ factors in the final tadpole.)

In Appendix~\ref{a.beta2proof}, we demonstrate that HPD reproduces the diagrammatic calculation  at $\mathcal{O}(\beta_{ij} \beta_{kl})$ $\forall \ i,j,k,l$. In particular, the proof in Appendix \ref{s.HPDvsfeynman} explicitly includes the 3‑loop daisy and superdaisy structures, the 3‑loop sunset tadpoles and the 3‑loop lollipop contributions.
In Appendix~\ref{s.HPDvsDR} we also explicitly show that the HPD calculation agrees exactly with the NLO DR calculation at order $\beta^2$, demonstrating the equivalence of DR and HPD in the high-temperature limit. However, unlike DR, the HPD calculation can be easily applied outside of the high-temperature regime. 

\subsubsection{Thermal Error Terms}
\label{sec. TET}
We proved that Hybrid Partial Dressing, as defined above for the two-field case, is accurate to   order $\beta^2$.
In fact, as illustrated in Figure~\ref{fig:HPDdiagrammaticproof}, HPD also correctly includes all  three-loop diagrams at $\mathcal{O}(\gamma^0)$ order, with the exception of the so-called basketball, shown below for the one-field $\phi^4$ case:
\begin{align}
V'_{\includegraphics[valign=c,scale=0.5]{ExtraDiagram/VAC_basketball_subscript.pdf}} \simeq 
-\frac{1}{12}
\includegraphics[valign=c,scale=1]{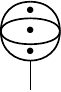} 
&\sim -\frac{1}{12} (\lambda \phi)  \lambda^2   I^\bullet_1(M) \mathcal{H}_4(M,M,M) \nonumber \\ & \approx  \frac{(\lambda \phi) \ \lambda^2 T^3}{4608\pi^3 M} = \frac{\lambda \phi T^2}{54} \alpha^{-1} \beta^3
\label{e.basketball} \ ,
\end{align}
where we neglect one of the overlapping momentum to be able to write the integral as the product of the 4-propagator sunset $\mathcal{H}_4$ integral and a single separate $I_1$ integral.\footnote{The leading contribution to $\mathcal{H}_4$ in the High-$T$ limit comes from zero-modes, meaning that the leading contribution has to be $I^\bullet_1(m)$ to conserve the Matsubara frequency at each vertex. We note that there is no closed form solution of this diagram even for the single $\phi^4$ theory at high temperature~\cite{Andersen:2000zn}.
}
This estimate of the basketball is therefore a good choice to quantify our thermal error term from neglected higher-order contributions in $\beta$.

For the two-scalar theory, 
we can again parametrically estimate the size of the four basketball-like diagrams with a similar estimate as in the Eq.~\eqref{e.basketball}. 
Considering the correct mass-dependence for each diagrams, this defines a thermal error term for our HPD calculation: 
\begin{align}
    \label{e.deltaVbasketballHPD}
    \delta V^{\mathrm{HPD}'}_{\includegraphics[valign=c,scale=0.5]{ExtraDiagram/VAC_basketball_subscript.pdf}} &\simeq  \frac{1}{12}(\lambda_1 \phi) I_1^\bullet(M_1)\left[\lambda_1^2 \mathcal{H}_4(M_1,M_1,M_1)+ 3\lambda_{12}^2  \mathcal{H}_4(M_1,M_2,M_2)\right] \nonumber \\ &+ \frac{1}{12}(\lambda_{12}\phi) I_1^\bullet(M_2)\left[3\lambda_{12}^2\mathcal{H}_4(M_2,M_1,M_1)+\lambda_2^2\mathcal{H}_4(M_2,M_2,M_2)\right] \nonumber \\
    &=\frac{T^3 \phi}{4608 \pi^3} \left[\frac{\lambda_1^3}{ M_1} + \frac{9 \lambda_1 \lambda_{12}^2}{M_1+2M_2} + \frac{9\lambda_{12}^3}{2M_1+M_2}+\frac{\lambda_{12} \lambda_2^2}{M_2}\right] \ ,
\end{align}
where the high-temperature limit was taken in the last line.

As our numerical studies in Section~\ref{sec:numeric} show, this is small for both of our benchmark points. This can be understood from the small expansion parameters $\beta_{ij}$, since parametrically, the error term is roughly 
\begin{align}
\nonumber
\delta V^{\mathrm{HPD}'}_{\includegraphics[valign=c,scale=0.5]{ExtraDiagram/VAC_basketball_subscript.pdf}}  &\sim
\frac{1}{54} (\lambda_{1}\phi \, T^2) \left( \, 
\alpha_{11}^{-1} \beta^3_{11} 
+ 9  
\beta_{12}^2 \max\left[ \alpha_{12}^{-1}  \beta_{12}, \frac12 \alpha_{12}^{-1} \beta_{21} \right] \right) \\
&+ \frac{1}{54} (\lambda_{12}\phi \, T^2) \left( 
9 \beta_{12}^2 \max\left[  \frac12 \alpha_{12}^{-1} \beta_{12}, \alpha_{12}^{-1} \beta_{21} \right] 
+ \alpha_{22}^{-1} \beta^3_{22} \right) \\
&\sim
\frac{\phi T^2}{54} \left[ \left( \lambda_{1}
\left( \alpha_{11}^{-1} \beta^3_{11} + 9  \alpha_{12}^{-1}\beta_{12}^3 \right) \right) 
+\lambda_{12} 
\left( \frac92 \alpha_{12}^{-1} \beta_{12}^3 + \alpha_{22}^{-1} \beta^3_{22} \right)  \right] \ ,
\end{align}
where $\alpha_{ij},\,\beta_{ij}$ are defined in 
Eqns.~(\ref{e.alphaijdefinitionNEW}-\ref{e.betaijdefinitionNEW}) and,  in the final line, we have assumed that $M_2 \gg M_1$ and hence $ \frac{\alpha_{12}^{-1} \beta_{12}}{ 2 \alpha_{12}^{-1} \beta_{21}} = \frac{M_2}{2M_1} > 1$. 
For both of our benchmark points, the largest quartic is $\lambda_{12}$, since that coupling radiatively generates the energy barrier needed for a strong first-order phase transition. The second term in square brackets above will therefore dominate.
Assuming again for simplicity that $m_1^2 = \lambda_{12} T^2/24$ during the phase transition, that second term becomes $\sim 0.01 (\lambda_{12})^{5/2} + 0.002 \lambda_{12}^{1/2} \lambda_{2}^2 (m_1/m_2)$, which is small in the high-temperature regime where the above simplification applies.

\section{RG Improved Hybrid Partial Dressing (RGHPD)} \label{sec:rgi}

In this section, we demonstrate that the one- and two-field HPD effective potentials in Equations~\eqref{e.VfullHPDphi1} and~\eqref{e.VfullHPDphi2} can be almost trivially RG-improved to define RGHPD,  simply by substituting the running couplings and evaluating them at $\mu_R^2(\phi) \sim (\pi T)^2 + M^2(\phi)$. This can be understood as interpolating between the dominant scales of the dynamics generating the effective potential in the high- and low-temperature regimes. We can then make the following conclusions:
\begin{itemize}
    \item RGHPD maintains a small $\mathcal{O}(\tilde \beta^3)$ error term, while also displaying a small scale variation $\mu \ d V'/d\mu = \mathcal{O}(\lambda^3)$, as expected from a perturbative calculation that captures all important two-loop effects in each regime.  
    \item RGHPD equivalent to NLO DR in the high-temperature regime. To the best of our knowledge, this is the first time that this level of accuracy has been formally and numerically demonstrated for an RGI 4D perturbative resummation scheme.\footnote{We comment briefly on earlier results for the now-defunct OPD procedure in Appendix~\ref{a.OPD}.} 
    The error terms of RGI Parwani are one order worse, at $\mathcal{O}(\beta^2, \lambda^2)$.
    \item Unlike NLO DR, RGHPD is valid at all temperatures, maintaining its small $\mathcal{O}(\tilde \beta^3)$ error term as $T/\phi \to 0$ and the potential exactly approaches the RG-improved two-loop zero-temperature Coleman-Weinberg potential. This makes it uniquely suited for the general study of both weak and strong first order phase transitions.

    \item As discussed in Section~\ref{sec:RGHPDdiagrammaticproof}, the small scale variation of RGHPD shows that RG-improvement does not introduce any overcounting mistakes, since the the important hard thermal loop contributions necessitating thermal resummation are parametrically distinct from the leading-log contributions resummed by RG improvement.

\end{itemize}

We briefly review multi-loop RG-improved effective potentials at zero temperature in Section~\ref{s.RGIzeroT}. 
This serves as a pedagogical review (since there are some well-known subtleties at higher loop order), establishes notation, explicitly demonstrates the $\mathcal{O}(\lambda^3)$ residual scale variation that we would like to maintain at finite temperature, and shows how the choice of $\mu_R$ which minimizes the scale variation is also the natural choice for the scale of the dominant physics generating the effective potential. This makes it obvious how to choose the scale in the finite-temperature case. 

We then study RG-improvement at finite temperature in Section~\ref{s.RGIfiniteT}: RGHPD in the one- and two-field case is discussed in Sections~\ref{s.singlescalarRGI} and~\ref{s.twoscalarRGI}, and compared to RGI Parwani in Section~\ref{s.RGIParwani} and NLO DR  in Section~\ref{s.RGINLODR}.

\subsection{Multi-Loop RG Improvement at Zero-Temperature}
\label{s.RGIzeroT}

We now review basic RG improvement in zero-temperature potentials, to demonstrate the parametric form of the residual scale variation for different choices of the fixed-order loop expansion and the order of RGEs used. This allows us to simply demonstrate in the next section that the finite-temperature HPD potential, once RG-improved in exactly the same way, has a residual scale variation that is of exactly the expected parametric size, clearly showing that the thermal resummation procedure, once consistently implemented in HPD, does not interfere with RG improvement in any way.

We consider first the single-scalar theory. Generalization to multiple scalars is mostly straightforward. 
In order to assess the scale variation of the effective potential, we formulate the Callan-Symanzyk equations carefully at each order. Define the operator
\begin{gather}
    \hat{\Gamma}^{(j)} \equiv \beta^{(j)} \partial_\lambda+\beta_{\nu^2}^{(j)} \nu^2 \partial_{\nu^2}-\gamma_\phi^{(j)} \phi \partial_\phi \ + \beta_\Lambda^{(j)} .
\end{gather}
where superscripts in round brackets indicate loop order (e.g. $\beta_\lambda^{(3)}$ is a three-loop beta function),
and we define $\Lambda$ to be the field-independent constant piece of the full potential.

The exact Callan-Symanzyk equation is simply $\mu \frac{d}{d \mu}\sum_{N=0}^\infty V^{(N)} = 0$. 
We can break this down at each separate loop order as
\begin{gather}
\mathrm{CS}_N
\equiv 
\mu \partial_\mu V^{(N)}+\sum_{k=0}^N \hat{\Gamma}^{(N-k)} V^{(k)}=0 \ ,
\label{eq.VexactRG}
\end{gather}
where each of the above two terms is $N$-loop order, defined not in reference to a particular coupling but by counting loop factors $\ell \equiv (1/16\pi^2)$ in scalar theories.

Note that, because $\hat \Gamma^{(0)} = 0$, the direct scale variation of the $N$-loop order potential (the first term) is canceled by the RG-improvement of the lower-order potential terms (the sum in the second term).
Explicitly, up to third loop order we have
\begin{eqnarray}
    \mathrm{CS}_1  &=& \hat{\Gamma}^{(1)} V^{(0)} + \mu \partial_\mu V^{(1)} =0\\
    \mathrm{CS}_2  &=& \hat{\Gamma}^{(1)} V^{(1)}+\hat{\Gamma}^{(2)} V^{(0)} + \mu \partial_\mu V^{(2)} = 0\\
    \mathrm{CS}_3 &=& 
    \hat{\Gamma}^{(1)} V^{(2)}+\hat{\Gamma}^{(2)} V^{(1)}+ \hat{\Gamma}^{(3)} V^{(0)} +  \mu \partial _\mu V^{(3)} = 0
    \label{e.orderbyorderCS}
\end{eqnarray}

The first step in RG-improving the effective potential is substituting the running parameters into the potential, which solves the Callan-Symanzyk equations up to a residual determined by $R$, the order of RG-equations used, and $L$, the loop order to which the effective potential is evaluated. The above loop-by-loop breakdown of the CS equations makes obvious what the size of the residual scale variation is for each choice of $R, L$: 
\begin{itemize}
    \item $L = 1, R = 0$: the fixed order one-loop potential has residual $-\hat{\Gamma}^{(1)} V^{(0)}$, corresponding to one-loop leading log scale variation, which we colloquially write as $\sim (\ell \ \mathrm{log})$. 
    \item $L = 0, R = 1$: substituting the one-loop running couplings into the tree-level potential gives you a dominant residual of $-\mu \partial_\mu V^{(1)}$, meaning the scale variation is one-loop order $\sim (\ell)$ with leading logs resummed.
    \item $L = 1, R = 1$: the dominant residual is the two-loop order $-(\hat{\Gamma}^{(2)} V^{(0)} + \mu \partial_\mu V^{(2)}) \sim (\ell^2 \log) + (\ell^2)$, with the first term referring to subleading logs, which could still spoil perturbativity if the logs are very large.
    \item $L = 1, R = 2$: the canonical method to RG-improve is to use beta functions at one higher order than the fixed-order potential, since the dominant residual of $-\mu \partial_\mu V^{(2)} \sim (\ell^2)$ no longer includes any potentially large logs.
    \item $L=2, R = 2$: this describes our two-loop HPD calculation, which we will RG-improve with two-loop RGEs. The dominant residual will be 
    $\hat{\Gamma}^{(3)} V^{(0)} +  \mu \partial _\mu V^{(3)} \sim (\ell^3 \log) + (\ell^3)$, corresponding to three-loop sub-sub-leading log size scale variation. 
\end{itemize}

Let us explicitly demonstrate this in the single-scalar case by substituting the two-loop running couplings (see Appendix~\ref{a.RG}) into the two-loop effective potential (see Appendix~\ref{a.potential}). Defining $L_m = \log m/\mu$ and the Claussen number Cl$_2 \approx 1.0149 $, the total scale variation is
\begin{align}
   \left. \mu \frac{\sum_{k = 0}^2V^{(k)}}{d\mu}\right|_\textrm{2-loop RG} &= \frac{\lambda^2}{(16\pi^2)^3}\left[\frac{1}{32} L_m^2\left(27 \lambda^2 \phi^4+56 \lambda \nu^2 \phi^2+20 \nu^4\right)\right. \nonumber\\  &\left.-\frac{1}{48} L_m\left(143 \lambda^2 \phi^4+296 \lambda \nu^2 \phi^2+68 \nu^4\right)\right. \nonumber \\ &\left.+\frac{1}{96}\left(277 \lambda ^2 \phi^4 + 568\lambda \nu^2 \phi^2 + 76 \nu^4 - 72\sqrt{3}\textrm{Cl}_2\left(\lambda^2\phi^4 + \frac{14}{9} \lambda \nu^2 \phi^2\right)\right)\right] ,
\end{align}
which is $\sim \ell^3 \log$ as required. For the single-scalar theory we can use the three-loop beta functions~\cite{Chung:1999xm} to demonstrate the vanishing of the sub-subleading log scale variation for $L = 2, R = 3$:

\begin{align}
   \left. \mu \frac{\sum_{k = 0}^2V^{(k)}}{d\mu}\right|_\textrm{3-loop RG} &=   \frac{\lambda^3}{(16\pi^2)^3} \left[\frac{1}{96}\left(277 \lambda ^2 \phi^4 + 568\lambda \nu^2 \phi^2 + 76 \nu^4\right) \right. \nonumber\\ & \left. - \frac{72\sqrt{3}}{96}\textrm{Cl}_2\left(\lambda^2\phi^4 + \frac{14}{9} \lambda \nu^2 \phi^2\right)  + \frac{29}{16}\frac{\lambda^3 \nu^2 \phi^2}{(16\pi^2)^3} + \left(\frac{49}{64} + \frac{\zeta_3}{2}\right) \frac{\lambda^4 \phi^4}{(16\pi^2)^3} \right]\nonumber \\ &= \frac{\lambda^3}{(16\pi^2)^3}\left(\nu^2 \phi^2 \left(\frac{371}{48} - \frac{7}{2\sqrt{3}}\textrm{Cl}_2\right)+ \lambda \phi^4 \left(\frac{701}{192}-\frac{3\sqrt{3}}{4} \textrm{Cl}_2 + \frac{\zeta_3}{2}\right)\right)
\end{align}
which as expected is $\sim \ell^3$.

The second step in the RG-improvement is the choice of the evaluation scale $\mu_R$ at which the running couplings are evaluated  in the potential. Intuitively, it makes sense that this $\mu_R$ should be at the scale of whatever dynamics generates the effective potential, i.e. the mass of fields in vacuum bubble loops. 
This is the correct choice, which can be seen in two ways. 

On the one hand, even after replacing the couplings with $\lambda(\mu_R)$, $\nu^2(\mu_R)$, the loop corrections still contain powers of explicit logarithms of the form $\log\big(m^2(\phi)/\mu_R(\phi)^2\big)$, which are minimized for $\mu_R^2 \sim m^2(\phi)$. Such a parametric choice therefore optimizes the convergence of our perturbative expansion.
On the other hand, one could explicitly require that the leading log terms in the effective potential to all orders is reproduced exactly by the running couplings, and this uniquely determines $\mu_R(\phi)^2 = m^2(\phi)$ (where one can use either tree-level or loop-corrected masses, and we will do the latter since this generalizes more obviously to finite temperature, see below).

In practice, any choice $\mu_R(\phi) = a  m(\phi)$, for $a \sim \mathcal{O}(1)$, gives a result that maintains leading-log accuracy even when $\lambda \log m^2(\phi)/\mu_0^2$ becomes large. 
Changing the coefficient $a$ introduces differences into the effective potential that are of the same order as higher-loop terms at subleading log order, which are not captured by RG improvement anyway. This makes it possible to assess the importance of these neglected higher-loop terms, and hence estimate the theory uncertainty on the calculation, by simply evaluating the potential with $\mu_R(\phi) = a m(\phi)$ for different choices of the coefficient $a$.

All of the above discussion generalizes straightforwardly to the two-scalar model.
The effective two-loop potential is again RG improved by substituting $\mu_0 \to \mu_R, \lambda_i \to \lambda_i(\mu_R)$ and analogously for the remaining parameters using two-loop RGEs from Appendix~\ref{a.RG}.
A complication arises in how exactly to define $\mu_R(\phi)$, since there is no longer a unique mass scale which generates the effective potential. 
In the fully general case with hierarchical masses, one has to carefully integrate out each degree of freedom and match the effective potentials at each threshold, see ~\cite{Manohar:2020nzp}. 
For simplicity, we adopt the common approximation of simply setting  $\mu_R(\phi) = a \ m_i(\phi_1)$, where $i$ is the ``more important'' field for generating the effective potential. For the cases we study, this is $\phi_2$ due to the sizable $\lambda_{12}$ coupling required for inducing a strong phase transition:
\begin{equation}
    V_\mathrm{eff}^\mathrm{RGzero}(\phi_1, \phi_2) \equiv V_\mathrm{eff}^\mathrm{zero}(\phi_1, \phi_2; \lambda_1 \to \lambda_1(\mu_R), \ldots, \mu_0 \to \mu_R = a m_2(\phi)) \ .
\end{equation}
This will introduce theoretical errors for the contributions from other fields, but we assume that their smaller coupling makes the resulting mistake sub-dominant compared to the scale variation of the contributions from the most important coupling. We leave the question of how to adapt the methods of~\cite{Manohar:2020nzp} to RG improve finite-temperature HPD effective potentials with hierarchical masses for future work.

\begin{figure}%
    \centering
    {{\includegraphics[width=9cm]{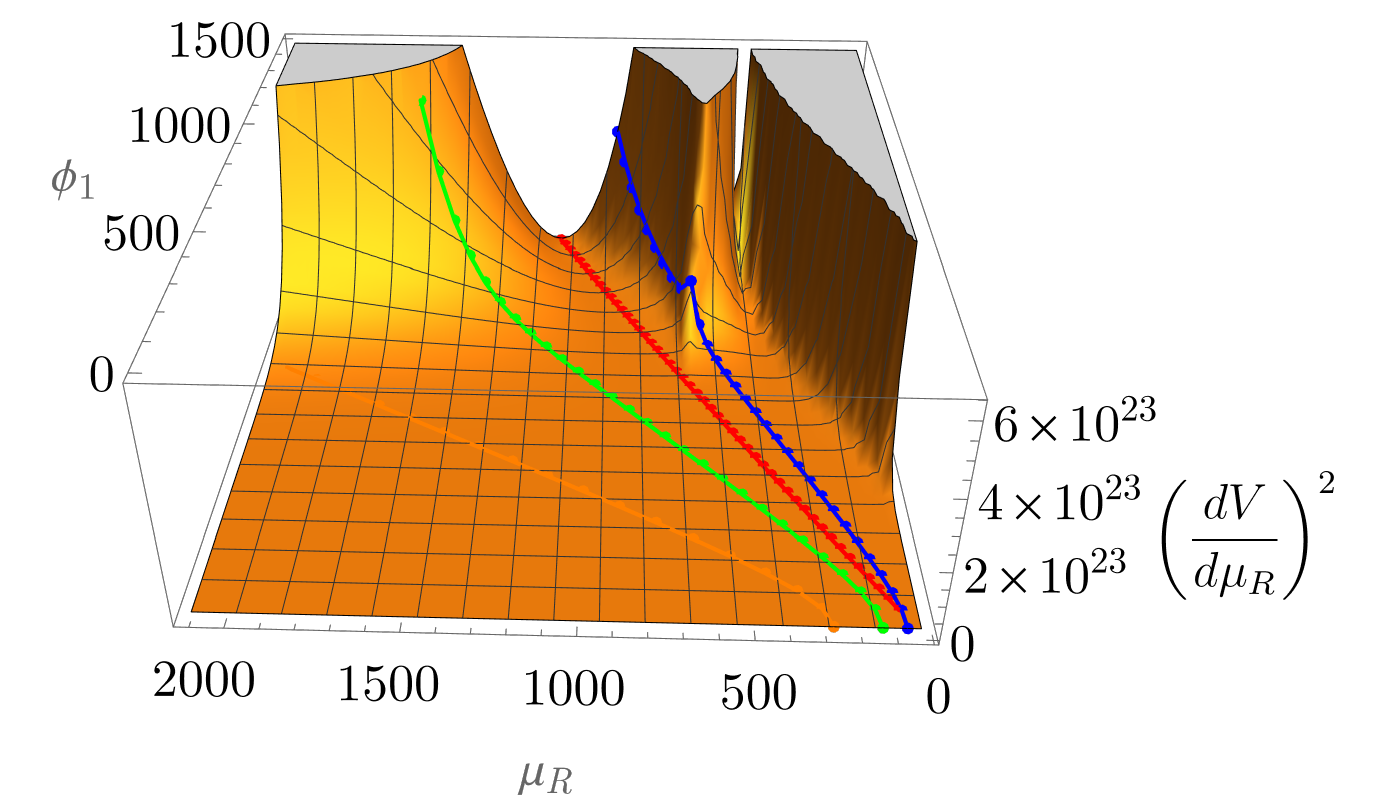} }}%
    {{\includegraphics[width=5.5cm]{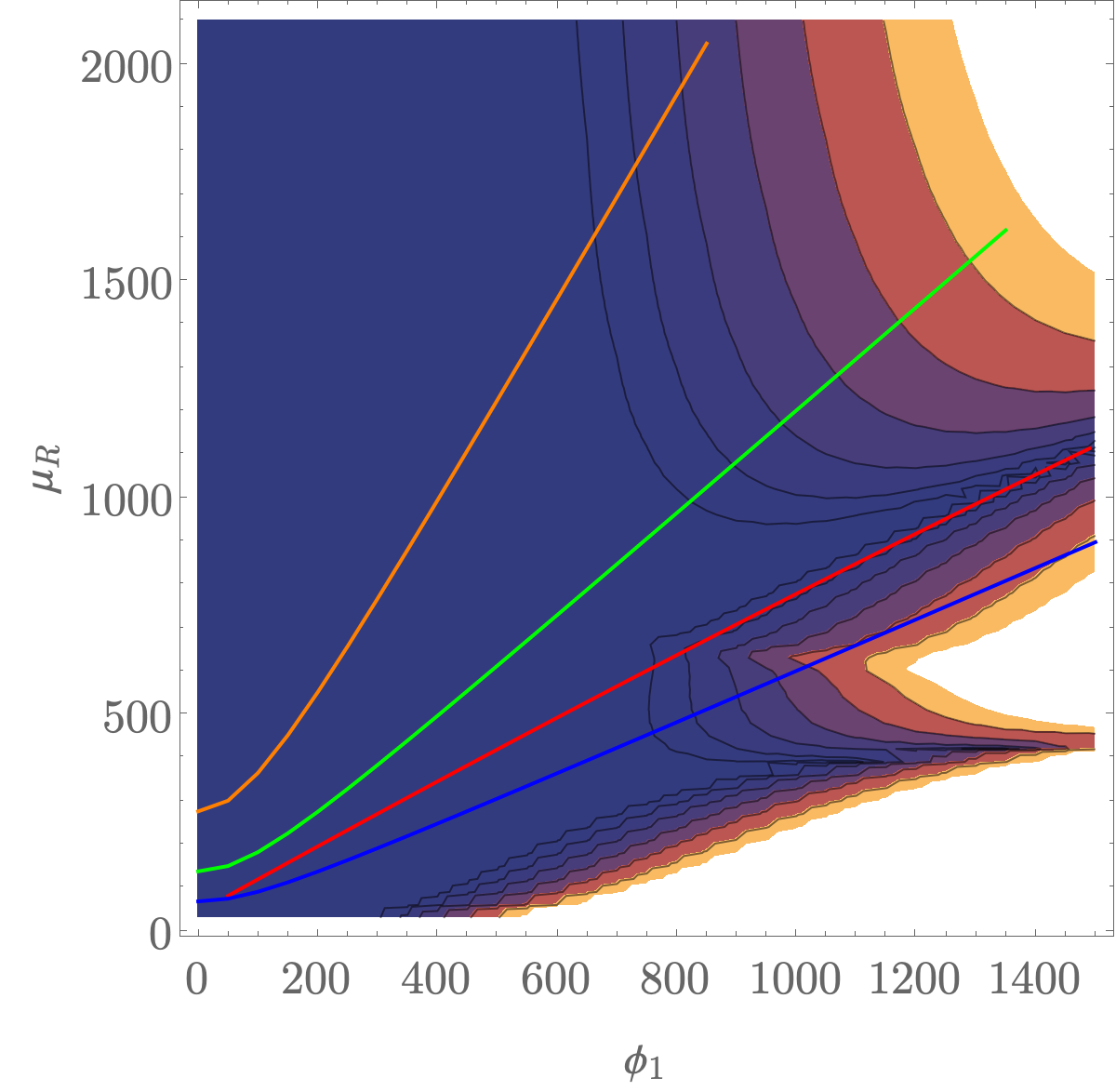} }}%
    \caption{We show the scale variation of the two-loop effective potential $(dV_\mathrm{eff}/d\mu_R)^2$ with two-loop RG improvement at zero temperature, as a function of $\phi_1$ and $\mu_R$, for Benchmark Point 2 (all energy units are GeV).
    The red line is the choice of $\mu_R(\phi)$ which exactly minimizes the scale variation, while the blue, green and orange lines correspond to $\mu_R = \frac{1}{2}m_2(\phi),m_2(\phi),2m_2(\phi)$  respectively.
    }%
    \label{fig:3D_zeroT}%
\end{figure}

In general, this choice of scale is a good assumption as long as the different particle masses are not too hierarchical and the subdominant couplings are sufficiently small. It appears to be satisfied for our BP1 benchmark point, but it is strained for BP2 due to a larger mass hierarchy. 
Even so, it is sufficient for our initial numerical studies. 
We show in Fig.~\ref{fig:3D_zeroT} a plot of the two-loop $(dV_\mathrm{eff}/d\mu_R)^2$ with two-loop running couplings, as a function of $\mu_R$ and $\phi$ for BP2 in the two-scalar model, at zero-temperature.
This clearly shows the expected valley of minimum scale variation running proportional to $\mu_R \sim m_2(\phi)$, demonstrating explicitly that choosing $\mu_R$ to be the energy scale of the physics which generates the effective potential also minimizes scale variation. This is well-known, but will uniquely determine (up to the usual $\mathcal{O}(1)$ factor) the choice of $\mu_R$ in the high-temperature case of the next section.

 \subsection{RG Improvement at Finite Temperature}
\label{s.RGIfiniteT}

We now study RG-improvement of the finite-temperature HPD effective potential, and compare it to NLO DR and Parwani. Note that in this context, we focus on the potential tadpole $V' = dV/d\phi$, since thermal potentials have temperature-dependent but $\phi$-independent terms that can have lower-order scale variation,  but which do not impact phase transition observables.

\subsubsection{RGHPD for single-scalar case}
 \label{s.singlescalarRGI}

Having shown that the HPD effective potential is accurate to $\mathcal{O}(\tilde \beta^2)$ (effectively two-loop) at all temperatures, 
we can now show that RG improving HPD is an almost trivial generalization of the zero-temperature case, except for the choice of renormalization scale:
\begin{equation}
    \label{e.VeffRGHPDonephi}
    V_\mathrm{eff}^\mathrm{RGHPD}(\phi, T) \equiv V_\mathrm{eff}^\mathrm{HPD}(\phi, T; \lambda \to \lambda(\mu_R), \ldots, \mu_0^2 \to \mu_R^2 )= (b \pi T/2)^2 + (a M(\phi))^2 ) \ .
\end{equation}
For simplicity, let us start with one-loop RGEs, and then we will also include two-loop to see what changes.
To show small scale variation, we consider the tadpole
\begin{gather}
    {V'_\mathrm{1+sun}} = \frac{\lambda \phi}{2} I_{1,\mathrm{th}}(M^2) + \frac{\lambda \phi M^2}{32\pi^2}\left(\log\frac{M^2}{\mu^2}-1\right) + \frac{d}{d\phi}(V_\textrm{sun}^\mathrm{full}(M^2)) \ .
\end{gather} 
$M^2$ is defined by the truncated HPD gap equation~\eqref{e.HPDMsqgapeqnbeyondhighT}, which we write suggestively without evaluating the thermal integral to make the derivation general for all temperatures:
\begin{gather}
    \label{e.HPDRGIgapeqn}
    M^2 = \nu^2 + \frac{\lambda \phi^2}{2}+ \frac{\lambda M^2}{32\pi^2}\left(\log\frac{M^2}{\mu^2}-1\right)+ \frac{\lambda}{2} I_{1,\mathrm{th}} (M^2) \ ,
\end{gather}
where $I_{1,\mathrm{th}} = T^4/(\pi^2) \frac{\partial J_B}{\partial m^2}$.
We first notice that
\begin{gather}
    \mu \frac{dM^2}{d\mu} \simeq \frac{\lambda^2 \phi^2}{16\pi^2} + 3 \frac{\lambda^2}{32\pi^2}I_{1,\mathrm{th}}(M^2)+ \mathcal{O}(\lambda^3)= \mathcal{O}(\lambda^2) \ .
\end{gather}
Let us now apply a CS operator to the tadpole:
\begin{align}
    \mu \frac{d{V_\mathrm{eff}^\mathrm{RGHPD}}'}{d\mu} =& \left(\beta_{\nu^2}\frac{\partial}{\partial \nu^2}+\beta_{\lambda}\frac{\partial}{\partial \lambda}\right)V'_\text{tree} +\left(\beta_{\nu^2}\frac{\partial}{\partial \nu^2}+\beta_{\lambda}\frac{\partial}{\partial \lambda}+ \mu \frac{\partial}{\partial \mu}\right)  V_{1+\mathrm{sun}}'
\end{align}
which gives
\begin{align} \label{e.CSsinglescalarfiniteTbreakdown}
    \mu \frac{d{V_\mathrm{eff}^\mathrm{RGHPD}}'}{d\mu}  =& \ \frac{\lambda \nu^2 \phi}{16\pi^2}+ \frac{\lambda^2\phi^3}{32\pi^2}  \\
    &+\frac{3\lambda^2 \phi M}{32\pi^2} \left(\frac{I_{1,\mathrm{th}}(M^2)}{M}\right)-\frac{\lambda \phi M^2}{16\pi^2}   \nonumber 
    \\&+\left(\beta_{\nu^2}\frac{\partial}{\partial \nu^2}+\beta_{\lambda}\frac{\partial}{\partial \lambda}\right) \left[\frac{\lambda M^2}{32\pi^2}\left(\log\frac{M^2}{\mu^2}-1\right)\right]\nonumber\\&+ \underbrace{\lambda \phi \left(\mu \frac{dM^2}{d\mu}\right)\left(-\frac{1}{2}I_{2,\mathrm{th}}(M^2)\right)}_{\mathcal{O}(\lambda^3)}
    + \mu\frac{\partial}{\partial \mu}\left(\frac{dV_\textrm{sun}^\mathrm{full }(M^2)}{d\phi}\right) \ . \nonumber 
\end{align}
where, from Appendix~\ref{a.sunsetlowT}, we have:
\begin{gather}
    \mu\frac{\partial}{\partial \mu}\left(\frac{dV_\textrm{sun}^\mathrm{full }(M^2)}{d\phi}\right)=     \mu\frac{\partial}{\partial \mu}\left(\frac{dV_\textrm{sun}^\mathrm{zero }(M^2)}{d\phi}\right)-\frac{\lambda^2\phi}{16\pi^2}I_{1,\mathrm{th.}}
    \label{e.scaleVarSunset}
\end{gather}
The third line of Eqn.~\eqref{e.CSsinglescalarfiniteTbreakdown} exactly cancels the first term in Eqn.~\eqref{e.scaleVarSunset}, up to  constant pieces in $\mu$ at order $1/\pi^4$, which are higher order.
Now, substituting $M^2$ directly from Eqn.~\eqref{e.HPDRGIgapeqn} (i.e. the first iteration of the gap equation), we get an exact cancellation between 
the second term of 
Eqn.~\eqref{e.scaleVarSunset}
and the second line of 
Eqn.~\eqref{e.CSsinglescalarfiniteTbreakdown}, giving a small $\mathcal{O}(\lambda^3/\pi^3)$ scale variation of the effective potential:
\begin{equation}
\mu \frac{d{V_\mathrm{eff}^\mathrm{RGHPD}}'}{d\mu} = \frac{3 \lambda^3 \phi}{64\pi^2} I_{1,\mathrm{th}}(M^2)I_{2,\mathrm{th}}(M^2) \stackrel{\text{High-$T$}}{=} \frac{\lambda^3 T^3 \phi}{2048 \pi^3 M}  \sim \mathcal{O}(\lambda^3/\pi^3) \ .
\label{e.singlefieldscalevariation}
\end{equation}
During the phase transition, $\beta \sim \lambda T/M \sim \sqrt{\lambda}$, meaning the above scale variation is $\sim (\lambda T^2) \beta^3 \phi$. Since we defined our $\beta$-expansion relative to the dominant thermal mass correction $\sim \lambda T^2$, this just tells us that the scale variation of the RG-improved potential is the same size as the fixed-order error term $\beta^3$, meaning it does not clash with the HPD thermal resummation procedure.

The zero-temperature scale variation for a two-loop potential with one-loop RGEs is expected to be two-loop sub-leading log size, i.e. $\sim (1/16\pi^2)^2 \log(M/\mu)$. By $\pi$-counting, this is subdominant to the dominant high-temperature scale variation above. It is now obvious what the addition of two-loop RGEs will buy us. The zero-temperature residual is now simply $\sim (1/16\pi^2)^2$, but this is even more subdominant to the leading thermal piece above.

The RGHPD effective potential automatically recovers the $T \to 0$ limit from Section~\ref{s.RGIzeroT}, provided $\mu_R \propto m(\phi)$ as $\phi \to \infty$, since the thermal contributions decouple and the gap equation does nothing for small $T$ beyond generating the figure-8 vacuum diagram. On the other hand, in the high-$T$ regime, it makes both physical sense and has been demonstrated in DR calculations~\cite{Gould:2021dzl,Croon:2020cgk} that the renormalization scale should be $\mu_R \sim \pi T$, since the potential is generated by thermal effects and all $\log(T/m)$ factors combine with $\log (m/\mu)$  to leave only $\log(T/\mu)$ terms surviving in the high-temperature expansion (at least at two-loop order).

We therefore choose $\mu_R(\phi)$ as the benchmark central choice of scale such that in the region where high-$T$ is a good approximation, the scale is $\sim \pi T$ and in the high-$T$ regime goes as $m(\phi)$. Thus we choose a smooth interpolation between two regimes, $\mu_R^2 = (\pi T/2)^2 + M^2(\phi)$, while scale variation can be explored by multiplying the temperature and mass term by varying $\mathcal{O}(1)$ coefficients $b, a$ as shown in Eqn.~\eqref{e.VeffRGHPDonephi}. In the next section, we demonstrate numerically how closely this physically motivated choice tracks the optimal $\mu_R(\phi)$ for minimizing the scale variation.

\subsubsection{RGHPD for multiple scalars}
 \label{s.twoscalarRGI}

\begin{figure}%
    \centering
    {{\includegraphics[width=9cm]{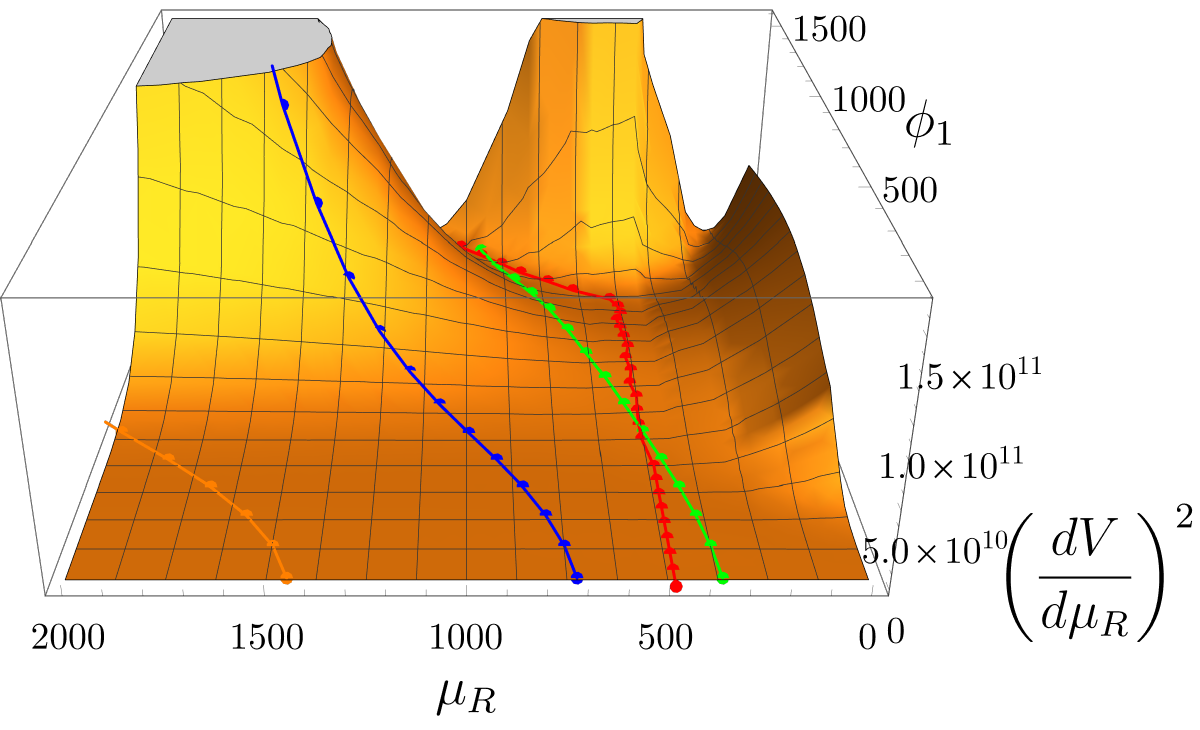} }}%
    {{\includegraphics[width=6cm]{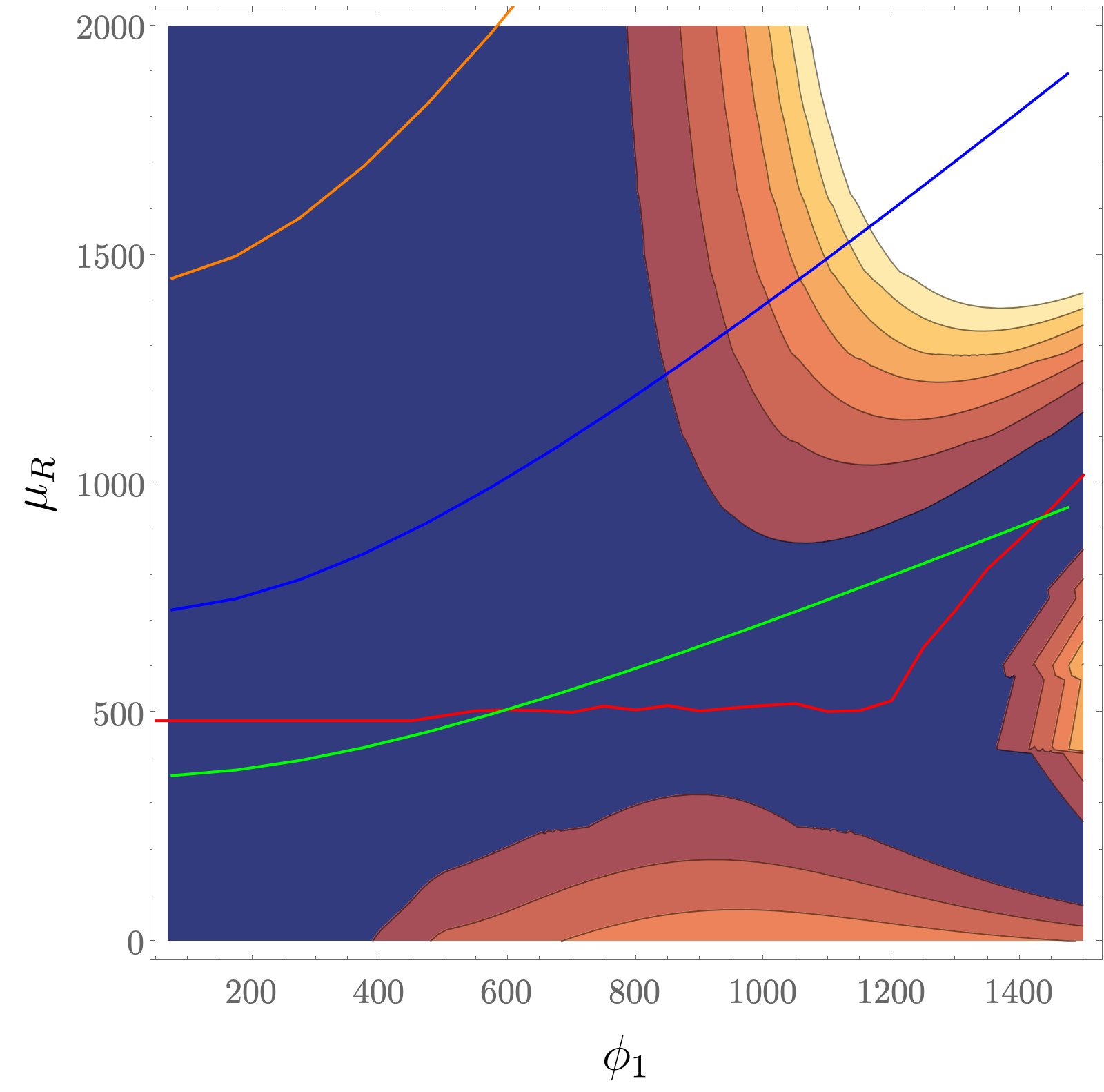} }}%
    \caption{
    \label{fig:3D_finiteT}
    We show the scale variation $(dV_\mathrm{eff}^\mathrm{HPD}/d\mu_R)^2$ of the RGHPD effective finite-temperature potential as a function of $\phi$ and $\mu_R$ for BP2, with $T = T_c= 439.97 \textrm{GeV}$ . We show $(dV_\mathrm{eff}^\mathrm{HPD}/d\mu_R)^2$ both in a 3D plot (left) and contour plot (right). The dashed line is the $\mu_R(\phi)$ of minimum scale variation, while the green, blue and orange lines corresponds to $\mu_R = \{\frac{1}{2},1,2\} \times\sqrt{(\pi T/2)^2+M_2^2(\phi_1)}$, respectively.
    Observe that the dashed line tracks the temperature at low VEV (higher temperatures) and is proportional to $M_2$ at higher VEV (low temperature limit). 
     }%
\end{figure}

The generalization to multiple scalars is trivial:
\begin{equation}
    \label{e.VeffRGHPDtwophi}
    V_\mathrm{eff}^\mathrm{RGHPD}(\phi_1, \phi_2 = 0, T) \equiv V_\mathrm{eff}^\mathrm{HPD}(\phi_1, \phi_2 = 0, T; \lambda_1 \to \lambda_1(\mu_R), \ldots, \mu_0 \to \mu_R ) \ ,
\end{equation}
where $\mu_R^2 =  (b \pi T/2)^2 +(a M(\phi))^2$.  The beta functions for our two-$\phi$ benchmark model are given in Appendix~\ref{a.RG}. The tadpole of the one-loop potential can be derived from Eqn.~\eqref{e.VfullHPDphi2}, setting $\phi_2=0$:
\begin{align}
  {V_\mathrm{eff}^\mathrm{RGHPD}}' =&\, V_{1,\mathrm{zero}}' (M_1) + V_{1,\mathrm{zero}}' (M_2) + V_{1,\mathrm{th}}'(M_1) + V_{1,\mathrm{th}}'(M_2) +V_\text{sun} (M_1,M_2)\nonumber\\
    =&\, \frac{\lambda_1 \phi_1 M_1^2}{32\pi^2}\left(\log\frac{M_1^2}{\mu_R^2}-1\right)+ \frac{\lambda_{12} \phi_1 M_2^2}{32\pi^2}\left(\log\frac{M_2^2}{\mu_R^2}-1\right) \nonumber\\ &+ \frac{\lambda_1 \phi_1}{2} I_{1,\mathrm{th}}(M_1) + \frac{\lambda_{12}\phi_1}{2} I_{1,\mathrm{th}}(M_2) +V_\textrm{sun}^\mathrm{full} (M_1,M_2)\ .
\end{align}
We again write the HPD gap equation in terms of explicit thermal integrals:
\begin{align}
    M_1^2 =&\, \nu_1^2 + \lambda_1 \frac{\phi_1^2}{2} + \frac{\lambda_{1}  M_1^2}{32\pi^2}\left(\log\frac{M_1^2}{\mu_R^2}-1\right)+\frac{\lambda_{12}  M_2^2}{32\pi^2}\left(\log\frac{M_2^2}{\mu_R^2}-1\right)\nonumber\\
    &+ \frac{\lambda_1}{2} I_{1,\mathrm{th}}(M_1)+ \frac{\lambda_{12}}{2} I_{1,\mathrm{th}}(M_2) \ , \\
    M_2^2 =&\, \nu_2^2 + \lambda_{12} \frac{\phi_1^2}{2} + \frac{\lambda_{12}  M_1^2}{32\pi^2}\left(\log\frac{M_1^2}{\mu_R^2}-1\right)+\frac{\lambda_{2}  M_2^2}{32\pi^2}\left(\log\frac{M_2^2}{\mu_R^2}-1\right)\nonumber\\
    &+ \frac{\lambda_2}{2} I_{1,\mathrm{th}}(M_2)+ \frac{\lambda_{12}}{2} I_{1,\mathrm{th}}(M_1) \ .
\end{align}
An exact repeat of the scale variation analysis from the single scalar case, using Appendix~\ref{a.sunsetlowT},  yields:
\begin{eqnarray}
    \mu_R \frac{d {V_\mathrm{eff}^\mathrm{RGHPD}}'}{d \mu_R} &=&
    I_{1,\mathrm{th}}\left(M_1\right) \left[\frac{3 \lambda _1^3 I_{2,\mathrm{th}}\left(M_1\right)}{64 \pi ^2}+\frac{3 \lambda _{12}^2 \lambda _1 I_{2,\mathrm{th}}\left(M_1\right)}{64 \pi ^2} + \right.
    \\ \nonumber 
    &&\left. \ \ \ \ \ \ \ \ \ \  \ \ \ \ \  \ \ 
    \frac{\lambda _{12}^2 \lambda _1 I_{2,\mathrm{th}}\left(M_2\right)}{64 \pi ^2}+
    \frac{\lambda _{12}^3 I_{2,\mathrm{th}}\left(M_2\right)}{16 \pi ^2}+\frac{\lambda _2 \lambda _{12}^2 I_{2,\mathrm{th}}\left(M_2\right)}{64 \pi ^2}\right] + 
    \\ \nonumber 
    && I_{1,\mathrm{th}}\left(M_2\right) \left[\frac{3 \lambda _{12}^3 I_{2,\mathrm{th}}\left(M_2\right)}{64 \pi ^2}+\frac{\lambda _1 \lambda _{12}^2 I_{2,\mathrm{th}}\left(M_1\right)}{16 \pi ^2}+
    \right.
    \\
    &&
    \left. \ \ \ \ \ \ \ \ \ \  \ \ \ \ \ \ \ 
    \frac{\lambda _1^2 \lambda _{12} I_{2,\mathrm{th}}\left(M_1\right)}{64 \pi ^2}+
    \frac{\lambda _1 \lambda _2 \lambda _{12} I_{2,\mathrm{th}}\left(M_1\right)}{64 \pi ^2}+\frac{3 \lambda _2^2 \lambda _{12} I_{2,\mathrm{th}}\left(M_2\right)}{64 \pi ^2}\right]\nonumber
    \\ \nonumber
    &\sim& \mathcal{O}(\lambda^3/\pi^3) \ ,
\end{eqnarray}
where in the last line we take $\lambda$ to include all types of quartic couplings.
The exact analogue of the single-field discussion below Eqn.~\eqref{e.singlefieldscalevariation} applies here as well. 
It is also clear that this reproduces the zero-temperature RGI Coleman-Weinberg potential, provided $\mu_R^2 =  (b \pi T/2)^2 +(a M(\phi))^2$,
where variation of the result for different values of $a, b \sim \mathcal{O}(1)$ will give an estimate of the theory uncertainty from scale variation. 
To demonstrate this choice of scale correctly interpolates between the scale of the dominant dynamics for the effective potential, we show scale variation for the BP2 benchmark point in the $\phi$-$\mu$ plane in Fig.~\ref{fig:3D_finiteT}.
We see that the $\mu(\phi)$ of minimum scale variation almost tracks  our choice of $\mu_R$ at small and high $\phi$. For intermediate values, there is modest deviation due to the hierarchy of scales $M_2 \gg M_1$, but this would be ameliorated by applying the methods of~\cite{Manohar:2020nzp} to RGHPD.

\subsubsection{Comparison with RG-improved Parwani resummation}
\label{s.RGIParwani}

For simplicity we limit our review of this analytical result  to the single-scalar case, and refer the reader to~\cite{Gould:2021oba, Curtin:2022ovx} for the analogous two-scalar derivation. We also focus on the most common Parwani resummation. Arnold-Espinosa and Full Dressing give parametrically equivalent results. 

Naively RG-improving  the Parwani resummed effective potential with $\overline{MS}$ running couplings will not correctly capture the log-structure at 2-loop level.
We perform the substitution 
$m^2(\phi)\rightarrow   m^2(\phi) + \frac{\lambda T^2}{24}$ in the thermal effective potential, substitute running couplings to obtain $V_\mathrm{eff}^\mathrm{RGPar}$, and apply the CS equation to observe
\begin{gather}
\label{e.dVdmuParwani}
    \mu \frac{d V_\mathrm{eff}^\mathrm{RGPar} (M^2)}{d\mu} \supset \frac{3\lambda^2}{16\pi^2}\frac{ \phi^2}{48}-\frac{\lambda^2}{16\pi^2}\frac{T^2 \phi^2}{48}-\frac{3}{2}\frac{MT}{12 \pi} \left(\frac{\lambda \nu^2}{16\pi^2}+ \frac{3\lambda^2 \phi^2}{32\pi^2}+ \frac{\lambda^2 T^2}{128\pi^2}\right) \ ,
\end{gather}
which clearly does not cancel, giving an $\mathcal{O}(\lambda^2)$ scale variation. Naive RG improvement of the Parwani effective potential therefore does not improve the accuracy  of the calculation.

Even though this is not commonly performed in Parwani or other non-partial-dressing schemes, it is interesting to examine whether adding the missing sunset potential contributions can ameliorate this large scale variation:
\begin{gather}
    \mu \frac{d V_\mathrm{sun}}{d\mu} = -\frac{\lambda^2 \phi^2T^2}{384\pi^2} + \frac{M T \lambda^2 \phi^2}{128\pi^3} \ .
\end{gather}
We observe the first term will cancel a term proportional to $T^2\phi^2$ in Eqn.~\eqref{e.dVdmuParwani}. However, there is a remaining scale dependence (after expanding $M= \sqrt{\nu^2 + \lambda \phi^2/2 + \lambda T^2/24}$ for small $\lambda$):
\begin{gather}
    \mu \frac{d(V_\textrm{eff}^\mathrm{RGPar} +V_\mathrm{sun})}{d\mu} = \frac{\lambda  \nu^3 T}{128 \pi ^3}+\frac{\lambda ^2 \nu \left(7 T^3+36 T \phi ^2\right)}{6144 \pi ^3}+O\left(\lambda ^3\right) \ ,
\end{gather}
which still gives a tadpole scale variation of $\mathcal{O}(\lambda^2)$.
Therefore, standard attempts of RG improving Parwani resummation 
by just substituting the running couplings still give $\mathcal{O}(\beta^2, \lambda^2)$ error terms, even if we explicitly add the sunset contribution. Attempts have been made to fix the large scale variation by changing the RGEs in a particular way (for example, see~\cite{Funakubo:2023eic}), but the ad-hoc nature of such procedures (in addition to the $\mathcal{O}(\beta^2)$ error term) clearly  demonstrate the benefits of using partial dressing in general, and our RGHPD procedure specifically.

\subsubsection{Comparison with NLO DR}
\label{s.RGINLODR}

Here we will review the analytical derivation of NLO DR scale variation to show it is parametrically the same as RGHPD. We again focus on the single-scalar case, the two-scalar analysis is analogous~\cite{Gould:2021oba, Curtin:2022ovx}. We can write the NLO DR effective potential as:
\begin{equation}
V_{\mathrm{DR}}= T\left[\frac{1}{2} m_3^2 \phi_3^2+\frac{1}{24} \lambda_3 \phi_3^4-\frac{\left(M_3^2\right)^{3/2}}{12 \pi}+\frac{\lambda_3 M_3^2}{128 \pi^2} -\frac{\lambda_3^2 \phi_3^2}{384 \pi^2} \left(1+2 \log \frac{\mu_3}{3 M_3}\right)\right] \ ,
\end{equation}
where we define the quantities:
\begin{align}
M_3^2 &= m_3^2+\frac{1}{2} \lambda_3 \phi_3^2 \ , \\
\lambda_3 &= T\left(\lambda-\frac{3}{32 \pi^2} \lambda^2 L_b\right) \ , \\
m_3^2 &= \nu^2+\frac{1}{24} \lambda T^2-\frac{1}{16 \pi^2}\left[\frac{1}{2} \lambda \nu^2 L_b+\frac{1}{48} \lambda T^2 L_b +\frac{1}{6} \lambda_3^2\left(c+\log \frac{3 T}{\mu_3}\right)\right] \ , \\
\phi_3 &= \frac{\phi}{\sqrt{T}} \ .
\end{align}
To make a fair comparison with RGHPD, we will consider scale variation with $\mu_3$, since that has larger scale variation then $\mu$. Applying the Callan-Symanzik equation with respect to $\mu_3$:
\begin{align}
\mu_3 \frac{d m_3^2}{d \mu_3} & =\mu_3 \frac{d M_3^2}{d \mu_3} =\frac{1}{96 \pi^2} \lambda^2 T^2 \ , \\ 
\mu_3 \frac{\partial V_{\mathrm{DR}}}{\partial \mu_3} & = -\frac{1}{192 \pi^2} \lambda^2 T^2 \phi^2 .
\end{align}
We get the remaining scale variation:
\begin{align}
\mu_3 \frac{d V_{\mathrm{DR}}}{d \mu_3} &= -\frac{1}{768 \pi^3} \lambda^2 T^3 M_3\\
\implies \mu_3 \frac{d V'_{\mathrm{DR}}}{d \mu_3} &= -\frac{1}{768 \pi^3} \frac{\lambda^3 T^3 \phi}{2 M_3} \sim \mathcal{O}(\lambda^3/\pi^3) \ , 
\end{align}
which is of the same $\mathcal{O}(\lambda^3)$ order as for RGHPD.

\section{Summary}
\label{sec:executivesummary}

We now provide a conceptual summary of (RG)HPD, as well as the final practical prescription for computing the RGHPD potential.

\subsection{Diagrammatic summary of (RG)HPD}

A diagrammatic summary of the HPD procdure for constructing ${V'}^\mathrm{HPD}_\mathrm{eff}$ is shown in Figure~\ref{fig:HPDsummary}. This is for a single-scalar $\phi^4$ model satisfying the $\mathbb{Z}_2$ $\phi \to - \phi$ symmetry for clarity, but generalizes readily to multiple scalars. The final potential is obtained by integrating the final tadpole $V^\mathrm{HPD}_\mathrm{eff} = \int d\phi {V'}^\mathrm{HPD}_\mathrm{eff}$.

\newcommand{\diagSUMMARY}[2][0pt]{\raisebox{#1}{\includegraphics[scale=1]{DiagSUM_HPD/#2.pdf}}}
\newcommand{\diagSUMMARYC}[2][0pt]{\raisebox{#1}{\includegraphics[valign=c,scale=1]{DiagSUM_HPD/#2.pdf}}}

\begin{figure}
\centering
\begin{tabular}{| l  l |}
\hline 
& \\[-12pt]
$\begin{aligned}
    &\textbf{1-loop tadpole:} \phantom{\diagSUMMARYC[0pt]{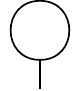}} \\
    &\textbf{2-loop sunset tadpole:} \phantom{\diagSUMMARYC{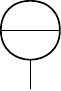}}
\end{aligned}$ &
$\begin{aligned}
    V_1' &= \left( \diagSUMMARYC{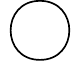} \right)' = \diagSUMMARYC[0pt]{TAD_Scalar1} \\
    V_\mathrm{sun}' &= \left( \diagSUMMARYC{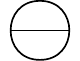} \right)' = \diagSUMMARYC{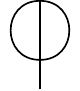} + \diagSUMMARYC{TADsunset_loop_Scalar1111}
\end{aligned}$ \\
& \\[-12pt]
\hline
& \\[-12pt]
\textbf{Gap equation:} &
$\begin{aligned}
  \diagSUMMARYC{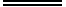} &= \diagSUMMARYC{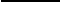} + \diagSUMMARY[1.5pt]{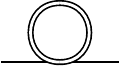} + \diagSUMMARYC{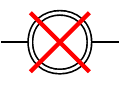} \\
  M^2 &= \left. {\color{red} \dfrac{\partial}{\partial\phi} }\,\dfrac{d}{d\phi}\, V_1\right|_{m^2 \to M^2}
\end{aligned}$ \\
& \\[-12pt]
\hline
& \\[-12pt]
\textbf{HPD effective potential:} &
$\begin{aligned}
  V_\mathrm{eff}^{\prime\,\mathrm{HPD}}
    &\equiv V_\mathrm{tree}' + \diagSUMMARYC{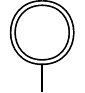} + \left( \diagSUMMARYC{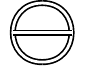} \right)' \\
    &= V'_\mathrm{tree} + V_1'\Big|_{m^2\to M^2} + V_\mathrm{sun}'\Big|_{m^2\to M^2}
\end{aligned}$ \\[30pt]
\hline
\end{tabular}
\caption{Diagrammatic summary of HPD.}
\label{fig:HPDsummary}
\end{figure}

The ingredients, in addition to the tree-level potential $V_\mathrm{tree}$, are the known analytical expressions for the one-loop potential $V_1'$ (see Eqns.~\eqref{e.Vcwtwophi} and~\eqref{e.VTH}) and the two-loop sunset $V_\mathrm{sun}$ (see Eqns.~\ref{e.sunsetOneScalatZeroT} and~\ref{e.singleScalarSun_FULL}) at zero  and finite temperature. 
A gap equation for the resummed mass $M^2$ is defined using only the one-loop effective potential. 
Importantly, explicitly $\phi$-dependent terms, i.e. terms involving the cubic couplings $\lambda \phi$, have to be deleted from the gap equation.
(This can be practically implemented by replacing a full $\phi$-derivative by a partial one.) 
This avoids including loops that depend on external momenta in the gap equation solution for $M^2$. 
The gap equation can be diagrammatically or numerically solved by straightforward iteration, and the solution is substituted into ${V'}^\mathrm{HPD}_\mathrm{eff} \equiv V'_\mathrm{tree} + \left[V'_1 + V'_\mathrm{sun}\right]_{m^2 \to M^2}$ to define the final HPD potential tadpole.

Substituting the resummed mass into the tadpole rather than the potential itself is required to correctly capture certain two, three, and higher-loop symmetry factors~\cite{Boyd:1993tz}. 
If loops that depend on external momenta had been retained in the gap equation, the substitution would generate diagrams which naively look like sunset contributions, but since the gap equation is evaluated at zero external momentum, these sunsets do not properly account for overlapping momenta in the two-loop diagram, introducing the need to manually insert corrective pre-factors which  become very complicated for general theories or beyond the high-temperature regime.
This difficulty is entirely circumvented by the ``hybrid method'' of deleting the $\phi$-dependent terms from the gap equation, allowing the gap equation to correctly capture joined-vacuum-bubble contributions to $M^2$ and nothing more, and then directly adding the missing diagrams at each order, which at two-loop is just the tadpoles descending from the sunset potential term. This makes the potential complete at $\mathcal{O}(\lambda^2)$ and $\mathcal{O}(\beta^2)$ order at zero- and finite-temperature, respectively.

We show that
RG-improvement is straightforward: Define ${V'}^\mathrm{RGHPD}_\mathrm{eff}$ by substituting running couplings $\lambda \to \lambda(\mu_R), \ldots$ and $\mu_0 \to \mu_R$ in the gap equation and ${V'}^\mathrm{HPD}_\mathrm{eff}$. Crucially, we find that the correct choice of renormalization scale in the effective potential is $\mu_R^2 \sim (\pi T)^2 + M^2(\phi, T)$, to capture the dominant leading log contributions to all loop orders at the mass scale of the most important physics generating the loop potential.

\newcommand{\diagPROOF}[2][0pt]{\raisebox{#1}{\includegraphics[scale=0.92]{DiagSUM_HPD/#2.pdf}}}
\newcommand{\diagPROOFC}[2][0pt]{\raisebox{#1}{\includegraphics[valign=c,scale=0.92]{DiagSUM_HPD/#2.pdf}}}
\newcommand{\orangeARROW}[1][0]{\overset{{\color{orange}\overset{\displaystyle\alpha}{\rotatebox{#1}{\ensuremath{\xrightarrow{\hspace{0.3cm}}}}}}}{+}}

\begin{figure}
\centering
\begin{tabular}{| l |}
\hline
 \\[-12pt]
\textbf{Gap equation:} \\
$\begin{aligned}
  \diagPROOFC{DressProg_scalar1} = {\color{blue}\underbrace{
    \diagPROOFC{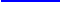} + \diagPROOF[1.5pt]{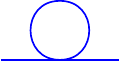}
  }_{\text{no iteration}}}
  + {\color{red}\underbrace{
    \diagPROOF[1.5pt]{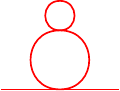} + \diagPROOF[1.5pt]{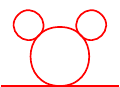} + \cdots + \diagPROOF[1.5pt]{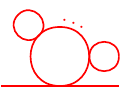}
  }_{\text{one iteration}}} + \ldots
\end{aligned}$ \\
 \\[-12pt]
\hline \\[-12pt]
\textbf{Potential Tadpole:} \\
{ \setlength{\tabcolsep}{2pt}
\begin{tabular}{ccccccccccccc}
 & & $\overbrace{\hphantom{\diagPROOFC{TAD_Scalar1}}}^{V_1'}$
 & & $\overbrace{\hphantom{\diagPROOFC{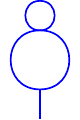}}}^{V_2'}$
 & & \multicolumn{7}{c}{$\overbrace{\hphantom{\hspace{8cm}}}^{\alpha\text{-enhanced }V_n'}$} \\[-6pt]
 \diagPROOFC{TAD_RESUM_LoopScalar1} & $\to$ & \diagPROOFC{TAD_Scalar1} & {\color{blue} $\orangeARROW$} & \diagPROOFC[7pt]{TAD_2loop_BLUE}& {\color{blue}$\orangeARROW$} & \diagPROOFC[7pt]{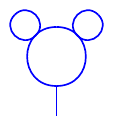}& {\color{blue}$+$} &  \multicolumn{3}{c}{{\color{blue}$\cdots$}} & {\color{blue}$+$} & \diagPROOFC[7pt]{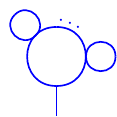} \\ 
 \\[-12pt]
  &  &  &  & & {\color{red}$\orangeARROW[-30]$} & \diagPROOFC{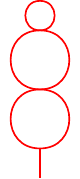}& {\color{red}$\orangeARROW$} &  \diagPROOFC{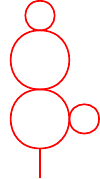} & \multirow{2}{*}{\raisebox{-1.3cm}{{\color{red}$+$}}} & \multirow{2}{*}{\raisebox{-1.3cm}{{\color{red}$\cdot \cdot \cdot$}}} & \multirow{2}{*}{\raisebox{-1.3cm}{{\color{red}$+$}}} & \multirow{2}{*}{\raisebox{-1cm}{\diagPROOFC{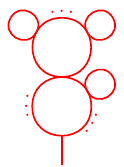}}} \\
  &  &  &  & &  &  & {\color{red}$\orangeARROW[-30]$} & \diagPROOFC{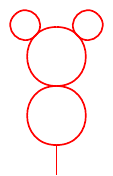} & & & & \\
  \diagPROOFC{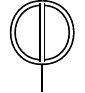} & $\to$ &  &  & \diagPROOFC{Lollipop_loop_Scalar111} & {\color{blue}$\orangeARROW$} & \diagPROOFC[6pt]{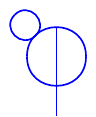} & {\color{blue}$+$}  &  \multicolumn{3}{c}{{\color{blue}$\cdots$}} & {\color{blue}$+$} & \diagPROOFC[6pt]{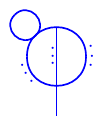} \\ 
  \diagPROOFC{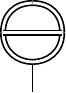} & $\to$ &  &  & \diagPROOFC{TADsunset_loop_Scalar1111} & {\color{blue}$\orangeARROW$} & \diagPROOFC[6pt]{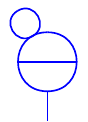}& {\color{blue}$+$}  &  \multicolumn{3}{c}{{\color{blue}$\cdots$}} & {\color{blue}$+$} & \diagPROOFC[6pt]{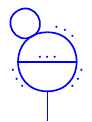}
\end{tabular}} \\[-12pt]
\\
\hline
\end{tabular}
\caption{ Illustration of how HPD captures all two-loop contributions to the effective potential tadpole, as well as all $\alpha$-enhanced hard thermal loop contributions at higher loop order to ensure validity at finite temperature, where $\alpha \sim \lambda T^2/m^2 \sim 1$ during the phase transition. At zero temperature, $V_1'$ and $V_2'$ are order $\lambda^1, \lambda^2$ relative to tree-level respectively. At high temperature, one (two) loop diagrams dressed with an arbitrary number of hard thermal loops are order $\beta^1, \beta^2$ respectively, with $\beta \sim \lambda T/m \ll 1$ during the phase transition.}
    \label{fig:HPDdiagrammaticproof}
\end{figure}

\begin{figure}[h]
\centering
\begin{tabular}{| l |}
\hline
 \\[-12pt]
\textbf{Gap equation:} \\
$\begin{aligned}
    \diagPROOFC{DressProg_scalar1} = 
  {\color{blue}\underbrace{
    \diagPROOFC{Prog_scalar1_BLUE} + \diagPROOF[1.5pt]{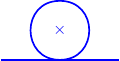} + \diagPROOF[1.5pt]{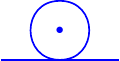}
  }_{\text{no iteration}}}
  + {\color{red}\underbrace{
    \diagPROOF[1.5pt]{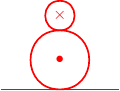} +  \cdots + \diagPROOF[1.5pt]{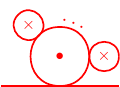}
  }_{\text{one iteration}}} + \ldots
\end{aligned}$ \\
 \\[-12pt]
\hline \\[-12pt]
\textbf{Potential Tadpole:} \\
{\setlength{\tabcolsep}{2pt}
\begin{tabular}{ccccccccccccc}
 & & $\overbrace{\hphantom{\diagPROOFC{TAD_Scalar1}}}^{V_1'}$
 & & $\overbrace{\hphantom{\diagPROOFC{TAD_2loop_BLUE}}}^{V_2'}$
 & & \multicolumn{7}{c}{$\overbrace{\hphantom{\hspace{8cm}}}^{\alpha\text{-enhanced }V_n'}$} \\[-6pt]
 \diagPROOFC{TAD_RESUM_LoopScalar1} & $\to$ & \diagPROOFC{TAD_Scalar1} & {\color{blue} $\orangeARROW$} & \diagPROOFC[7pt]{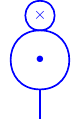}& {\color{blue}$\orangeARROW$} & \diagPROOFC[7pt]{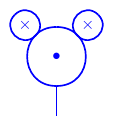}& {\color{blue}$+$} &  \multicolumn{3}{c}{{\color{blue}$\cdots$}} & {\color{blue}$+$} & \diagPROOFC[7pt]{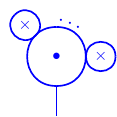} \\
 \\[-10pt]
 & & & {\color{blue} $+$} & \diagPROOFC[7pt]{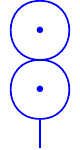} & {\color{blue} $\orangeARROW$} & \diagPROOFC[7pt]{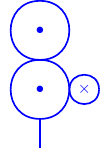} & & & & & & \\
 \\[-10pt]
  &  &  &  & & {\color{red}$\orangeARROW[-30]$} & \diagPROOFC{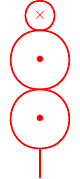}& {\color{red}$\orangeARROW$} &  \diagPROOFC{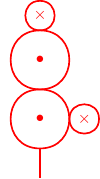} & \multirow{2}{*}{\raisebox{-1.3cm}{{\color{red}$+$}}} & \multirow{2}{*}{\raisebox{-1.3cm}{{\color{red}$\cdot \cdot \cdot$}}} & \multirow{2}{*}{\raisebox{-1.3cm}{{\color{red}$+$}}} & \multirow{2}{*}{\raisebox{-1cm}{\diagPROOFC{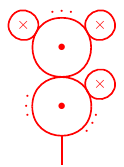}}} \\
  &  &  &  & &  &  & {\color{red}$\orangeARROW[-30]$} & \diagPROOFC{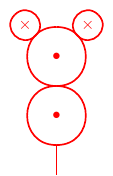} & & & & \\
  \diagPROOFC{Lollipop_RESUM_Scalar111} & $\to$ &  &  & \diagPROOFC{Lollipop_loop_Scalar111} & {\color{blue}$\orangeARROW$} & \diagPROOFC[6pt]{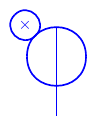} & {\color{blue}$+$}  &  \multicolumn{3}{c}{{\color{blue}$\cdots$}} & {\color{blue}$+$} & \diagPROOFC[6pt]{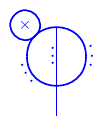} \\ 
  \diagPROOFC{TADsunset_RESUM_Scalar1111} & $\to$ &  &  & \diagPROOFC{TADsunset_loop_Scalar1111} & {\color{blue}$\orangeARROW$} & \diagPROOFC[6pt]{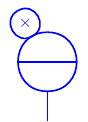}& {\color{blue}$+$}  &  \multicolumn{3}{c}{{\color{blue}$\cdots$}} & {\color{blue}$+$} & \diagPROOFC[6pt]{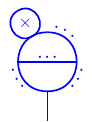}
\end{tabular}} \\[-12pt]
\\
\hline
\end{tabular}
\caption{Same as Figure~\ref{fig:HPDdiagrammaticproof}, but explicitly for the high temperature case, with zero- and non-zero Matsubara mode vacuum bubbles indicated explicitly with  $\bullet$ and $\times$ respectively in some diagrams or parts thereof. Vacuum bubbles with a single propagator are dominated by the non-zero mode (hard thermal loop) $\sim T^2$, while for vacuum bubbles with $j \geq 2$ propagators, the zero mode gives the dominant $\sim T m^{3-2j}$ contribution. }
\label{fig:HPDdiagrammaticproofhighT}
\hspace{5mm}
\end{figure}

\subsection{Illustration of (RG)HPD validity to $\mathcal{O}(\beta^2)$}
\label{sec:RGHPDdiagrammaticproof}

We now illustrate how (RG)HPD correctly captures all two-loop  effective potential terms at zero and finite temperature, and how at high temperature, the gap equation resums hard thermal loops without introducing overcounting complications with RG improvement. 
We ignore all $\mathcal{O}(1)$ parameters in this diagrammatic summary.

Figure~\ref{fig:HPDdiagrammaticproof} (top) shows what diagrams are included in the resummed mass $M^2$ without any iteration (blue) and after the first iteration (red) of the gap equation. 
The bottom shows the corresponding diagrams included in the tadpole potential upon substituting that gap equation solution into $V'_1 + V'_\mathrm{sun}$. The diagrams included 'by hand' are shown in black. Diagrams generated from the $m^2 \to M^2$ substitution are shown in blue and red, indicating whether they are generated from the no-iteration or one-iteration part of the gap equation solution. 
It is clear that all two-loop diagrams are generated, though the only addition from mass resummation is the blue double-bubble diagram in the first row. (Incidentally, this is exactly the diagram for which the symmetry factor is not correctly included if $M^2$ is instead substituted into the potential $V$ rather than the tadpole.) At zero temperature, ${V'}_\mathrm{eff}^\mathrm{HPD}$ therefore contains all $\mathcal{O}(\lambda^2)$ contributions.
Importantly, all versions of the one- and two-loop diagrams which are generated by adding a hard thermal loop, which add a multiplicative factor of $\alpha \sim 1$ during the phase transition, indicated with orange arrows, are also generated correctly.  Importantly, the triple vacuum bubble diagram in the second row is only included if the gap equation is iterated at least once.

The importance of this diagram is more readily illustrated by explicitly separating vacuum bubbles into their zero- and non-zero-mode contributions in Figure~\ref{fig:HPDdiagrammaticproofhighT}. During the phase transition, hard thermal loop resummation effectively resums all subsequent diagrams in a given row into the first diagram of that row. The one-loop contributions are then $\mathcal{O}(\beta^1)$, while the two-loop contributions, represented by the double-zero-mode-bubble and the sunset-type graphs, are order $\mathcal{O}(\beta^2)$. This ensures the HPD effective potential includes all terms up to $\beta^2$ order. One can also notice that at $\mathcal{O}(\gamma^0)$, with $\gamma \equiv \phi^2/T^2$, all three-loop contributions are correctly included except basketball-type diagrams, see Eqn.~\eqref{e.basketball}. This motivates our use of the basketball tadpole as the thermal error term to estimate the effect of neglected higher-order contributions.

We can also understand why RG improvement does not lead to any overcounting mistakes with the mass resummation. For example, solving the one-loop RGEs for the running parameters and performing the substitution $\lambda \to \lambda(\mu_R)$ (and other parameters) and $\mu_0 \to \mu_R$ in any complete fixed-order $n$-loop effective potential will simply reproduce that same $n$-loop potential plus leading-log contributions to all loop orders $\geq n$. Since the HPD effective potential includes all complete one- and two-loop contributions, the only place where RG improvement could introduce overcounting errors is in the partial higher-loop contributions generated by substituting  the gap equation solution into the tadpole. At zero-temperature, those contributions are higher-order, and so is the overcounting error introduced by RG-improving a partial higher-loop contribution. This does not impact the accuracy of a two-loop RG-improved calculation. At high temperature, the only higher-loop contributions that matter at order $\beta^2$ are one- and two-loop diagrams dressed with non-zero-mode vacuum bubbles, and furthermore, only the hard $\sim T^2$ pieces of those hard thermal loops are required for the resummation, since only those contributions have ratio $\alpha^n \sim 1$ to the corresponding two-loop zero-mode diagram. The higher-loop zero-temperature leading-log contributions added by RG-resummation, even if they correspond to identical diagrammatic topologies, cannot overcount the same contribution, since those leading logs are parametrically distinct from the $\sim T^2$ hard thermal loops. 
This argument persists even in the intermediate regime outside of the high-temperature limit, and for higher-order RG improvement.

With this diagrammatic perspective in mind, we simply can sum up HPD in the following way (at least for scalar theories): take a modified two-loop potential tadpole, with the two-vacuum bubble contribution deleted; solve a modified one-loop gap equation, which omits $\phi$-dependent terms to consistently resum vacuum-bubble-type mass contributions; and sub the gap equation solution into the tadpole. This yields the standard two-loop potential, which can be RG improved in the usual way, plus a set of higher order diagrams which is irrelevant at zero temperature but implements thermal mass resummation at higher temperature to maintain the required two loop accuracy during a phase transition. 

This description, in addition to perhaps demystifying how to consistently generalize thermal effective potentials out of the high-temperature regime, suggests two important lessons. First, one might naively think that the HPD construction of modifying the two-loop potential and then substituting in the solution of a modified one-loop gap equation could be circumvented by simply substituting a propagator with a resummed  one-loop mass correction into the two-loop potential. We can immediately see that this will lead to overcounting errors (even if the resummed mass is substituted into the tadpole to avoid the symmetry factor mistakes identified for Full Dressing by~\cite{Boyd:1993tz}): if the double-bubble tadpole is included, it will overcount the double-bubble tadpole generated by substituting the one-loop thermal mass into the one-loop tadpole (see Figure~\ref{fig:HPDdiagrammaticproof}). 
The HPD construction is maximally simple while avoiding such mistakes. 
Second, this diagrammatic understanding immediately suggests a way to extend HPD beyond $\mathcal{O}(\beta^2)$: write down the three-loop potential, delete the contributions which are not generated by the gap equation (joined one-loop bubbles, or sunset-type diagrams with a joined bubble), and then substitute in the solution to the same modified one-loop gap equation.

While this manuscript was in preparation, a recent analysis~\cite{Navarrete:2025yxy} presented another way of organizing a finite-temperature effective potential calculation and perform thermal resummation without invoking the high-temperature approximation. The authors computed fixed-order contributions to a given loop order; separated them into  soft (zero-mode) and hard (non-zero mode) contributions; and substituted the one-loop hard thermal mass correction (evaluated without high-temperature approximation) into the soft part. This is simply another way of organizing the calculation in such a way that the one-loop mass correction is kept consistently even at low or zero temperature (where it smoothly decouples). The calculation can also be straightforwardly RG-improved in the same way. At two-loop order, it is therefore entirely equivalent to the RG-improved Hybrid Partial Dressing method developed in this paper. However, our (RG)HPD method will be significantly simpler to generalize and apply to BSM phenomenology studies of the phase transition, since it involves a smaller number of pre-computed expressions and by its nature avoids having to separate diagrams into their distinct Matsubara contributions.

\subsection{Practical recipes for computing the HPD and RGHPD effective potentials}\label{sec. NumPres}

We summarize here the exact steps one must take to compute $V_\mathrm{eff}^\mathrm{HPD}(\phi_1, \phi_2 = 0, T)$ and $V_\mathrm{eff}^\mathrm{RGHPD}(\phi_1, \phi_2 = 0, T)$ for the two-scalar theory defined in Eqn.~\ref{e.lagrangian}, starting with  a set of physical matching conditions like our BP1, BP2 benchmark points defined in Table.~\ref{tab:bm}. Either will give reliable predictions with accurate theoretical uncertainties self-consistently captured by scale variation. 
RGHPD will be more precise in the high-temperature regime, but for $m \gtrsim T$ and strong phase transitions, the presence of modest mass hierarchies may necessitate generalizing the methods of Ref.~\cite{Manohar:2020nzp} for fully consistent RG improvement. In that case, we found HPD to have smaller theoretical uncertainties.

Note that if desired for convenience, the below procedures can be applied with lower-order results, e.g. using only the one-loop effective potential without the zero-temperature sunset, or one-loop RGEs. The resulting lower precision will be self-consistently reflected in larger scale variation, and may be acceptable depending on the application. The applicability to all temperatures will be retained.

We now supply the HPD and RGHPD recipe in full below. The redundant overlap between the two procedures is  retained for clarity. We work in dimensional regularization and the $\overline{\mathrm{MS}}$ 
    renormalization scheme.

\subsubsection*{HPD recipe:}

\begin{enumerate}
       
    \item Construct the fixed-order two-loop effective potential $V_\mathrm{eff}^\mathrm{zero}(\phi_1, \phi_2)$. One can use many pre-computed results, see Appendix~\ref{a.potential}.\footnote{Alternatively, HPD can be consistently used at zero temperature using Eqns.~\eqref{e.Vcwtwophi} and~\eqref{e.sunsetTwoScalatZeroT} for the one-loop and sunset tadpole respectively.\label{foot.HPDzerotemp}} 
    This will define the theory completely for a given choice of input parameters $\lambda_{1, 2, 12}(\mu_0), \nu_{1,2}(\mu_0)$, where we fix $\mu_0 = a m_{2,\mathrm{phys}}$ for some choice of $a \sim \mathcal{O}(1)$.
    
    \item We then perform a two-loop matching procedure described in Appendix~\ref{a.matching}, finding input parameters $\lambda_{1, 2, 12}(\mu_0), \nu_{1,2}(\mu_0)$ that reproduce the pole masses, VEV $\langle \phi_1 \rangle = v$, and zero-momentum quartics $\lambda_{2, 12}$ for a given benchmark point in Table~\ref{tab:bm}.
    The theory is now fitted to the desired physical benchmark parameters. Note that these parameters and the predictions for the potential shape far away from the vacuum $v$ carry a small residual $a$-dependence, representing the zero-temperature scale variation of the RG-improved effective potential.
    \item We are now ready to evaluate the HPD effective potential $V_\mathrm{eff}^\mathrm{HPD}(T, \phi_1; a)$, with $\phi_2 = 0$,  for a given temperature $T$ and $\phi_1$ VEV. The explicit dependence on $a$ encodes scale dependence.

    We start by numerically solving the HPD gap equations by iteration, see Eqn.~\eqref{e.HPDtwoscalargapeqn}, for the given $\phi, T$. Specifically, labeling each iterative solution to the gap equation by index $r$ and starting with $M^2_{i,r = 0}(\phi, T) = m^2_{i}(\phi)$, the RHS of the gap equations can be taken to define the next step in the iteration, i.e. $M_{i, r + 1}^2 = \mathrm{RHS}(M_{i, r})$ for $i = 1,2$.
    Iterate until the solution converges, and for a given temperature, repeat this for a grid of $\phi_1$ values in some range of interest $(0, \phi_{1, max})$. 

    This gives us $M_{1,2}(\phi, T)$ for the current temperature $T$ and choice of $a$.

    \item We can now numerically evaluate $V_\mathrm{eff}^\mathrm{HPD}(T, \phi_1; a)$ for the chosen $T, a$
    by substituting the numerical solutions for the resummed effective massees $M_{1,2}(\phi, T)$ from the previous step into Eqn.~\eqref{e.VfullHPDphi2}.

    \item One can then perform the usual phase transition analysis by evaluating the potential for different $T$ to find the critical temperature $T_{c}$ where the minimum at the origin and the second minimum at $v = v_c$, if it exists, are degenerate.
    
\end{enumerate}
In our numerical investigations, we investigate the scale dependence of HPD by varying $a \in (1/2, 2)$ in $\mu_0 = a  m_{2,\mathrm{phys}}$. 

\subsubsection*{RGHPD recipe:}

\begin{enumerate}
\setcounter{enumi}{-1}
    \item Obtain the two-loop RGEs for all Lagrangian parameters, including the vacuum energy. We use existing results for the two-scalar theory, see Appendix~\ref{a.RG}.
    
    \item Construct the fixed-order two-loop effective potential $V_\mathrm{eff}^\mathrm{zero}(\phi_1, \phi_2)$. One can use many pre-computed results, see Appendix~\ref{a.potential}.\textsuperscript{\ref{foot.HPDzerotemp}} 
    Define the RG-improved potential $V_\mathrm{eff}^\mathrm{RGzero}(\phi_1, \phi_2; a)$ 
    by replacing all parameters by the running parameters obtained by solving the RGEs, and replacing $\mu_0$ by $\mu_R = a \ m_{2,\mathrm{phys}}(\phi_1)$ for some choice of $a \sim \mathcal{O}(1)$.  This will define the theory completely for a given choice of input parameters $\lambda_{1, 2, 12}(\mu_0), \nu_{1,2}(\mu_0)$, where we fix $\mu_0 = m_{2,\mathrm{phys}}$. 
    
    \item We then perform a two-loop matching procedure described in Appendix~\ref{a.matching}, finding input parameters $\lambda_{1, 2, 12}(\mu_0), \nu_{1,2}(\mu_0)$ that reproduce the pole masses, VEV $\langle \phi_1 \rangle = v$, and zero-momentum quartics $\lambda_{2, 12}$ for a given benchmark point in Table~\ref{tab:bm}.
    The theory is now fitted to the desired physical benchmark parameters. Note that these parameters and the predictions for the potential shape far away from the vacuum $v$ carry a small residual $a$-dependence, representing the zero-temperature scale variation of the RG-improved effective potential.
    \item We are now ready to evaluate the RGHPD effective potential $V_\mathrm{eff}^\mathrm{RGHPD}(T, \phi_1; a, b)$, with $\phi_2 = 0$, for a given temperature $T$ and $\phi_1$ VEV. The explicit dependence on $a$ and $b$ encodes scale dependence from $\mu_R^2 = \mu_R^2(\phi, T; a, b) = (b \pi T/2)^2 + (a M_2(\phi_1))^2$. 

    We start by numerically solving the HPD gap equations by iteration, see Eqn.~\eqref{e.HPDtwoscalargapeqn}, for the given $\phi, T$, where all parameters are replaced by the running couplings $\lambda \to \lambda(\mu_R(\phi, T)), \ldots$. Specifically, labeling each iterative solution to the gap equation by index $r$ and starting with $M^2_{i,r = 0}(\phi, T) = m^2_{i}(\phi)$, the RHS of the gap equations can be taken to define the next step in the iteration, i.e. $M_{i, r + 1}^2 = \mathrm{RHS}[M_{i, r}; \mu_R^2 = (b \pi T/2)^2 + (a M_{2,r}(\phi_1))^2]$ for $i = 1,2$.
    Note that at each iterative step, the gap equation is self-consistently evaluated at scale $\mu_R(\phi, T; a, b)$ with the current effective masses substituted in. 
    Iterate until the solution converges, and for a given temperature, repeat this for a grid of $\phi_1$ values in some range of interest $(0, \phi_{1, max})$.  

    This gives us $M_{1,2}(\phi, T)$ for the current temperature $T$ and choice of $a, b$. 

    \item We can now numerically evaluate $V_\mathrm{eff}^\mathrm{RGHPD}(T, \phi_1; a, b)$ for the chosen $T, a, b $
    by substituting running couplings $\lambda(\mu_R(\phi,T; a, b)), \ldots$ and the numerical solutions for the resummed effective massees $M_{1,2}(\phi, T)$ from the previous step into Eqn.~\eqref{e.VfullHPDphi2}.

    \item One can then perform the usual phase transition analysis by evaluating the potential for different $T$ to find the critical temperature $T_{c}$ where the minimum at the origin and the second minimum at $v = v_c$, if it exists, are degenerate.
    
\end{enumerate}
In our numerical investigations, we investigate the scale dependence of RGHPD, by performing the above steps for $a, b$ taking all possible combination of values $\frac{1}{2}, 1, 2$, and show the variation from the default $(a, b) = (1, 1)$. 

When we numerically perform the above HPD or RGHPD procedures, we discard the imaginary parts of the RHS of the gap equation at each iterative step, as well as the imaginary part of the (RG)HPD effective potential in Step 5. We found that this never generates numerical problems with convergence of the gap equation solution.
These imaginary parts do not in fact signal a breakdown of perturbation theory, but instead are an artifact of using a constant $\phi_1$ in the effective action which defines the effective potential~\cite{Weinberg:1987vp,Ai:2020sru}. Physically, a constant field profile in such a region is unstable: quantum fluctuations drive the system toward a more stable configuration, and the imaginary part of the potential quantifies the decay rate of a field held fixed at a constant value there~\cite{Weinberg:1987vp}. It therefore carries no information about the physical configuration, and evaluating the effective action on a profile that self-consistently solves the equations of motion removes it altogether~\cite{Garbrecht:2015oea}. This justifies simply discarding the imaginary part of the potential in our calculation.

\begin{figure}%
    \centering
    \subfloat[\centering BP1 RG evolution of masses and couplings. Note that $\lambda_1 \sim \mathcal{O}(10^{-3})$, so the vertical axis has been rescaled.]{{\includegraphics[width=13cm]{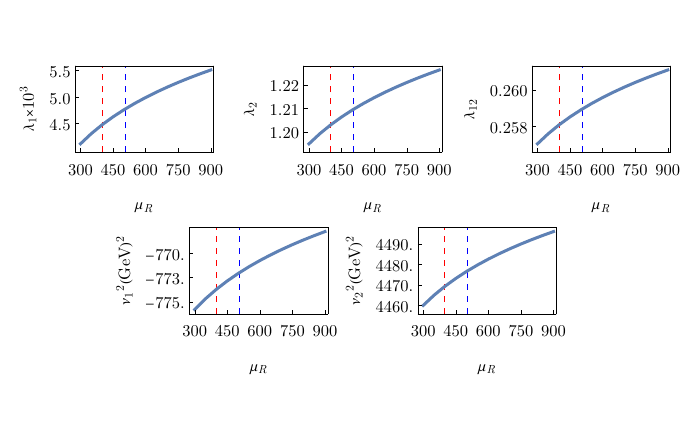} }}%
    \qquad
    \subfloat[\centering BP2 RG evolution of masses and couplings]{{\includegraphics[width=13cm]{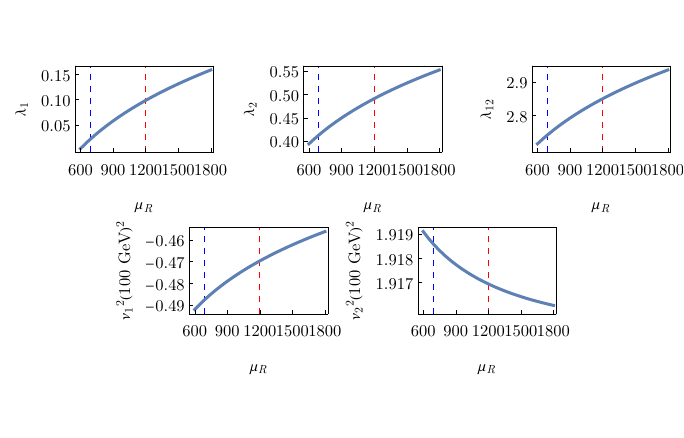} }}%
    \caption{ Two-loop RG evolution of masses couplings of the two-scalar toy model for both BP1 and BP2. The red vertical line indicates matching scale (which we set to the mass of the singlet) and blue vertical line indicates scale $\pi T_c/2$. 
    }%
    \label{fig:rge_ev}%
\end{figure}

\begin{figure}%
    \centering
    \subfloat[\centering $M_1(\phi_1, T_c)$ for BP1]{{\includegraphics[width=6.5cm]{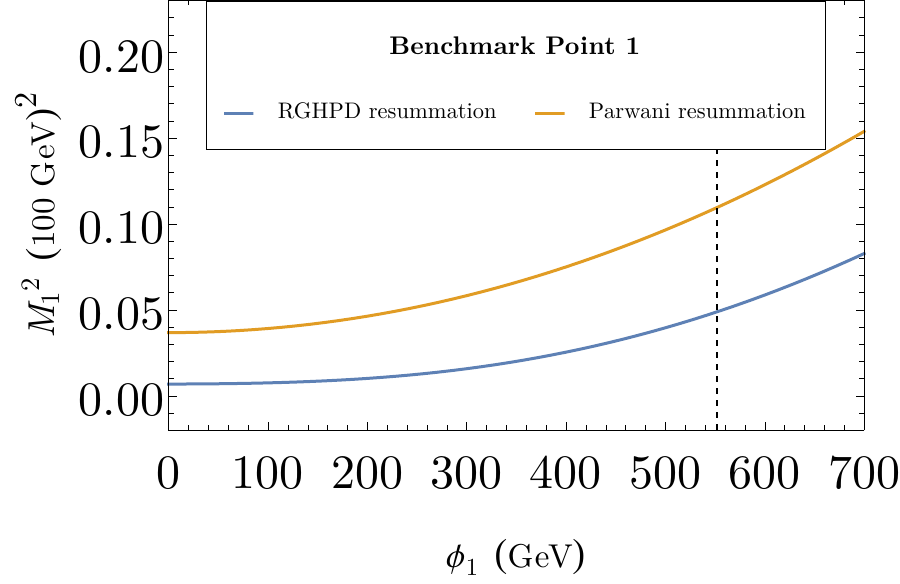} }}%
    \subfloat[\centering $M_2(\phi_1, T_c)$ for BP1]{{\includegraphics[width=6.5cm]{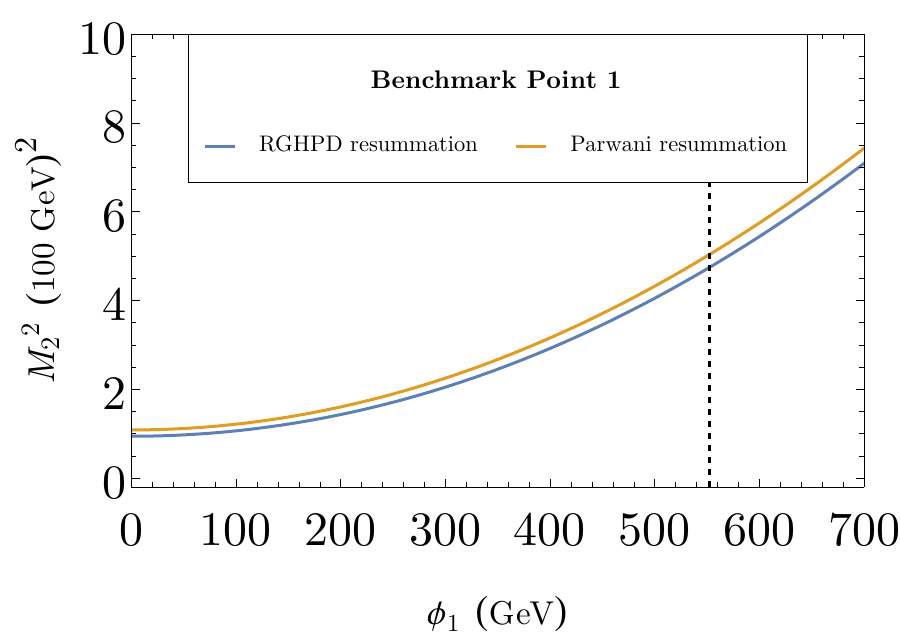} }}%
    \qquad
    \subfloat[\centering 
    $M_1(\phi_1, T_c)$ for BP2]{{\includegraphics[width=6.5cm]{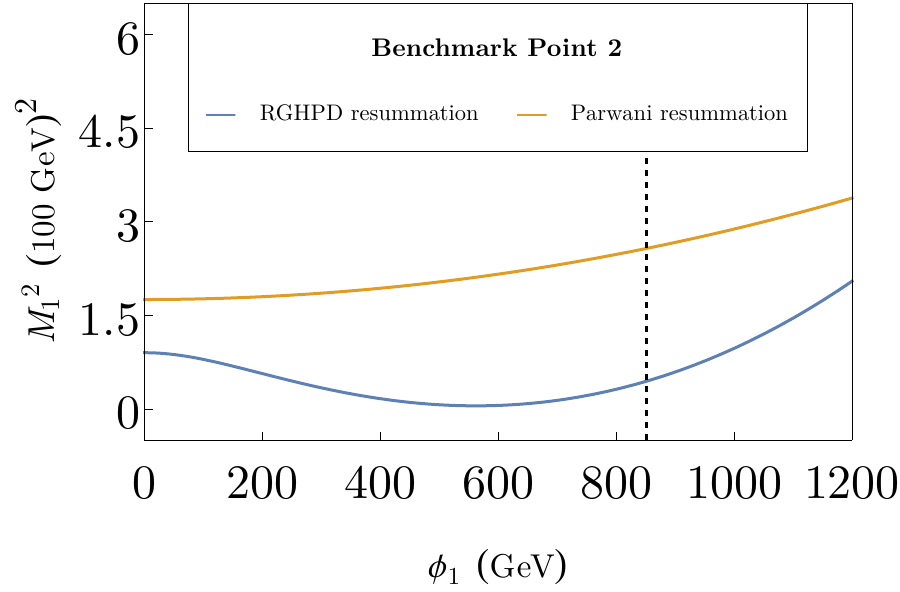} }}%
    \subfloat[\centering 
    $M_2(\phi_1, T_c)$ for BP2]{{\includegraphics[width=6.5cm]{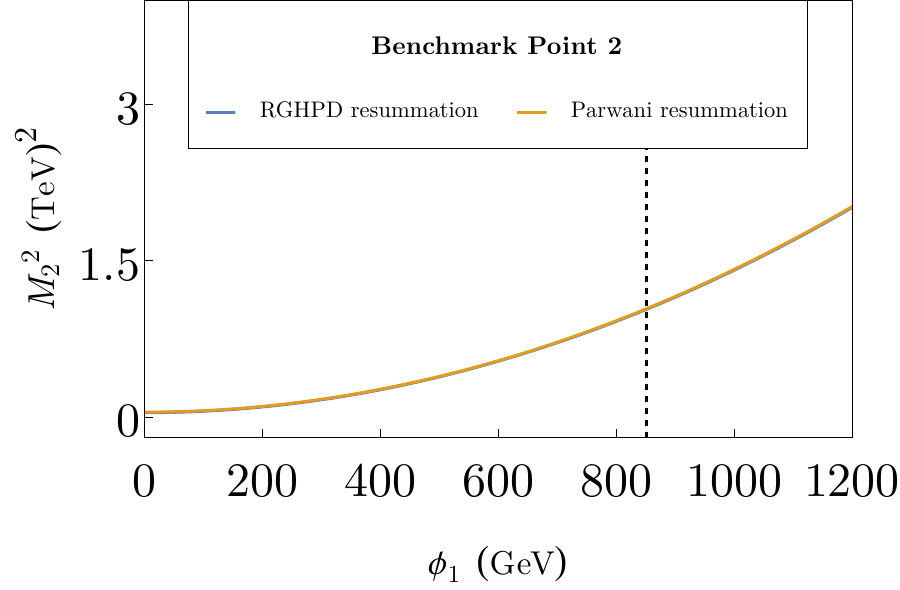} }}%
    \caption{ Resummed thermal mass $M_i$ at $T = T_c$ for BP1 and BP2, with the iterative solution from the RGHPD gap equation  (for default $a = b = 1$) shown in blue, and the standard Parwani result shown in orange. The dashed vertical line indicates $\phi_{1,c}$.
    The differences are significant (by $\mathcal{O}(1)$ factors) for the lighter scalar $\phi_1$.
    }%
    \label{fig:gap}%
\end{figure}

\section{Numerical Results} \label{sec:numeric}

We now compute the HPD and RGHPD effective potentials 
for the two benchmark points BP1 and BP2 from Table~\ref{tab:bm} of the two-scalar toy model. The procedure is summarized in Section~\ref{sec. NumPres}.
This is then used to compute the  gravitational wave signal from the first order phase transition.

We compare HPD and RGHPD to Dimensional Reduction and Parwani resummation throughout, analyzing both the central predictions and the variation from scale variation and  thermal error terms. 
When summarizing theoretical uncertainties, we express the total uncertainty on observable $X$ as the relative difference between the upper and lower error band, i.e. $(X_{max} - X_{min})/X_{min}$.
For the NLO DR calculation, the first three steps are identical to the procedure from Section~\ref{sec. NumPres} to match the zero-temperature theory to the given benchmark point. We then use DRalgo to obtain the effective potential following Section~\ref{s.dimensionalreduction}, which automatically includes the effects of running couplings. Note that in perfect correspondence to RGHPD, NLO DR also has two $\mathcal{O}(1)$ parameters which capture the scale dependence, $a$ for the scale dependence of the RGI zero-temperature effective potential, and $b$ for the scale dependence of the corresponding prediction of the high-temperature potential evaluated at scale $\mu_3 = b \pi T/2$.

\begin{figure}
    \hspace*{-6mm}
    \centering
    \begin{tabular}{cc}
    {{\includegraphics[width=7.5cm]{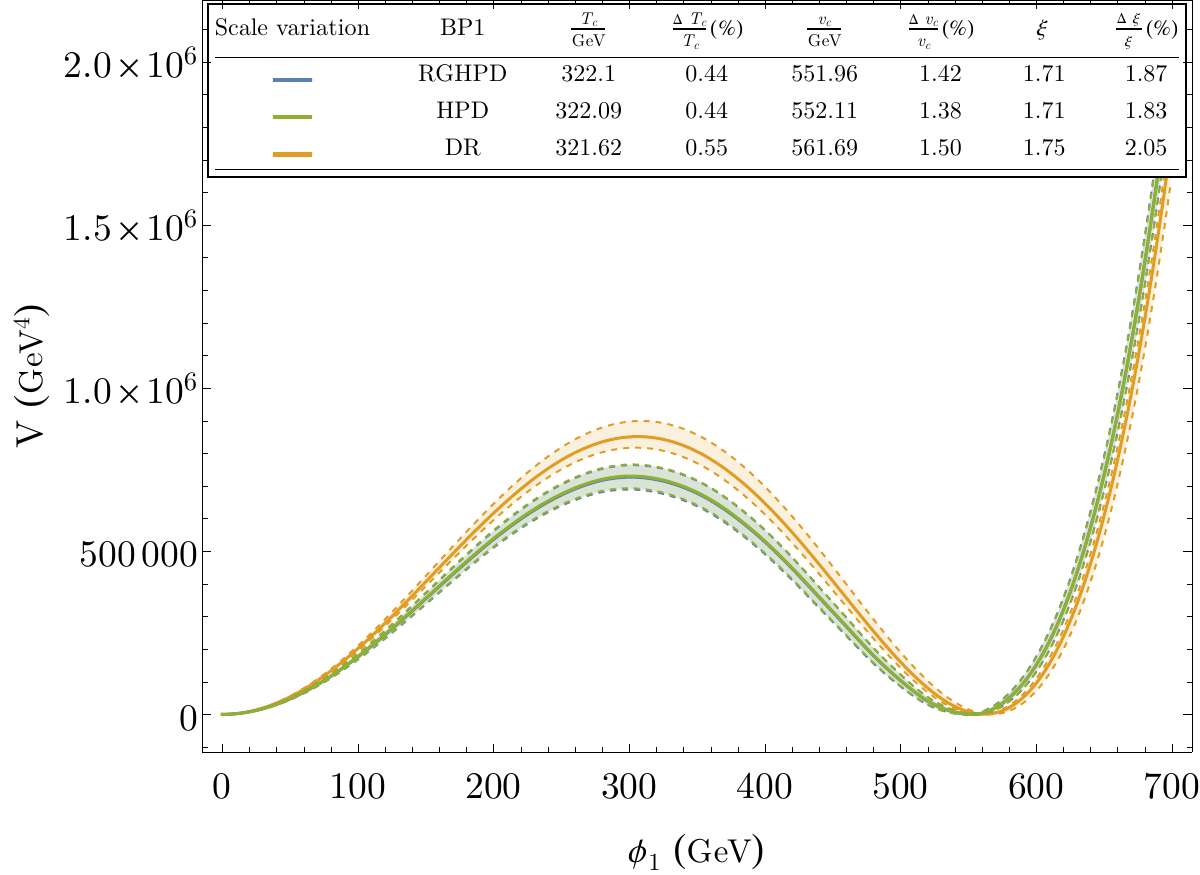} }}%
    &
    {{\includegraphics[width=7.5cm]{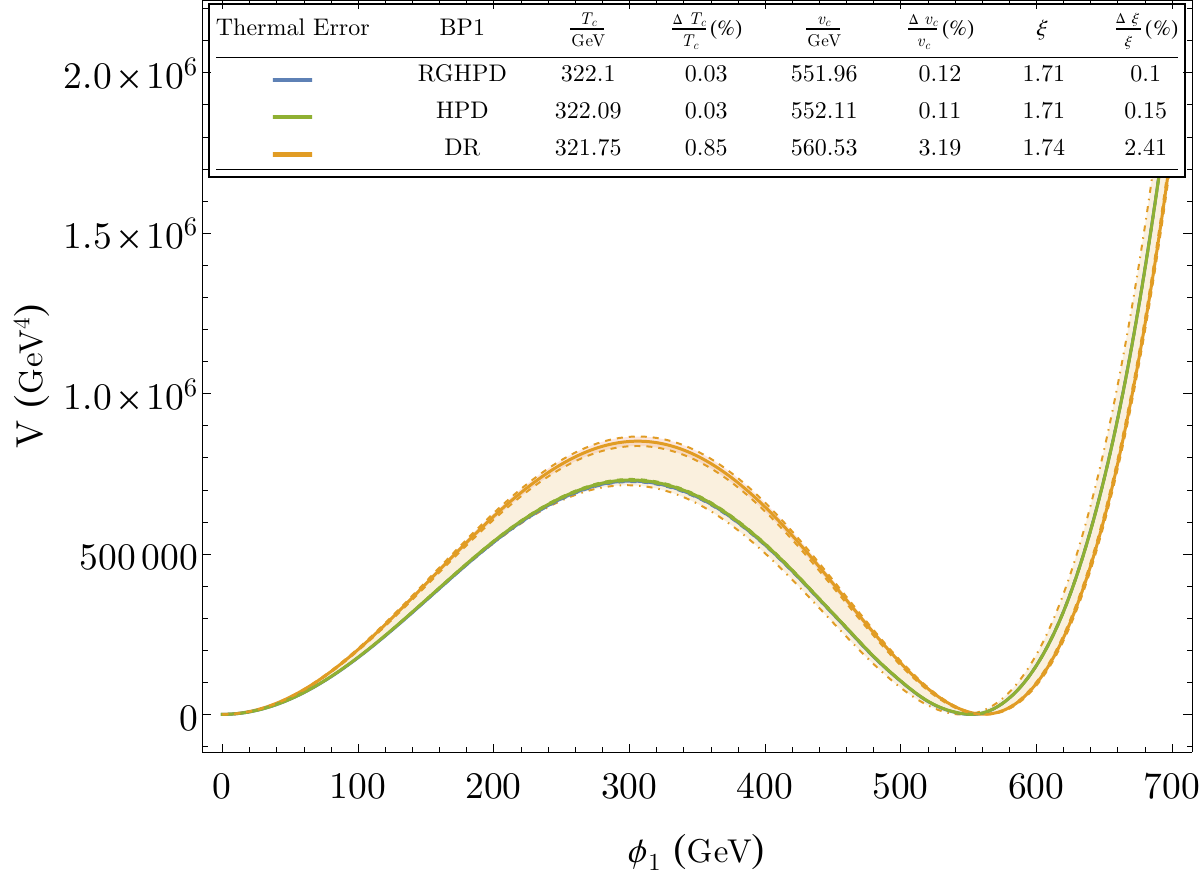} }}%
    \\
{{\includegraphics[width=7.5cm]{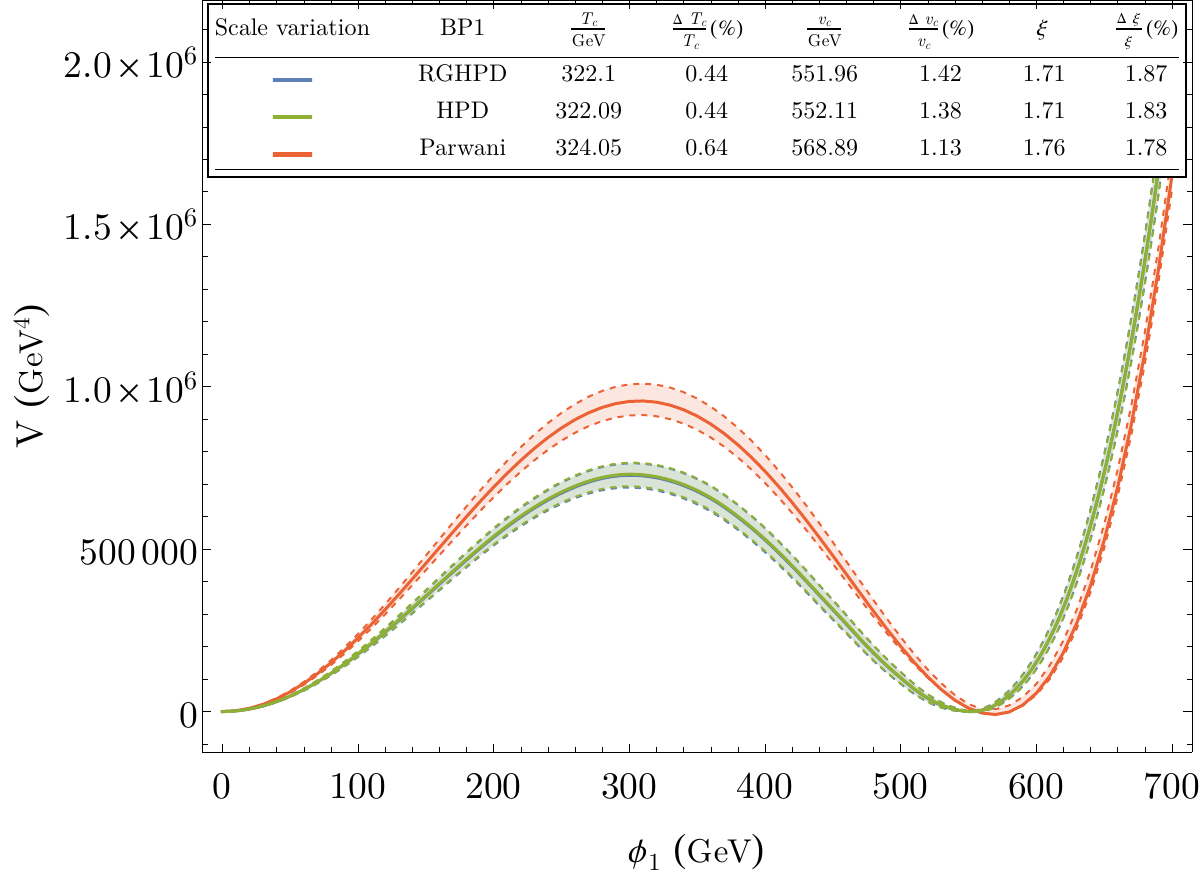} }}%
    &
    {{\includegraphics[width=7.5cm]{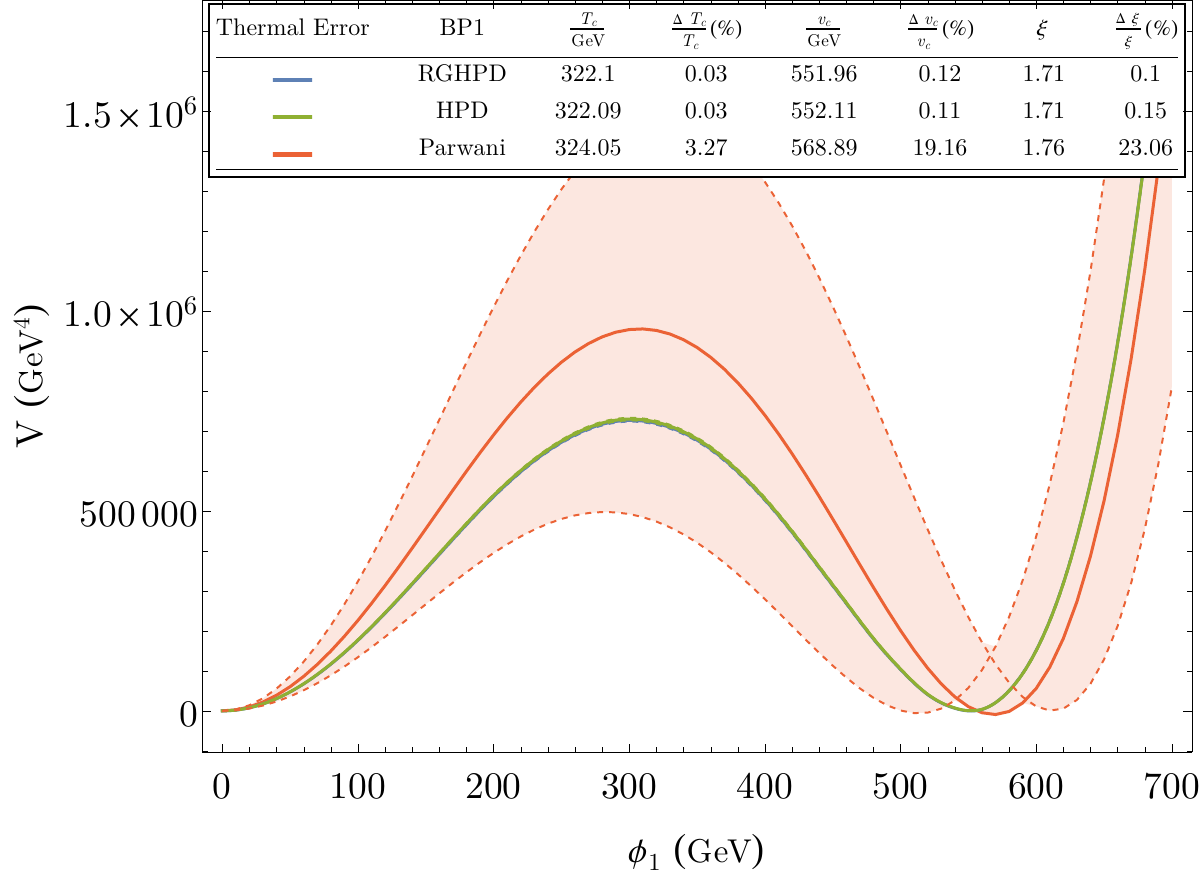} }}%
    \end{tabular}
    \caption{
    Effective potential  $V_\mathrm{eff}(\phi_1, T_c)$ for the BP1 benchmark of the two-scalar toy model, where the high-temperature approximation applies at $T_c$. 
    We compare (RG)HPD to HPD and Dimensional Reduction (top) and RGHPD to Parwani resummation (bottom). Bands indicate theoretical uncertainty from scale variation (left) and thermal error terms (right). For DR, the dashed error band corresponds to the effect of $\phi^6$ operators, while the dot-dashed band shows the impact of the high-temperature approximation on the two-loop sunset alone. 
    Note that all potentials are consistent with each other, taking respective theoretical uncertainties into account.
     } %
    \label{fig:BM1_pot_shape}%
\end{figure}

 \begin{figure}
    \centering
\includegraphics[width=1\linewidth]{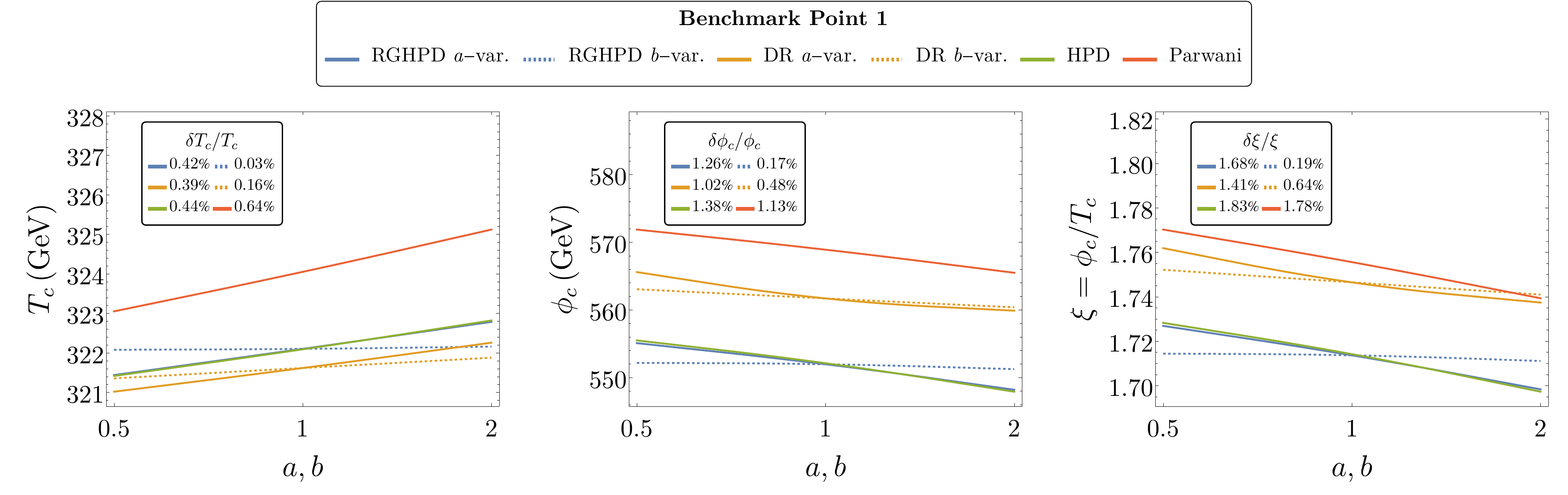}
    \caption{
    Scale variation of $T_c$ (left), $\phi_c$ (middle) and $\xi = \phi_c/T_c$ (right) for BP1 in HPD, RGHPD, DR and Parwani resummation. This does not show the larger thermal errors of DR and Parwani. Note that in this plot, the the $a$ and $b$ are varied separately, whereas in the scale variation Fig.~\ref{fig:BM1_pot_shape} the error presented is the total error accumulated by varying both.
    }
    \label{fig:RGIOPD_BM1_scale}
\end{figure}

\begin{figure}%
    \hspace*{-6mm}
    \centering
    \begin{tabular}{cc}    {{\includegraphics[width=7.5cm]{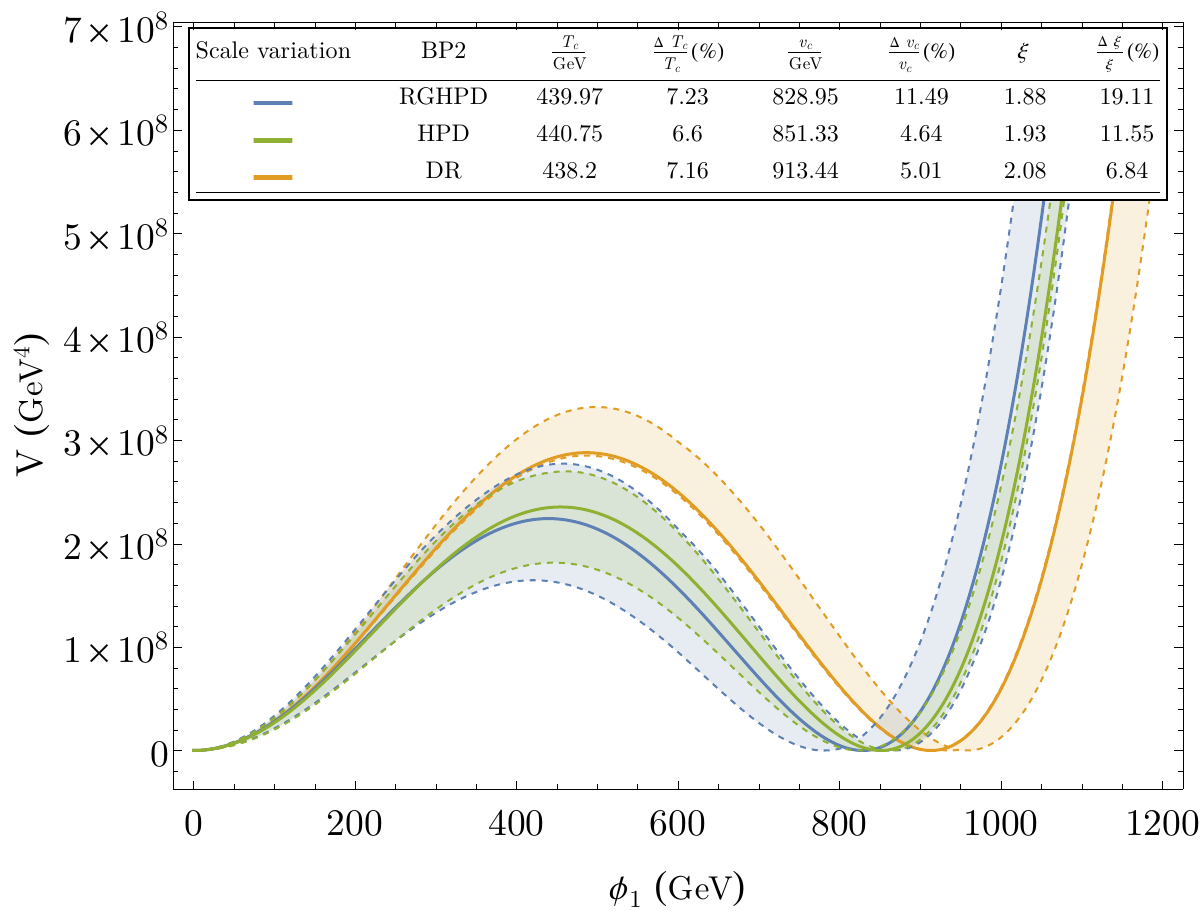} }}%
    &
    {{\includegraphics[width=7.5cm]{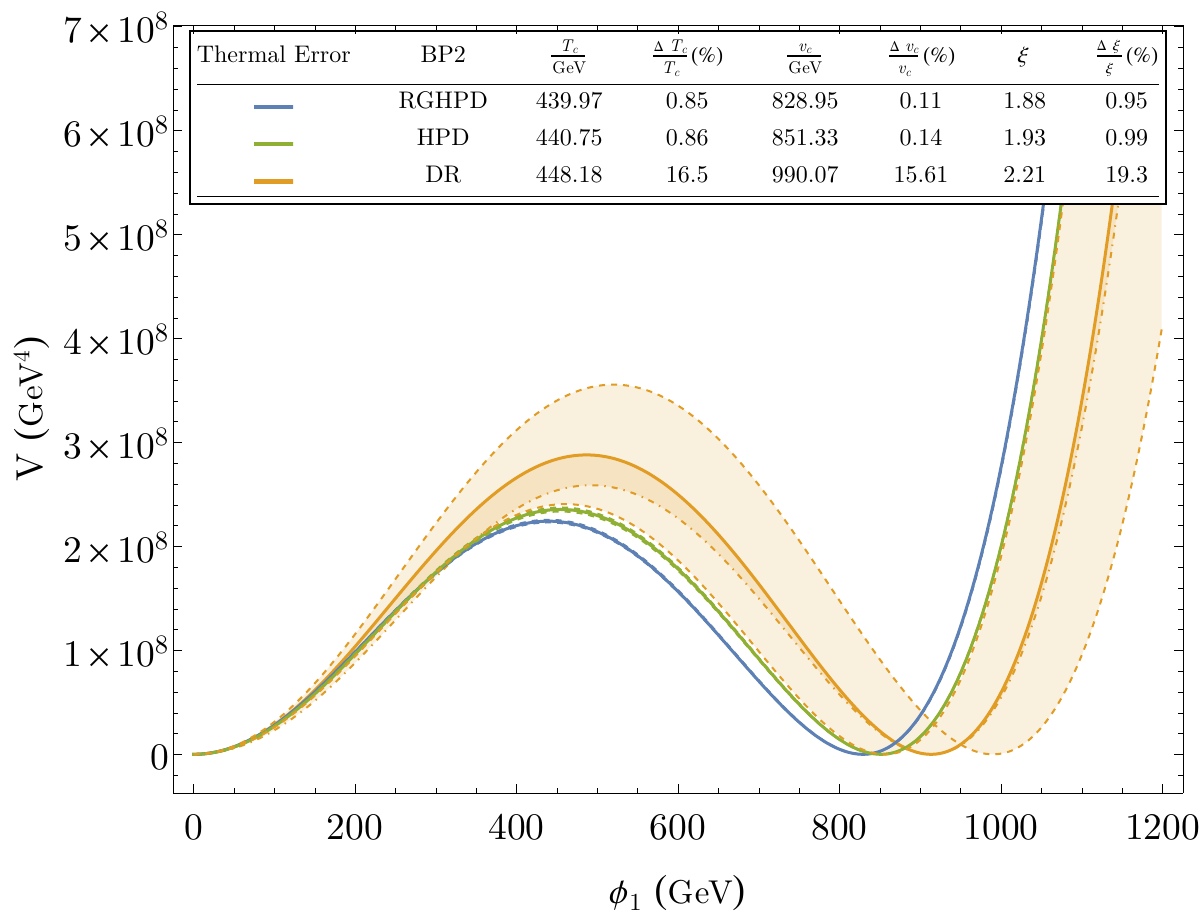} }}%
    \\
    {{\includegraphics[width=7.5cm]{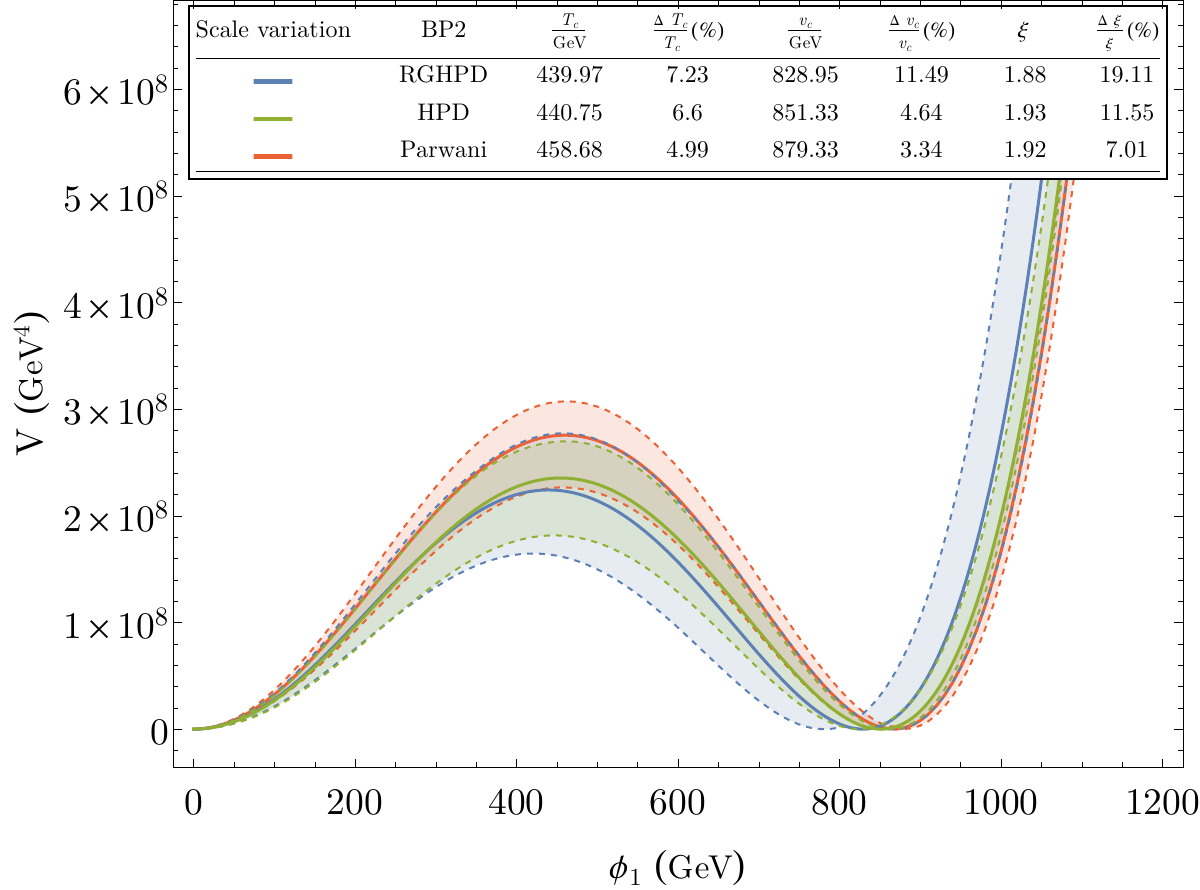} }}%
     &
     {{\includegraphics[width=7.5cm]{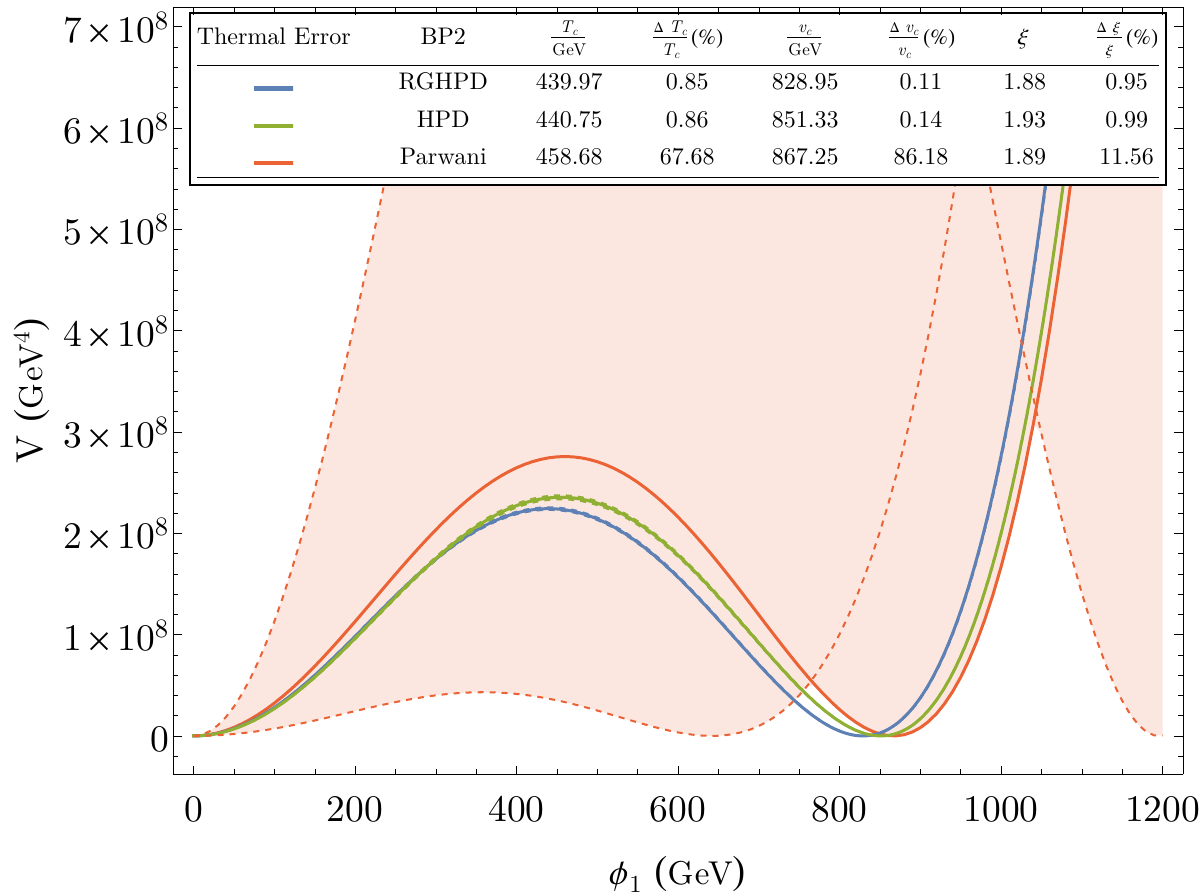} }}%
    \end{tabular}
    \caption{
    Effective potential  $V_\mathrm{eff}(\phi_1, T_c)$ for the BP2 benchmark of the two-scalar toy model, where the high-temperature limit does not apply.
    We compare (RG)HPD to HPD and Dimensional Reduction (top) and RGHPD to Parwani resummation (bottom). Bands indicate theoretical uncertainty from scale variation (left) and thermal error terms (right). For DR, the dashed error band corresponds to the effect of $\phi^6$ operators, while the dot-dashed band shows the impact of the high-temperature approximation on the two-loop sunset alone. 
    Note that all potentials are consistent with each other, taking respective theoretical uncertainties into account, but the central predictions for DR and Parwani systematically over-estimate the size of the potential barrier, and hence the gravitational wave signal, due to neglected Boltzmann suppressions. 
    } %
    \label{fig:BM2_pot_shape}%
\end{figure}

\begin{figure}
    \centering
    \includegraphics[width=1\linewidth]{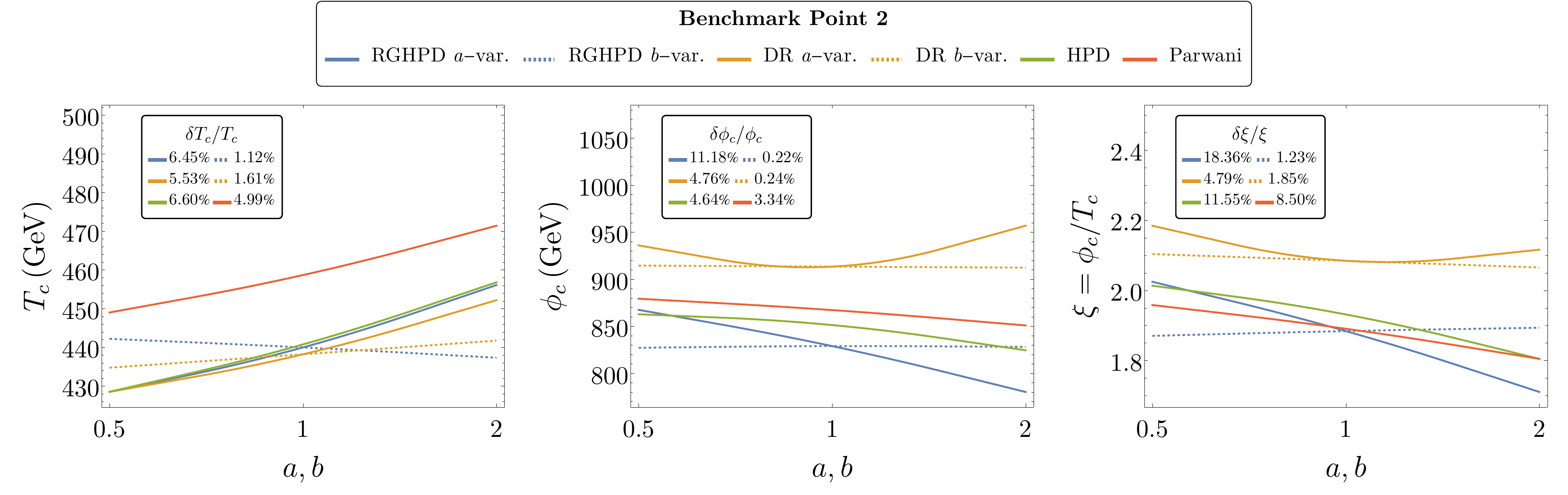}
    \caption{
    Scale variation of $T_c$ (left), $\phi_c$ (middle) and $\xi = \phi_c/T_c$ (right) for BP2 in HPD, RGHPD, DR and Parwani resummation. This does not show the much larger thermal errors of DR and Parwani.
    The higher scale variation of RGHPD compared to HPD is an artifact of the $M_2/M_1$ hierarchy, motivating incorporating the methods of~\cite{Manohar:2020nzp}.}
    \label{fig:BM2_scale}
\end{figure}

First, we  show the RG evolution of the couplings and masses for both BP1 and BP2 in Fig~\ref{fig:rge_ev}. For our choice of benchmarks, running is not to be a large effect for BP1. The couplings in BP2 run by at most $\sim 10\%$ between the matching scale and the critical temperature.

Figure.~\ref{fig:gap} compares the resummed masses $M_i^2$ for BP1 and 2 in RGHPD to Parwani. The differences are highly pronounced for the lighter scalar $\phi_1$. This also demonstrates that BP2 starts to exit the high-temperature regime as $\phi_1 \to \phi_c$, with $M_2(\phi_c) \sim 2.5 T_c$.

The effective potential at the critical temperature for the BP1 benchmark, where the high-temperature approximation applies, is shown in Figure~\ref{fig:BM1_pot_shape}.
The bands on the left show the theoretical uncertainty from scale variation (varying $a$ and $b$ between 0.5 and 2, see Section~\ref{sec. NumPres}), with minimal difference between HPD and RGHPD for this weakly coupled phase transition. 
The bands on the right show the theoretical uncertainty from adding or subtracting the thermal error terms: Eqn.~\eqref{e.deltaVTParwani} for Parwani, Eqn.~\eqref{e.deltaVbasketballHPD} for (RG)HPD, 
and for DR, Eqn.~\eqref{e.deltaVbasketballDR} (dashed boundaries) for the dim-6 error term and Eqn.~\eqref{e.deltaVsunDR} for the difference between the high-$T$ and full thermal sunset. 
These thermal errors are crucial in assessing the real precision of a calculation, since only they demonstrate the extreme unreliability of the Parwani effective potential. Also note the surprisingly large error introduced by the high-temperature approximation of the sunset in DR, to the point where this supplies the dominant theoretical uncertainty, which would be underestimated from the $\phi^6$ thermal error term alone. 

The variation of the most important phase transition parameters $T_c$, $\phi_c$ and $\xi = \phi_c/T_c$ as a function of scale choice is shown in Figure.~\ref{fig:RGIOPD_BM1_scale}, with $\sim \%$ uncertainty on $\xi$. This  demonstrates clearly that RGHPD and NLO DR have equivalent scale variation in the high-$T$ limit, as we have shown analytically.

BP2 is a more strongly coupled theory with a stronger first-order phase transition that starts to strain the high-temperature approximation as $\phi \to \phi_c$, with $M_2(\phi_c) \sim 2.5 T_c$. In particular, we chose a point where the singlet is light near origin compared to the thermal scale, but as phase transition proceeds, it becomes heavy near $\phi_c$. The effective potential with theoretical uncertainties, and scale variation of phase transition observables, are shown in Figures~\ref{fig:BM2_pot_shape} and~\ref{fig:BM2_scale} respectively.  
Even though the theoretical uncertainties are higher than for the weakly coupled BP1, the predictions of HPD and RGHPD are still of sufficient precision to be useful, with $\sim 5\%$ theoretical error on $\xi$. 
Note that HPD appears to have  smaller scale variation than the RG-improved RGHPD potential. This stems  from our choice of renormalization scale, $\mu_R^2 = (b \pi T/2)^2 + (a M_2(\phi))^2$, which introduces spurious logs given the large $M_2/M_1$ hierarchy of BP2 for parts of the $\phi_1$ phase transition, see Figure~\ref{fig:gap}. This motivates future work on applying the methods for correct two-scale RG improvement by~\cite{Manohar:2020nzp} to the HPD effective potential. In the interim, it may be advisable to use HPD instead of RGHPD for strong phase transitions with large mass hierarchies.

DR has smaller theoretical errors from scale variation, but this is spurious, and examining the impact of neglected $\phi^6$ contributions on the right makes clear that DR has much larger theoretical uncertainties in this intermediate-temperature regime. 
This is even more evident with the Parwani calculation, which has similar scale variation to HPD but enormous thermal error terms on the right. 
With all theoretical uncertainties taken into account, the (RG)HPD, DR and Parwani calculations are all consistent with each other, but the DR and Parwani central predictions systematically overestimate the barrier height, since Boltzmann suppressions of plasma degrees of freedom which radiatively generate the barrier but decouple as $\phi \to \phi_c$
are neglected. 
It is therefore clear that the (RG)HPD calculation is the more accurate one in this regime, and this would be even more so the case for supercooled transitions.

\begin{figure}%
    \centering
    \subfloat[\centering BP1]{{\includegraphics[width=7.5cm]{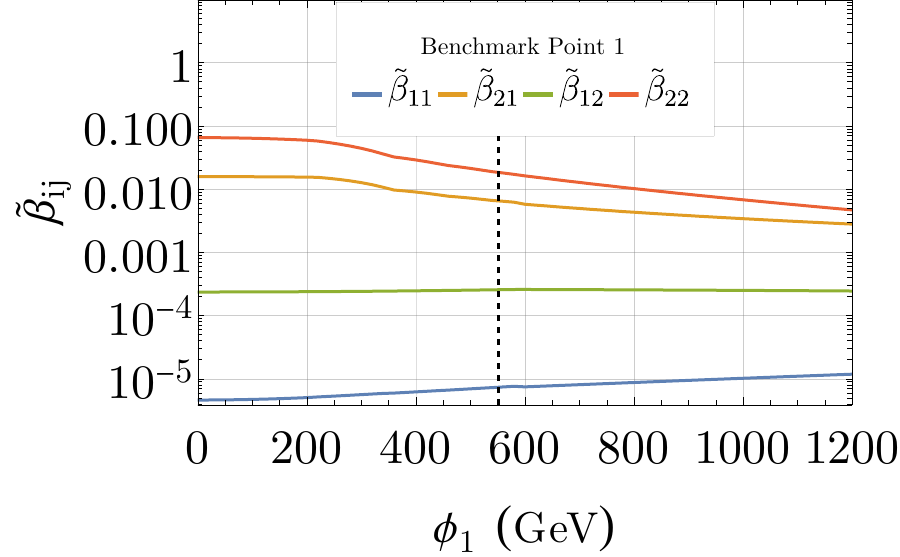} }}%
    \subfloat[\centering BP2]{{\includegraphics[width=7.5cm]{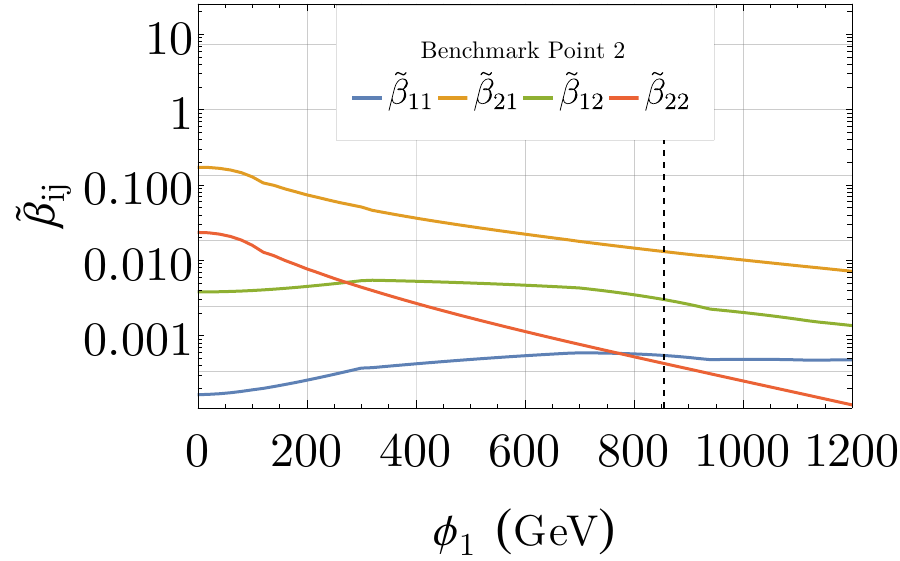} }}%
    \caption{
    The 4 expansion parameters $\tilde{\beta}_{ij}$ as defined in Eqn.~\eqref{e.betaijdefinitionALLT} at the critical temperature, demonstrating the HPD calculation is under tight perturbative control. The vertical dashed line indicates $\phi_c$.
    } %
    \label{fig:betas}%
\end{figure}

The small numerical variation of our results due to scale variation and the thermal error term clearly demonstrate that the HPD and RGHPD perturbative expansions are under control. This is also evident in the small sizes of the $\tilde{\beta_{ij}} \ll 1$ expansion parameters defined in Eqn.~\eqref{e.betaijdefinitionALLT}, shown in Figure~\ref{fig:betas}.
Note that, as discussed in Section~\ref{s.IRproblemFTQFT}, the small size of $\tilde \beta_{ij}$ is not directly related to the numerical uncertainty on phase transition observables, since the the latter rely on tree vs loop cancellations which amplify numerical errors. 

\begin{figure}%
    \centering
    \subfloat{{\includegraphics[width=1\linewidth]{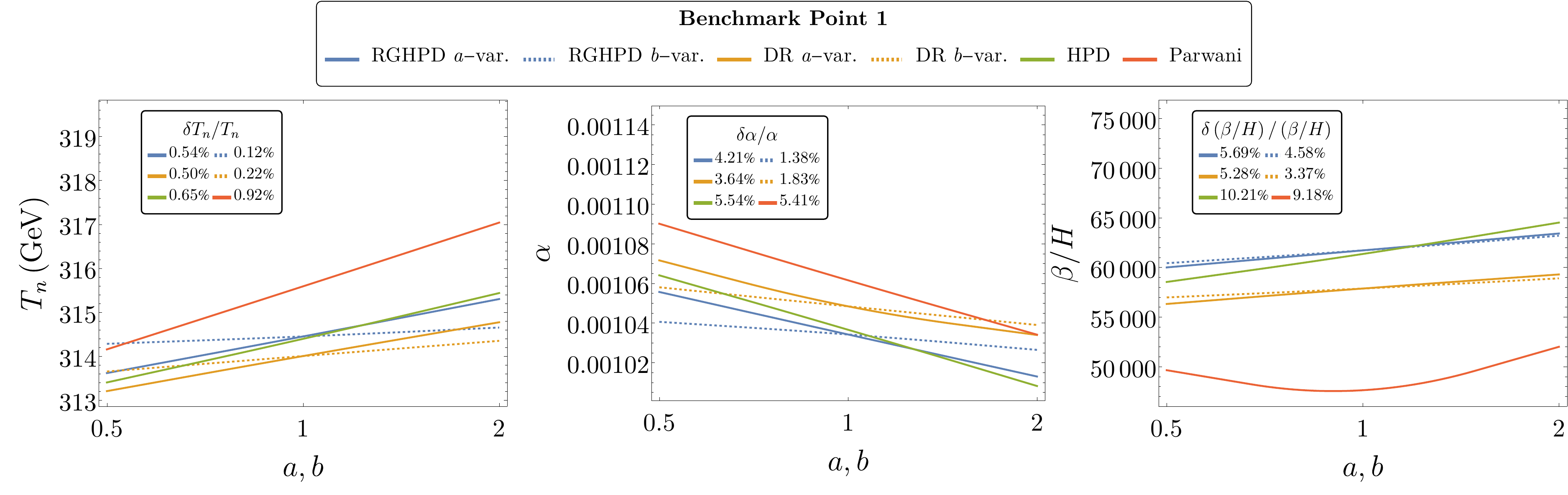} }}%
    \qquad
    \subfloat{{\includegraphics[width=1\linewidth]{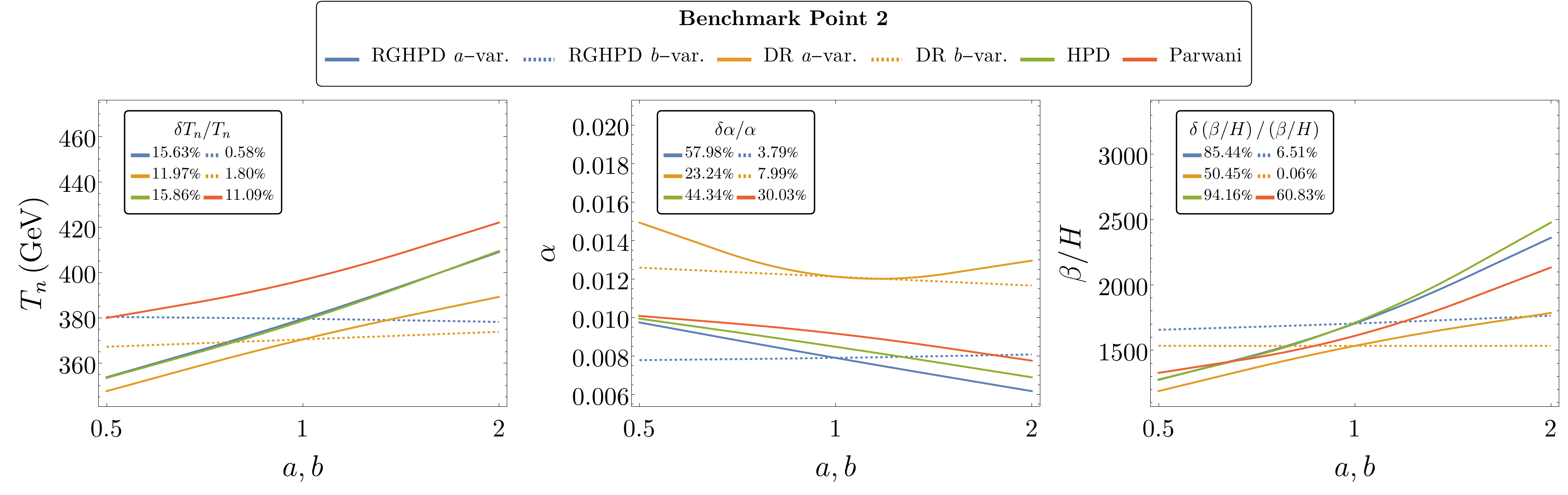} }}%
    \caption{We show scale variation of nucleation temperature (left), trace anomaly normalized by the critical density (middle) and inverse lifetime of transition (right) for BP1 (top) and BP2 (bottom). For BP1 we see that RGHPD and DR have comparable precision as expected. For BP2 we see that HPD and RGHPD have seemingly larger scale variation than DR, but DR has larger theoretical uncertainties from the high-$T$ approximation coming under strain, as parameterized by its variation when adding dim-6 operators.
    }%
    \label{fig:grav_waves_obs}%
\end{figure}

\begin{figure}%
    \centering
    \subfloat{{\includegraphics[width=1\linewidth]{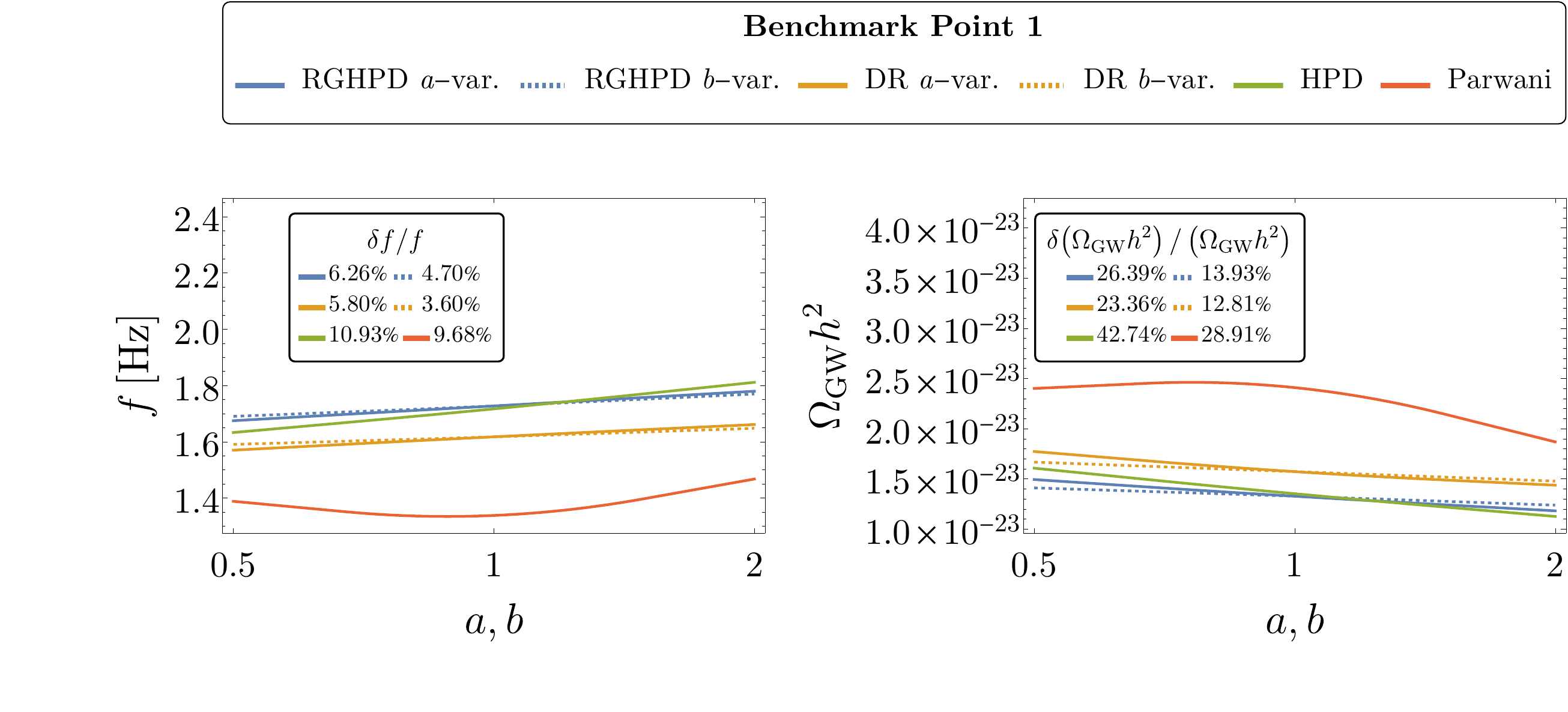} }}%
    \qquad
    \subfloat{{\includegraphics[width=1\linewidth]{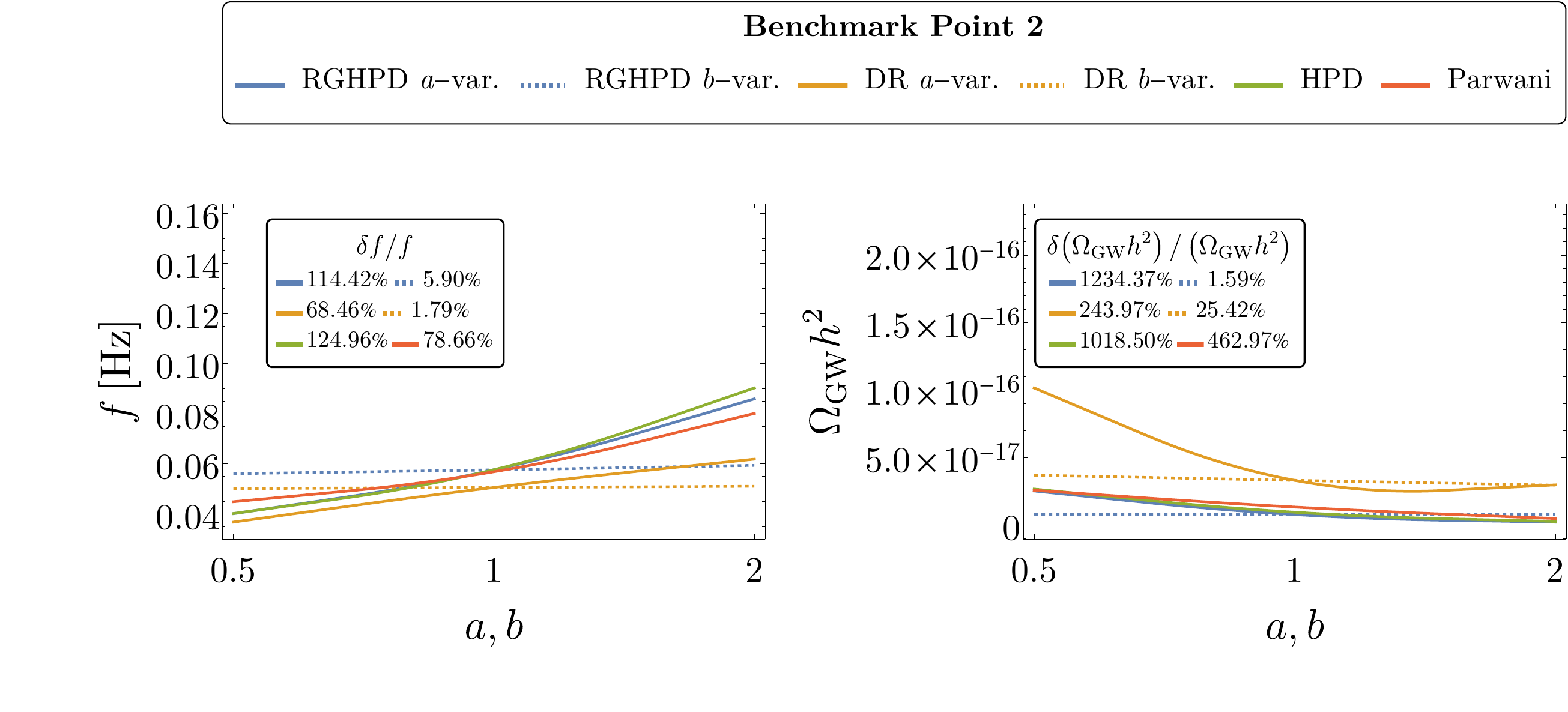} }}%
    \caption{We show scale variation of peak frequency (left) and amplitude (right) of gravitational waves for BP1 (top) and BP2(bottom). We see that for BP1, RGHPD and DR can predict amplitude within $\sim 20\%$. For the more strongly coupled BP2, HPD predicts the amplitude within an order of magnitude. The DR and Parwani theoretical errors are underestimated by scale variation, they both have larger errors than HPD once the thermal error term is taken into account.    
    }%
    \label{fig:grav_waves_freq_amp}%
\end{figure}


\begin{figure}%
\centering
\begin{tabular}{c}
\hspace*{-45pt}\includegraphics[width=0.9\textwidth]{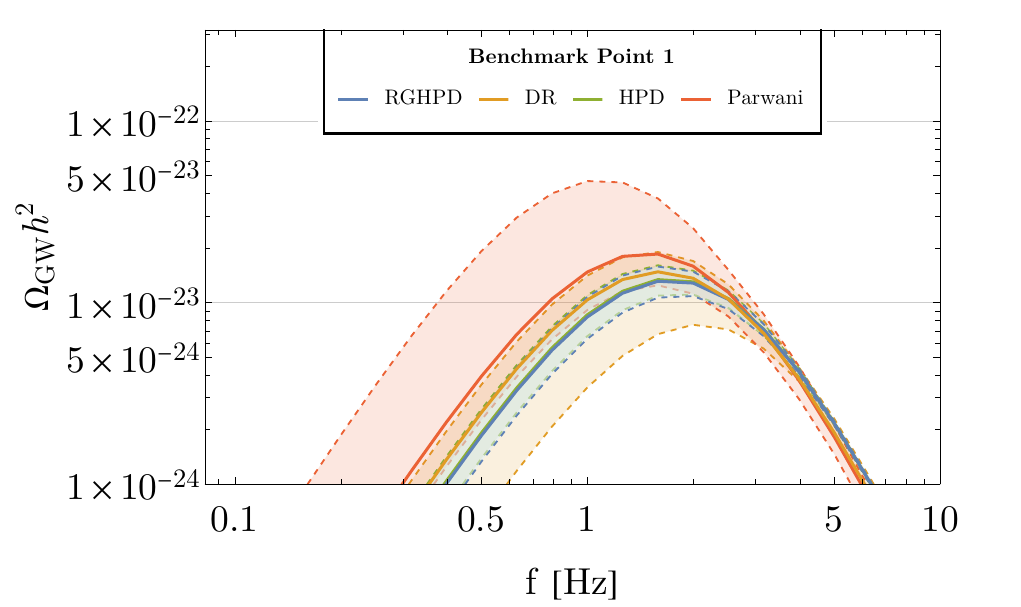} \\
\\
\includegraphics[width=0.9\textwidth]{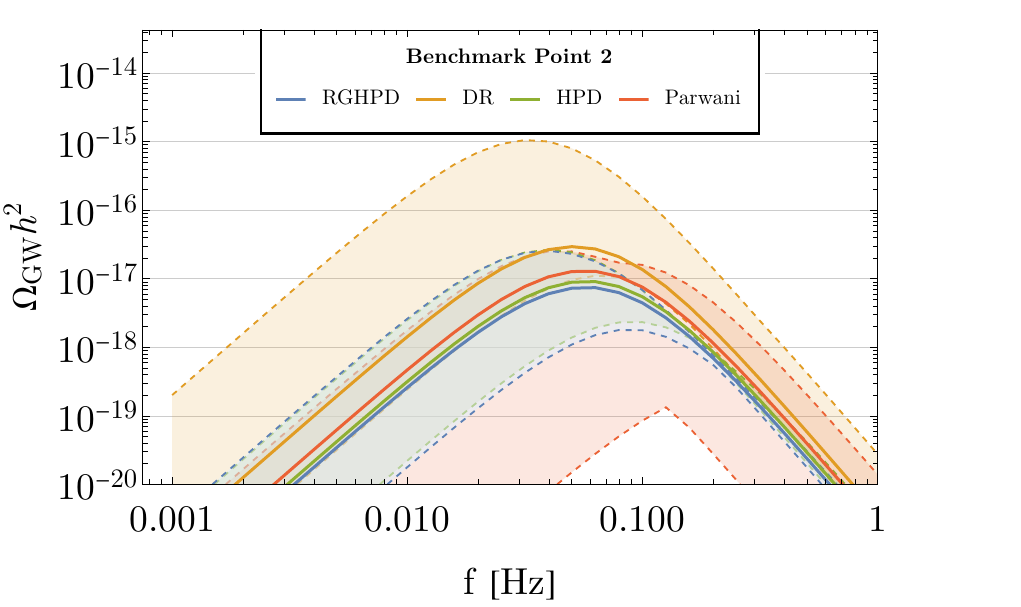} 
\end{tabular}
    \caption{
    Predicted spectrum of stochastic gravitational waves from the strong phase transition of BP1 (top) and BP2 (bottom), comparing RGHPD, DR and Parwani.
    }%
    \label{fig:GW_Sig}%
\end{figure}

We follow prescription described in Appendix~\ref{s.gravwave} to compute the spectra of stochastic gravitational waves produced by these first order phase transitions of BP1 and BP2, as well as relevant physical observables: nucleation temperature $T_n$, trace anomaly normalized by the critical density $\alpha$, and inverse lifetime of transition  $\beta/H$. (Note that our gravitational wave calculation is highly simplified and very much not cutting-edge, but this is sufficient for demonstrating the relative impact of different effective potential calculations.) The dependence of these computed phase transition parameters on the choice of scale is shown in Fig~\ref{fig:grav_waves_obs}. The corresponding predictions, and the scale variations of the peak amplitude and frequency for RGHPD, DR and Parwani are shown in Figure~\ref{fig:grav_waves_freq_amp}. We also show predicted spectrum of gravitational waves, including the total uncertainty band (defined as the largest range of outcomes derived from stacking scale variation and thermal error terms), in Figure~\ref{fig:GW_Sig}.

For BP1, all methods but Parwani have largest uncertainty coming from scale variation. RGHPD and HPD are almost equivalent and have theoretical uncertainty of $\sim 30\%$ on the final GW amplitude, about half the uncertainty of DR. Parwani has $\sim 100\%$ errors even for the weakly coupled BP1 phase transition.

For the more strongly coupled BP2, HPD has an uncertainty of about a factor of 10 on the GW amplitude, while DR systematically overestimates the amplitude (due to the neglect of Boltzmann suppression on radiative contributions to the barrier) and has a larger theoretical uncertianty, about a factor of 50. 
Various cancellations make the Parwani GW error band of similar size as HPD -- the theoreticaly uncertainty is more apparent in the large variation of the potential -- but this coincidence is not robust in more general theories.
It is interesting to note that the (RG)HPD theoretical uncertainty decreases dramatically at the high-frequency part of the GW spectrum, while both DR and Parwani maintain their large uncertainty.

Our numerical results lead us to expect that HPD would maintain its reliability and accuracy even for very strongly coupled or supercooled phase transitions, where both Parwani and DR would fail completely. 
All the numerical results and theoretical uncertainties are summarized in Table~\ref{t.summary}.

\section{Conclusions}
\label{s.conclusions}

\begin{table}
\centering
\scriptsize
\begin{tabular}{|c|c|c|c|c|c|}
\hline
Benchmark & Quantity & RGHPD & HPD & Parwani & DR \\ \hline

\multirow[c]{5}{*}{\begin{tabular}{c} \\ \\ Benchmark \\ Point 1 \end{tabular}}
& $\delta T_c/T_c$ & \begin{tabular}{c}
$0.44\%$ \\
$(0.03\%)$
\end{tabular} & \begin{tabular}{c}
$0.44\%$ \\
$(0.03\%)$
\end{tabular} & \begin{tabular}{c}
$0.64\%$ \\
$(3.27\%)$
\end{tabular}& \begin{tabular}{c}
$0.55\%$ \\
$(0.85\%)$
\end{tabular} \\ \cline{2-6}

& $\delta\phi_c/\phi_c$ & \begin{tabular}{c}
$1.42\%$ \\
$(0.12\%)$
\end{tabular} & \begin{tabular}{c}
$1.38\%$ \\
$(0.11\%)$
\end{tabular}& \begin{tabular}{c}
$1.13\%$ \\
$(19.16\%)$
\end{tabular}& \begin{tabular}{c}
$1.50\%$ \\
$(3.19\%)$
\end{tabular} \\ \cline{2-6}

& $\delta \xi/\xi$ &\begin{tabular}{c}
$1.87\%$ \\
$(0.1\%)$
\end{tabular}&\begin{tabular}{c}
$1.83\%$ \\
$(0.15\%)$
\end{tabular} & \begin{tabular}{c}
$1.78\%$ \\
$(23.06\%)$
\end{tabular} & \begin{tabular}{c}
$2.05\%$ \\
$(2.41\%)$
\end{tabular} \\ \cline{2-6}

& $\delta\Omega_{GW}/\Omega_{GW}$ &\begin{tabular}{c} 
$29.84\%$\\
$(3.33\%)$
\end{tabular} & \begin{tabular}{c} $42.74 \%$\\ $(3.52\%)$ \end{tabular}& \begin{tabular}{c}$28.91\%$\\$(126.08\%)$\end{tabular} & \begin{tabular}{c}$26.64\%$ \\$(55.03\%)$ \end{tabular} \\ \cline{2-6}

& $\delta f_{GW}/f_{GW}$ & \begin{tabular}{c}7.83\%\\ $(0.52\%)$\end{tabular}  & \begin{tabular}{c}10.93\%\\ $(0.55\%)$\end{tabular} &  \begin{tabular}{c}9.68\%\\ $(26.04\%)$\end{tabular} & \begin{tabular}{c}6.83\%\\ $(17.47\%)$ \end{tabular}\\ \hline

\multirow[c]{5}{*}{\begin{tabular}{c} \\ \\ Benchmark \\ Point 2 \end{tabular}}
& $\delta T_c/T_c$ &\begin{tabular}{c}
$7.23\%$ \\
$(0.85\%)$
\end{tabular} &\begin{tabular}{c}
$6.6\%$ \\
$(0.86\%)$
\end{tabular}  &\begin{tabular}{c}
$4.99\%$ \\
$(67.68\%)$
\end{tabular}   &\begin{tabular}{c}
$7.16\%$ \\
$(16.5\%)$
\end{tabular}  \\ \cline{2-6}

& $\delta\phi_c/\phi_c$ &\begin{tabular}{c}
$11.49\%$ \\
$(0.11\%)$
\end{tabular}  &\begin{tabular}{c}
$4.64\%$ \\
$(0.14\%)$
\end{tabular}  &\begin{tabular}{c}
$3.34\%$ \\
$(86.18\%)$
\end{tabular}  &\begin{tabular}{c}
$5.01\%$ \\
$(15.61\%)$
\end{tabular}  \\ \cline{2-6}

& $\delta \xi/\xi$ &\begin{tabular}{c}
$19.11\%$ \\
$(0.95\%)$
\end{tabular}  &\begin{tabular}{c}
$11.55\%$ \\
$(0.99\%)$ 
\end{tabular}  & \begin{tabular}{c}
$7.01\%$ \\
$(11.56)$
\end{tabular}  &\begin{tabular}{c}
$6.84\%$ \\
$(19.3\%)$
\end{tabular}  \\ \cline{2-6}

& $\delta\Omega_{GW}/\Omega_{GW}$ & \begin{tabular}{c}$1234.37\%$\\ $(11.92\%)$ \end{tabular} & \begin{tabular}{c}$1018.50\%$\\ $(11.84\%)$\end{tabular}  &\begin{tabular}{c}$245.29\%$\\ $(894.51\%)$ \end{tabular} &\begin{tabular}{c}$462.97\%$\\ $(969.33\%)$ \end{tabular} \\  \cline{2-6}

& $\delta f_{GW}/f_{GW}$ & \begin{tabular}{c}$114.57\%$\\ $(3.67\%)$ \end{tabular} & \begin{tabular}{c}$124.96\%$\\ $(3.54\%)$\end{tabular}  &\begin{tabular}{c}$78.66\%$\\ $(183.67\%)$ \end{tabular} & \begin{tabular}{c}$68.48\%$\\ $(16.81\%)$\end{tabular}  \\ \hline

\end{tabular}
\caption{
\label{t.summary}
Summary of theoretical uncertainties of phase transition parameters from the effective potential and resulting gravitational wave observables for BP1 and BP2 from RGHPD, HPD, Parwani and Dimensional Reduction.  Table entries show scale variation uncertainty, while the variation from thermal error terms is shown in brackets. Note that the combined uncertainty is roughly the product of these two uncertainty factors, which is most important for the combined error on $\delta \Omega_{GW}/\Omega_{GW}$ for DR in BP2, which is about a factor of 50 compared to HPD's factor of 10, see Figure~\ref{fig:GW_Sig} (bottom).
}
\end{table}

Understanding symmetry breaking in the early universe, whether in the electroweak sector or general scalar or dark sectors in various extensions of the SM, remains a central challenge for both theory and experiment. Strong first-order phase transitions could be detected via their produced stochastic gravitational wave background.
In the case of electroweak symmetry breaking, this would provide an independent probe of the shape of the Higgs potential, which could be correlated with future collider  measurements of Higgs self couplings~\cite{Cepeda:2019klc,DiMicco:2019ngk,deBlas:2019rxi,DiVita:2017eyz,DiVita:2017vrr,Curtin:2014jma,Huang:2016cjm,Katz:2014bha,Morrissey:2012db} to 
conclusively determine the physics of electroweak symmetry breaking and possible connections to baryogenesis. 
In the case of general scalar sectors, either weakly coupled to or at much higher masses than the SM, such a detection would provide a window into dynamics that may be completely inaccessible to colliders in the foreseeable future. 

Realizing this discovery potential requires reliable methods of computing the effective scalar potential $V_\mathrm{eff}(\phi,T)$ to at least second order in $\beta \sim \lambda T/m$. 
The simplest and most common method for computing the effective potential for phase transitions, Daisy resummation, is only correct to first order in $\beta$ and is hence unsuitable. 
Dimensional reduction is under excellent theoretical control and is highly accurate, but it is limited by the high-temperature expansion, and studying the most general phase transitions, in particularly strong and supercooled transitions, requires a framework which is valid at arbitrary temperatures. 
Furthermore, general exploration of BSM scenarios would be greatly facilitated by a calculational scheme that is easy to implement for a variety of theories.

In this paper, we presented (RG-improved) Hybrid Partial Dressing, (RG)HPD, which combines the simplicity of Daisy Resummation, relying only on one-loop analytical or pre-computed two-loop expressions, with the two-loop precision and accuracy of NLO Dimensional Reduction, without ever relying on the high-temperature approximation. With small $\mathcal{O}(\beta^3, \lambda^3)$ error terms, it allows for the straightforward and rigorously theoretically controlled calculation of precise gravitational wave predictions from weak, strong and super-cooled first-order phase transitions. 

We derived (RG)HPD in the context of pure scalar theories, but the method is readily generalized to theories with scalar cubic couplings, multiple scalar VEVs, fermions, and gauge bosons. The corresponding analysis is currently underway and will be presented in a forthcoming publication, which should make (RG)HPD a standard tool for the broad BSM exploration of first-order phase transition.

\subsection*{Acknowledgments}
The authors are especially grateful to Philip Schicho for many in-depth conversations and help with DRAlgo. We also thank Bryce Friesen, Keegan Humphrey, Philipp Klose, Stefan Stelzl and Carlos Tamarit for useful discussions. 
The work of RB, DC and AR was in part supported by Discovery Grants from the Natural Sciences
and Engineering Research Council of Canada, the
Canada Research Chair program, 
the Ontario Early Researcher Award, and the University of Toronto McLean Award.
The work of RB was additionally supported by the Canada Graduate Research Scholarship Doctoral award.
The work  of AR was additionally supported by the HQP Pooled Resources program of the Arthur B. McDonald Canadian Astroparticle Physics Research Institute. The work of ML was in part supported by the Natural Sciences and Engineering Research
Council of Canada. The work of JR was supported by the U.S. Department of Energy under grant contract number DE-FG02-05ER41367.

\subsection*{AI Disclosure}
In our analysis and the preparation of this manuscript, Claude and ChatGPT were used in the following way: to assist with simple coding segments, in particular to optimize plots, all of which were human-verified; to accelerate  the creation of certain nontrivial latex figures; to assist with many literature searches; and to provide feedback on draft versions of the paper to help identify inconsistencies, editing mistakes, etc.

\appendix

\section{Known Zero-Temperature Results for Scalar Theories}

\subsection{RG equations}
\label{a.RG}
Up to two loop, the beta functions for arbitrary theories can be computed with modern tools~\cite{Sartore:2020gou}.
Define:
\begin{align}
    \beta_X &= \mu \frac{d X}{d\mu} = \frac{1}{(4\pi)^2}\beta_X^{(1)}+\frac{1}{(4\pi)^4}\beta_X^{(2)}+
    \frac{1}{(4\pi)^4}\beta_X^{(3)}
\end{align}
 For the single-scalar theory, 
$\beta$ functions and anomalous dimensions are given below through 3-loop order from~\cite{Chung:1999xm}.
\begin{align}
    \beta_\lambda^{(1)} &=3\lambda^2\\
    \beta_\lambda^{(2)} &=-\frac{17}{3}\lambda^3\\
    \beta_\lambda^{(3)} &=\left(\frac{145}{8}+ 12 \zeta_3\right)\lambda^4\\
    \beta_{\nu^2}^{(1)} &=\lambda \nu^2\\
    \beta_{\nu^2}^{(2)} &=-\frac{5}{6}\lambda^2\nu^2\\
    \beta_{\nu^2}^{(3)} &= \frac{7}{2} \lambda^3 \nu^2\\
    \gamma_{\phi}^{(1)} &= 0\\
    \gamma_{\phi}^{(2)} &= \frac{1}{12}\lambda^2\\
    \gamma_{\phi}^{(3)} &= -\frac{1}{16} \lambda^3
\end{align}

For the two-scalar theory the one- and two-loop beta functions are: 
\begin{align}
    \beta^{(1)}_{\lambda_1}&=3 \lambda_1^2+3 \lambda_{12}^2 \\
    \beta^{(2)}_{\lambda_1}&=-\frac{17}{3} \lambda_1^3-5 \lambda_1 \lambda_{12}^2-12 \lambda_{12}^3 \\
    \beta^{(1)}_{\lambda_2}&=3 \lambda_2^2+3 \lambda_{12}^2 \\
    \beta^{(2)}_{\lambda_2}&=-\frac{17}{3} \lambda_2^3-5 \lambda_{12}^2 \lambda_2-12 \lambda_{12}^3 \\
    \beta^{(1)}_{\lambda_{12}}&=\lambda_1 \lambda_{12}+\lambda_{12} \lambda_2+4 \lambda_{12}^2 \\
    \beta^{(2)}_{\lambda_{12}}&=-6 \lambda_1 \lambda_{12}^2-6 \lambda_{12}^2 \lambda_2-\frac{5}{6} \lambda_1^2 \lambda_{12}-\frac{5}{6} \lambda_{12} \lambda_2^2-9 \lambda_{12}^3\\
    \beta^{(1)}_{\nu^2_{1}}&=\lambda_1 \nu^2_{1}+\lambda_{12}\nu^2_{2} \\
    \beta^{(2)}_{\nu^2_{1}}&=-\frac{5}{6} \lambda_1^2 \nu^2_{1}-\frac{1}{2} \lambda_{12}^2\nu^2_{1}-2 \lambda_{12}^2 \nu^2_{2} \\
    \beta^{(1)}_{\nu^2_{2}}&=\lambda_{12} \nu^2_{1}+\lambda_2 \nu^2_{2} \\
    \beta^{(2)}_{\nu^2_{2}}&=-2 \lambda_{12}^2\nu^2_{1}-\frac{5}{6} \lambda_2^2 \nu^2_{2}-\frac{1}{2} \lambda_{12}^2 \nu^2_{2}
\end{align}
At two-loop order, there is also field anomalous dimension (one-loop RGEs are 0):
\begin{align}
    \gamma_{X} &= \frac{1}{(4\pi)^4} \gamma_X^{(2)}\\
    \gamma_{\phi_1}^{(2)} &=\frac{1}{12}\lambda_1^2 + \frac{1}{4} \lambda_{12}^2\\
    \gamma_{\phi_2}^{(2)} &=\frac{1}{12}\lambda_2^2 + \frac{1}{4} \lambda_{12}^2
\end{align}

\subsection{One and Two-loop potential in scalar theories}
\label{a.potential}

Here we present known results for vacuum one and two-loop potential in scalar theories we consider in this paper. Starting with the single scalar theory, the effective potential up to two-loop order is given by~\cite{Chung:1999xm,Martin:2002iu}:
\begin{align}
    V_\textrm{eff}^{\mathrm{zero}}&=V^{(0)} + \frac{1}{16\pi^2} V^{(1)}+\frac{1}{(16\pi^2)^2} V^{(2)}\\
    V^{(1)} &= \frac{m^4}{4}\left(\log \frac{m^2}{\mu^2}-\frac{3}{2}\right)\\
    V^{(2)} &= V_\mathrm{fig-8} + V_\mathrm{sunset}\\
    V_\mathrm{fig-8}&= \frac{\lambda}{8} f_{SS}(m^2,m^2)\\
    V_\mathrm{sunset} &= -\frac{1}{12} (\lambda \phi)^2 f_{SSS} (m^2,m^2,m^2)\label{e.sunsetOneScalatZeroT}
\end{align}
where $V^{(0)}$ is tree level potential and:
\begin{align}
    f_{SS}(x,y) &= xy (\log(x/\mu^2)-1)(\log(y/\mu^2)-1)\\
    f_{SSS}(x,y,z)& = \frac{1}{2}(x-y-z)\log(y/\mu^2)\log(z/\mu^2)+\frac{1}{2}(y-x-z) \log(x/\mu^2) \log(z/\mu^2)\nonumber \\&+\frac{1}{2}(z-x-y)\log(x/\mu^2)\log(y/\mu^2)
 +2 x \log(x/\mu^2)+2 y \log(y/\mu^2)+2 z \log(z/\mu^2)\nonumber \\ &-\frac{5}{2}(x+y+z)-\frac{1}{2} \xi(x, y, z) \\
\xi(x, y, z)&=  R\{2 \log [(z+x-y-R) / 2 z] \log [(z+y-x-R) / 2 z]-\log (x / z) \log (y / z)\nonumber \\
& \left.-2 \operatorname{Li}_2[(z+x-y-R) / 2 z]-2 \operatorname{Li}_2[(z+y-x-R) / 2 z]+\pi^2 / 3\right\}\\
R &= \left(x^2 + y^2 + z^2 - 2xy-2xz-2yz\right)^{1/2}
\end{align}
For the two-scalar theory, the one and two loop potential is given by the sum of the below terms:
\begin{align}
    V^{(1)} &= \frac{m_1^4}{4}\left(\log \frac{m_1^2}{\mu^2}-\frac{3}{2}\right)+\frac{m_2^4}{4}\left(\log \frac{m_2^2}{\mu^2}-\frac{3}{2}\right) \\
    V_\mathrm{fig-8}&= \frac{\lambda_1}{8} f_{SS}(m_1^2,m_1^2)+ \frac{\lambda_2}{8} f_{SS}(m_2^2,m_2^2)+\frac{3\lambda_{12}}{4} f_{SS}(m^2,m^2)\\
    V_\mathrm{sunset} &= -\frac{1}{12} (\lambda_1 \phi_{1})^2 f_{SSS} (m_1^2,m_1^2,m_1^2)-\frac{1}{4}(\lambda_{12} \phi_{1})^2 f_{SSS} (m_1^2,m_2^2,m_2^2)\label{e.sunsetTwoScalatZeroT}
\end{align}

\subsection{Two-loop matching}
\label{a.matching}

Using one-loop and two-loop results, we construct our effective potential. Then one can define matching conditions:
\begin{align}
     V_\mathrm{eff}^{\mathrm{zero}\prime}(v) &=0\\
     m_{1,\mathrm{phys.}}^2 &= \left.\frac{\partial^2  V_\mathrm{eff}^\mathrm{zero}}{\partial \phi_1^2}\right|_{(\phi_1,\phi_2)\rightarrow (v,0)}\\
    m_{2,\mathrm{phys.}}^2 &= \left.\frac{\partial^2  V_\mathrm{eff}^\mathrm{zero}}{\partial \phi_2^2}\right|_{(\phi_1,\phi_2)\rightarrow (v,0)}\\
     \lambda_{1,\mathrm{phys.}} &= \left.\frac{\partial^4  V_\mathrm{eff}^\mathrm{zero}}{\partial \phi_1^4}\right|_{(\phi_1,\phi_2)\rightarrow (v,0)}\\
    \lambda_{2,\mathrm{phys.}} &= \left.\frac{\partial^2  V_\mathrm{eff}^\mathrm{zero}}{\partial \phi_2^4}\right|_{(\phi_1,\phi_2)\rightarrow (v,0)}\\
    \lambda_{12,\mathrm{phys.}} &= \left.\frac{\partial^4  V_\mathrm{eff}^\mathrm{zero}}{\partial \phi_1^2\phi_2^2}\right|_{(\phi_1,\phi_2)\rightarrow (v,0)}
\end{align}
This procedure gives us $\overline{MS}$ parameters from starting physical parameters. Note that we chose this particular renormalization condition to avoid dealing with two-loop momentum dependence in the mass matching, and can do the matching directly from the effective potential.

\section{Thermal Integrals} \label{thermalint}

\subsection{Thermal Function: High and Low temperature limit}
\label{thermalint_lowhighTlimits}

\begin{figure}
    \centering
    \includegraphics[width=0.8\linewidth]{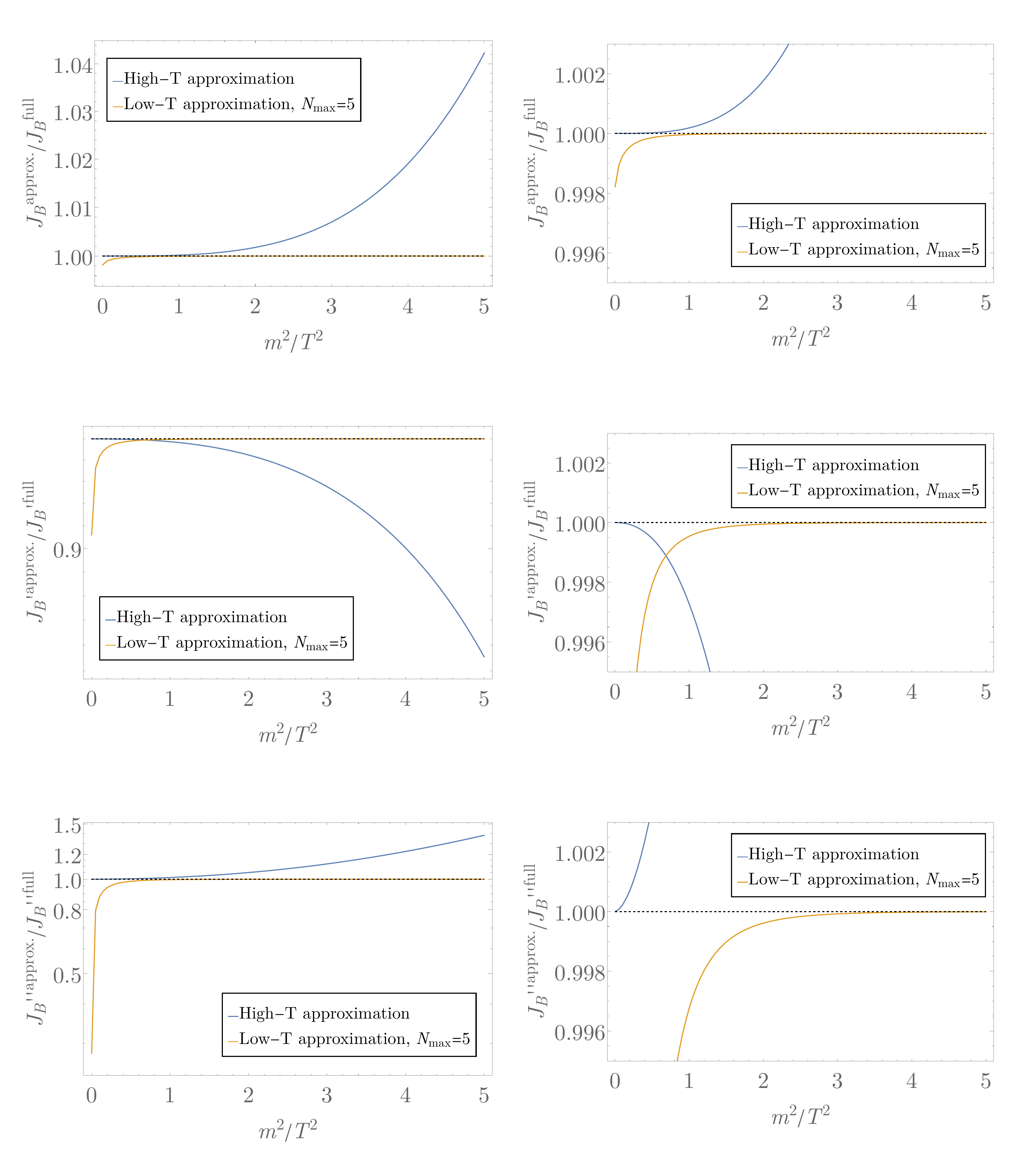}
    \caption{We show the ratio of the high/high-$T$ expansions of the bosonic thermal functions (first row) and the first and second derivatives (second and third row) to the full integral, where the latter is expanded to $N_\mathrm{max} = 5$. The left plots are the same as on the right, but with a zoomed-in vertical range to clearly illustrate the deviation from the full thermal integrals.
    We observe that high-$T$ (high-$T$) approximation is valid to within $0.1\%$ for $y^2 = m^2/T^2 < 2$,  but there is no way to join the two approximations without introducing discontinuities.}
    \label{fig:hightlowt1}
\end{figure}

The bosonic thermal function defines the thermal effective potential at one loop. It is given in terms of an integral defined in Eqn.~\eqref{eq: Tf}. It evaluates to an infinite sum, 
\begin{gather}
    J_B (y^2) = -\sum_{n=1}^{\infty} \frac{y^2}{n^2} K_2 (n y)\label{eq:bess} \ ,
\end{gather}
where $K_2$ are modified Bessel functions of second kind. This expansion has a useful and commonly used high temperature expansion:
\begin{gather}
    \label{e.JBhighT}
    J_B(y^2) \simeq - \frac{\pi^2}{45} + \frac{\pi^2}{12} y^2 - \frac{\pi}{6} (y^2)^{3/2} - \frac{1}{32}y^4 \log \frac{y^2}{a_b}  \quad \text{for }  y^2 \ll 1  \ ,
\end{gather}
where $a_b = 16 \pi^2 \exp(3/2 - 2 \gamma_E)$ and $\gamma_E$ is Euler-Mascharoni constant. Similarly, to capture the asymptotic behavior of the bosonic thermal function for large values of $y$ (ie. the low temperature limit), we can keep only first few terms from the sum: 
\begin{align}
    \label{e.JBlowT}
    J_B(y^2) &\simeq \sum_{n=1}^{N_{max}} \frac{y^2}{n^2} K_2 (n y) \quad \text{for } y^2 \gg 1 \ .
\end{align}
Previous partial dressing calculations (e.g. Ref.~\cite{Curtin:2016urg,Curtin:2022ovx})  choose $N_{max} \sim 3 - 5$ and define a piecewise thermal potential and its derivatives by joining Eqns.~\eqref{e.JBhighT} and~\eqref{e.JBlowT} around $y^2 \sim 1$.
However, even the small discontinuities this piecewise definition introduces into the potential derivatives, see Fig.~\ref{fig:hightlowt1}, can generate unphysical artifacts in numerical solutions of the gap equation. 

\begin{figure}
    \centering
    \includegraphics[width=0.8\linewidth]{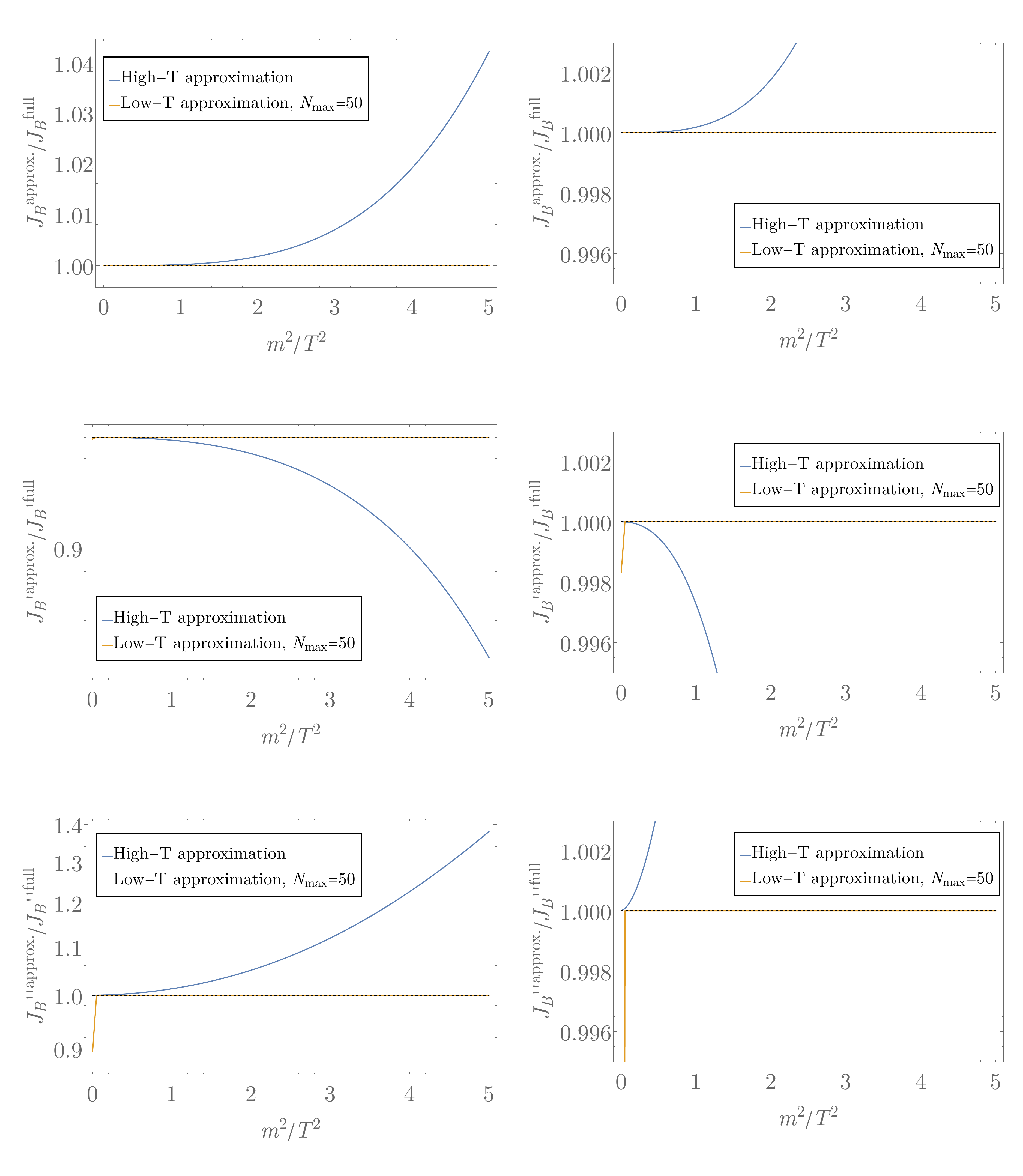}
    \caption{
    Same as Fig.~\ref{fig:hightlowt1}, except $N_\mathrm{max} = 50$ in the high-$T$ expansion, which lets the Bessel sum match the full thermal integral and its derivatives to better than 0.1\% accuracy even as $y^2 \to 0$.}
    \label{fig:hightlowt2}
\end{figure}

To solve this problem, we observe that one can simply use Eq.~\eqref{e.JBlowT} with larger $N_{max} \sim 50$. We show $J_B$ and its derivatives in Fig.~\ref{fig:hightlowt2}. This provides a fast analytical evaluation of $J_B$ and its derivatives that we can use in both the low- and high-temperature regime throughout our analysis. 

\subsection{Thermal Loop Integrals} \label{thermalint_loopints}
We now define various important thermal loop integrals that are used throughout our analysis. The following integrals include the zero-temperature contributions: 
\begin{alignat}{2}
\label{e.I0}
    I_0(m) &= - &&T \sum_{n=-\infty}^{\infty}\int\frac{d^3p}{(2\pi)^3} \log(p^2 + m^2 + (2\pi n T)^2) \ , \\
    \label{e.Ij}
    I_j(m) &= &&T \sum_{n=-\infty}^{\infty}\int\frac{d^3p}{(2\pi)^3} \frac{1}{(p^2 + m^2 + (2\pi n T)^2)^j} \ .
\end{alignat}
It is useful to separate zero ($\bullet$) and non-zero ($\times$) matsubara mode loops:
\begin{alignat}{2}
    I^\bullet_0(m) &= - &&T \int\frac{d^3p}{(2\pi)^3} \log(p^2 + m^2)   \ ,\\
    I^\bullet_j(m) &= &&T \int\frac{d^3p}{(2\pi)^3} \frac{1}{(p^2 + m^2)^j}  \ ,\\
    I^\times_0(m) &= - &&T \sum_{n\neq 0}\int\frac{d^3p}{(2\pi)^3} \log(p^2 + m^2 + (2\pi n T)^2) \ , \\
        I^\times_j(m) &= &&T \sum_{n\neq 0}\int\frac{d^3p}{(2\pi)^3} \frac{1}{(p^2 + m^2 + (2\pi n T)^2)^j} \ .
\end{alignat}
In Table.~\ref{t:thlo}, we show the first few thermal loop integrals $I_j$ that we will use when evaluating diagrams, explicitly calculated in high-temperature approximation including the zero-$T$ contributions \cite{Boyd:1993tz}.
\begingroup
\setlength{\tabcolsep}{10pt} 
\renewcommand{\arraystretch}{1.3} 
\begin{table}
\begin{center}
\caption{Thermal scalar loop integrals where $L_R \equiv \log\left(\frac{\mu_0^2}{16 \pi^2 e^{-2\gamma_E}T^2}\right)$}
\label{t:thlo}
\begin{tabular}{|| c |c|c|c|c|c ||}
\hline
$j$ & 0 & 1 & 2 & 3 & 4 \\
\hline
$I^\bullet_j(m)$
    & $ \frac{T m^3}{6 \pi} $
    & $ -\frac{T m}{4 \pi} $
    & $ \frac{T}{8 \pi m} $
    & $ \frac{T}{32 \pi m^3} $
    & $\frac{T}{64 \pi m^5}$
     \\
\hline
$I^\times_j(m)$
    & $ \frac{\pi^2 T^4}{45}-\frac{T^2 m^2}{12} + \frac{m^4 L_R}{32 \pi^2}  $
    & $ \frac{T^2}{12} - \frac{m^2 L_R}{16 \pi^2}  $
    & $ \frac{L_R}{16 \pi^2}  $
    & 0
    & 0
    \\
\hline 
\end{tabular}
\end{center}
\end{table}
\endgroup

We also introduce the two-propagator integral with two different masses \cite{Bahl:2024ykv}:
\begin{align}
    I_2(m_1,m_2) &= T \sum_{n=-\infty}^{\infty}\int\frac{d^3p}{(2\pi)^3} \frac{1}{(p^2 + m_1^2 + (2\pi n T)^2) (p^2 + m_2^2 + (2\pi n T)^2)} \\
   &= \frac{I_1(m_2) - I_1(m_1)}{m_1^2-m_2^2}  \label{e.I2diffmass} \\ 
   &\approx  \frac{T}{4\pi (m_1 +m _2)} + \frac{L_{R}}{16 \pi^2} \ . 
\end{align}
where the high-temperature limit was taken in the last line. The three-propagator integral with two different masses follows
\begin{align}
    I_3(m_1,m_2, m_2) &= T \sum_{n=-\infty}^{\infty}\int\frac{d^3p}{(2\pi)^3} \frac{1}{(p^2 + m_1^2 + (2\pi n T)^2) (p^2 + m_2^2 + (2\pi n T)^2)^2} \\
   &= - \frac{\partial}{\partial m_2^2} I_2(m_1,m_2)
   \label{e.I3diffmass} \\ 
   &\approx  \frac{T}{8\pi} \frac{1}{m_2 (m_1 +m _2)^2} \ ,
\end{align}
where the high-temperature limit was taken in the last line. 

Instead of splitting the thermal loop integral in zero and no-zero Matsubara modes, we can also split $I_j$ in zero-temperature and thermal only contribution $I_{j,\mathrm{th}}$. We actually did this in Eq.~\eqref{eq:V1split-ZeroandT} for the 1-loop effective potential. We get $I_{0,\mathrm{th}}$, the thermal only part of $I_0$ defined in Eq~\eqref{e.I0}, exactly like we get $V_{1,\mathrm{th}}$ in Eq.~\eqref{e.VTH} up to the symmetry factor of the potential 
\begin{equation}
    I_{0,\mathrm{th}} = - \frac{T^4}{\pi^2} \int_0^\infty dx\, x^2 \log\left[ 1 - e^{-\sqrt{x^2+m^2/T^2}} \right] = - \frac{T^4}{\pi^2} J_B(m^2/T^2) \ ,
\end{equation}
where $J_B(m^2/T^2)$ is the bosonic thermal function define in Eq.~\eqref{eq: Tf}. We get $I_{1,\mathrm{th}}$ and $I_{2,\mathrm{th}}$ from the derivative of $I_{0,\mathrm{th}}$ with respect to $m^2$, as we get Eq.~\eqref{e.Ij} from Eq.~\eqref{e.I0}
\begin{alignat}{4}
    I_{1,\mathrm{th}} &= - &&\frac{\partial I_{0,\mathrm{th}}}{\partial m^2} &&= &&\frac{T^4}{\pi^2} \frac{\partial J_B }{\partial m^2} \label{e.I1th} \ , \\
    I_{2,\mathrm{th}} &= &&\frac{\partial^2 I_{0,\mathrm{th}}}{\partial (m^2)^2} &&= - &&\frac{T^4}{\pi^2} \frac{\partial^2 J_B }{\partial(m^2)^2} \label{e.I2th} \ . 
\end{alignat}

In Table.~\ref{t:thloNOzeroT}, we show the first few thermal loop integrals $I_{j,\mathrm{th}}$ without zero-$T$ contributions, explicitly calculated in high-temperature approximation. Removing zero-$T$ contributions comes down to $L_R \to L_{R,\mathrm{th}} = \log\left(\frac{M^2}{a_b T^2}\right)$ in Table~\ref{t:thlo}.
\begingroup
\setlength{\tabcolsep}{10pt} 
\renewcommand{\arraystretch}{1.3} 
\begin{table}[h]
\begin{center}
\caption{Thermal scalar loop integrals without zero-$T$ contributions \\ where $L_{R,\mathrm{th}} = \log\left(\frac{M^2}{a_b T^2}\right)$}
\label{t:thloNOzeroT}
\begin{tabular}{|| c |c|c|c|c|c||}
\hline
$j$ & 0 & 1 & 2 & 3 & 4\\
\hline
$I^\bullet_{j,\mathrm{th}}(m)$
    & $ \frac{T m^3}{6 \pi} $
    & $ -\frac{T m}{4 \pi} $
    & $ \frac{T}{8 \pi m} $
    & $\frac{T}{32 \pi m^3}$
    & $\frac{T}{64 \pi m^5}$
     \\
\hline
$I^\times_{j,\mathrm{th}}(m)$
    & $ \frac{\pi^2 T^4}{45}-\frac{T^2 m^2}{12} + \frac{m^4 L_{R,\mathrm{th}}}{32 \pi^2}  $
    & $ \frac{T^2}{12} - \frac{m^2 L_{R,\mathrm{th}}}{16 \pi^2}  $
    & $ \frac{L_{R,\mathrm{th}}}{16 \pi^2}  $
    & 0 & 0 
     \\
\hline 
\end{tabular}
\end{center}
\end{table}
\endgroup

Similarly for the two-propagator integral with two different masses, removing the zero-$T$ contributions in the high-temperature, it yields
\begin{align}
    I_{2,\mathrm{th}}(m_1,m_2)  =  \frac{T}{4\pi (m_1 +m _2)} + \frac{L_{R,\mathrm{th}}}{16 \pi^2} \ .
\end{align}

\begin{figure}[!ht]
\subfloat[$\mathcal{H}_3$\label{subfig-1:SunsetIntDiag}]{\includegraphics[valign=c,scale=1]{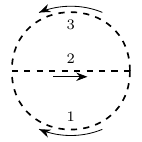}}
    \hfill
    \subfloat[$\mathcal{H}_4$\label{subfig-2:SunsetIntDiag}]{\includegraphics[valign=c,scale=1]{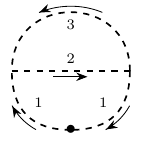}}
    \hfill
    \subfloat[$\mathcal{H}_5^{(1)}$\label{subfig-3:SunsetIntDiag}]{\includegraphics[valign=c,scale=1]{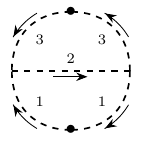}}
    \hfill
    \subfloat[$\mathcal{H}_5^{(2)}$\label{subfig-4:SunsetIntDiag}]{\includegraphics[valign=c,scale=1]{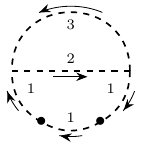}}
    \caption{Diagrammatic representation of the sunset integrals in Eqs.~\ref{eq:H3}–\ref{eq:H52}, with mass indices indicated on each propagator. See the corresponding equations for the explicit momentum dependence.}
    \label{fig:SunsetIntDiag}
  \end{figure}

We also need to define the sunset integrals described in a diagrammatic way in Fig.~\ref{fig:SunsetIntDiag}. The general forms of the three-, four-, and five‑propagator sunset integrals needed for our calculations are given below
\begin{align}
    \mathcal{H}_3 (m_1,m_2,m_3) &= T^2 \SumInt_{K,P} \frac{1}{(K^2+m_1^2)((P+K)^2 + m_2^2)(P^2+ m_3^2)}  \label{eq:H3} \ , \\
    \mathcal{H}_4 (m_1,m_2,m_3) &= T^2 \SumInt_{K,P} \frac{1}{(K^2+m_1^2)^2((P+K)^2 + m_2^2)(P^2+ m_3^2)} \label{eq:H4} \ ,\\
    \mathcal{H}_5^{(1)} (m_1,m_2,m_3) &= T^2 \SumInt_{K,P} \frac{1}{(K^2+m_1^2)^2 ((P+K)^2 + m_2^2)(P^2+ m_3^2)^2}  \label{eq:H51} \ ,\\
    \mathcal{H}_5^{(2)} (m_1,m_2,m_3) &=  T^2 \SumInt_{K,P} \frac{1}{(K^2+m_1^2)^3((P+K)^2 + m_2^2)(P^2+ m_3^2)} \ ,\label{eq:H52}
\end{align}
 where we define the euclidean four-momenta $P = (\omega_p, p)$ and  $K = (\omega_k, k)$ with the bosonic matsubarra modes $\omega_p, \omega_k$, and $\SumInt_{K} = \sum_{\omega_k} \int \frac{d^3 k}{(2\pi^3)}$. 

In the high-temperature approximation, we get the following leading results, which exclude the zero-temperature contributions \cite{Parwani:1991gq, Arnold:1992rz, Smet:2001un, Ekstedt:2020qyp}
\begin{align}
    \mathcal{H}_3 (m_1,m_2,m_3) &= -\frac{T^2}{32 \pi^2} \left( \log \left[ \frac{(m_1 + m_2 + m_3)^2}{9 \mu T} \right] + 1.65 \right) \label{eq:H3highT} \ ,\\
    \mathcal{H}_4 (m_1,m_2,m_3) &= \frac{T^2}{32\pi^2}\frac{1}{m_1(m_1+m_2+m_3)} \label{eq:H4highT} \ ,\\ 
    \mathcal{H}_5^{(1)} (m_1,m_2,m_3) &= \frac{T^2}{64\pi^2} \frac{1}{m_1 m_3 (m_1+m_2+m_3)^2} \label{eq:H51highT} \ ,\\
    \mathcal{H}_5^{(2)} (m_1,m_2,m_3) &= \frac{T^2}{128\pi^2} \frac{2 m_1 + m_2 + m_3}{m_1^3 (m_1+m_2+m_3)^2} \label{eq:H52highT} \ ,
\end{align}
where $\mathcal{H}_4$,  $\mathcal{H}_5^{(1)}$ and $\mathcal{H}_5^{(2)}$ are calculated from $\mathcal{H}_3$ by taking derivatives with respect to $m_1^2$ and $m_3^2$.
We emphasize that this leading high-temperature term can be simply understood as one of the sunset sub-loops being a thermal loop (and keeping it's $T^2$ piece), and the other loop being a zero-temperature loop. These expressions do not contain purely zero-temperature contributions. Therefore, they turn off when taking $T\to0$.

The sunset integral expressions \eqref{eq:H4highT}, \eqref{eq:H51highT} and \eqref{eq:H52highT} include the overlapping momenta. To determine the order in $\beta$ of a given diagram, we will neglect the overlapping momenta between the two loops by treating, as shown in Figure \ref{fig:SunsetIntDiag}, the lower propagator(s) as the first loop and the middle and upper propagators as the second loop. This allows us to systematically establish the order in $\beta$ of the lollipop and sunset tadpole diagrams of Sections \ref{appBETA:secLollipop} and \ref{appBETA:secSunset} using the parameters defined in Equations \eqref{AppBETA:parameters} and \eqref{AppBETA:MIXparameters}. Neglecting the overlapping momenta, these expressions become
\begin{alignat}{2}
    \mathcal{H}_4 (m_1,m_2,m_3) &\approx I^\bullet_2(m_1) I^\bullet_2 (m_2, m_3) &&= \frac{T^2}{32 \pi^2}\frac{1}{ m_1 (m_2+m_3)} 
    \label{eq:H4highT_NOoverlapping} \ ,\\ 
    \mathcal{H}_5^{(1)} (m_1,m_2,m_3) &\approx I^\bullet_2(m_1) I_3^\bullet (m_2, m_3, m_3) &&= \frac{T^2}{64 \pi ^2}\frac{1}{m_1 m_3 (m_2+m_3)^2}
    \label{eq:H51highT_NOoverlapping} \ ,\\
    \mathcal{H}_5^{(2)} (m_1,m_2,m_3) &\approx
    I^\bullet_3(m_1) I_3^\bullet (m_2, m_3) &&= \frac{T^2}{128 \pi ^2} \frac{1}{m_1^3 (m_2+m_3)} \ ,
    \label{eq:H52highT_NOoverlapping}
\end{alignat}
where the relevant thermal integrals are defined above.

\subsection{Sunset thermal integral outside of the high-$T$ approximation}
\label{a.sunsetlowT}

In the previous section we showed expressions for thermal sunset, written in the the high-$T$ approximation. However, as one of our goals is to be able to do calculations outside of the range of validity of this approximation, we would want to be able to generalize sunset thermal integral. To do this, one utilizes Saclay decomposition~\cite{Smet:2001un}. In single $\phi^4$ theory, the sunset thermal integral is given by:
\begin{align}
    V_\mathrm{sun} &= -\frac{\lambda^2 \phi^2}{12} \left(G_0(m^2) + G_1(m^2,T) + G_2 (m^2,T)\right)
\end{align}
where each subscript corresponds to number of Bose-Einstein factors. Thus, $G_0$ is temperature independent piece (zero-temperature sunset, as presented in~\ref{a.potential}. The second piece is the `zero $\times$ thermal' as it has one factors of Bose-Einstein factor. We can write it as:
\begin{align}
    G_1(m^2,T) &= 3 \int \frac{d^3 q_1}{(2\pi)^3}\frac{d^3 q_2}{(2\pi)^3} \frac{n_B(q_1)}{8 E_{q_1}E_{q_2}E_{q_3}} \left(\frac{2}{E_{q_1}+ E_{q_2}+E_{q_3}}+\frac{2}{-E_{q_1}+ E_{q_2}+E_{q_3}}\right)
\end{align}
where $q_3 = - (q_1+q_2)$ and $E_q = \sqrt{q^2+m^2}$. After performing $q_1$ integral, and removing the divergences, we get~\cite{Smet:2001un}:
\begin{align}
        G_1(m^2,T) &=\frac{3}{(4\pi)^2} I_1^\mathrm{th} (m^2) \left(\log\frac{\mu^2}{m^2}+2\right)\nonumber \\&+ \frac{3}{4 (2\pi)^4} \int_0^\infty dq_1 \frac{q_1n_B(q_1)}{E_{q_1}} \times \int_0^\infty \frac{dq_2}{E_{q_2}}\left(q_2\log \left|\frac{X_+}{X_-}\right|-4q_1\right)\\
        &\equiv\frac{3}{(4\pi)^2} I_1^\mathrm{th} (m^2) \left(\log\frac{\mu^2}{m^2}+2\right) + \mathcal{X}^{(1)}
        \label{e.G1sunset}
\end{align}
where $X_\pm = (E_{q_1} +E_{q_2} + E_{q_1\pm q_2})^2\times(-E_{q_1} +E_{q_2} + E_{q_1\pm q_2})^2$, and $I_1^\mathrm{th.}$ is 1-loop thermal function. Note, that this sunset has the same UV structure as the high-$T$ approximation. $G_2$ similarly can be evaluated as:
\begin{gather}
    G_2^{(1)}(m^2, T) = \frac{3}{4 (2\pi)^4} \int_0^\infty dq_1 \frac{q_1n_B(q_1)}{E_{q_1}} \times \int_0^\infty dq_2 \frac{q_2n_B(q_2)}{E_{q_2}} \log \left|\frac{Y_+}{Y_-}\right|
    \label{e.G2sunset}
\end{gather}
where $Y_\pm = (E_{q_1} +E_{q_2} + E_{q_1\pm q_2})^2\times(-E_{q_1} +E_{q_2} + E_{q_1\pm q_2})^2 \times (E_{q_1} -E_{q_2} + E_{q_1\pm q_2})^2\times(E_{q_1} +E_{q_2} - E_{q_1\pm q_2})^2$. The integrals in $G_1$ and $G_2$ need to perform numerically. 

\begin{figure}
    \centering
    \includegraphics[width=0.7\linewidth]{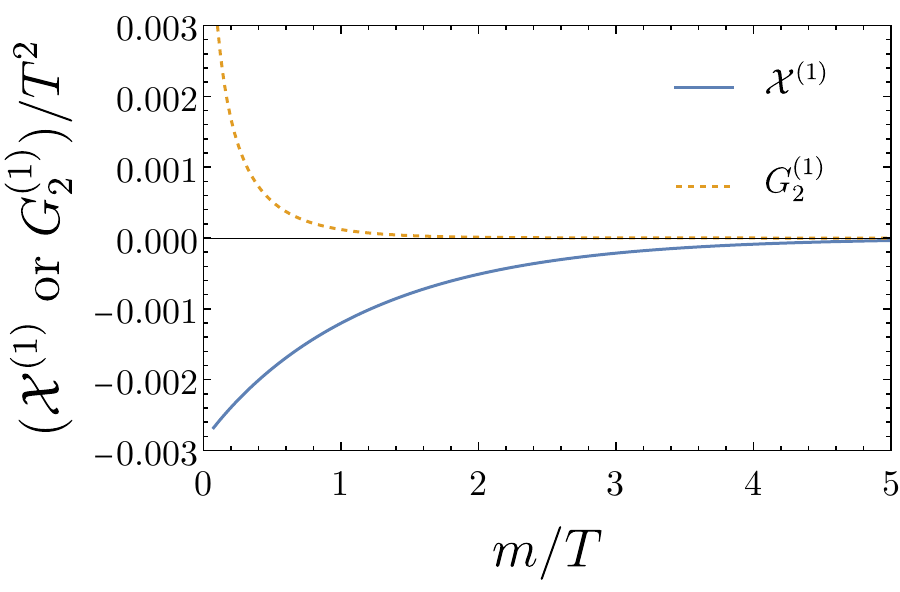}
    \caption{ Plot of numerical parts of single-scalar sunset from Eqns.~\eqref{e.G1sunset} and~\eqref{e.G2sunset}. Note that both decay to zero as $m/T\rightarrow \infty$.}
    \label{fig:sunset_single}
\end{figure}

\begin{figure}%
    \hspace*{-6mm}
    \centering
    \begin{tabular}{cc}    {{\includegraphics[width=7.5cm]{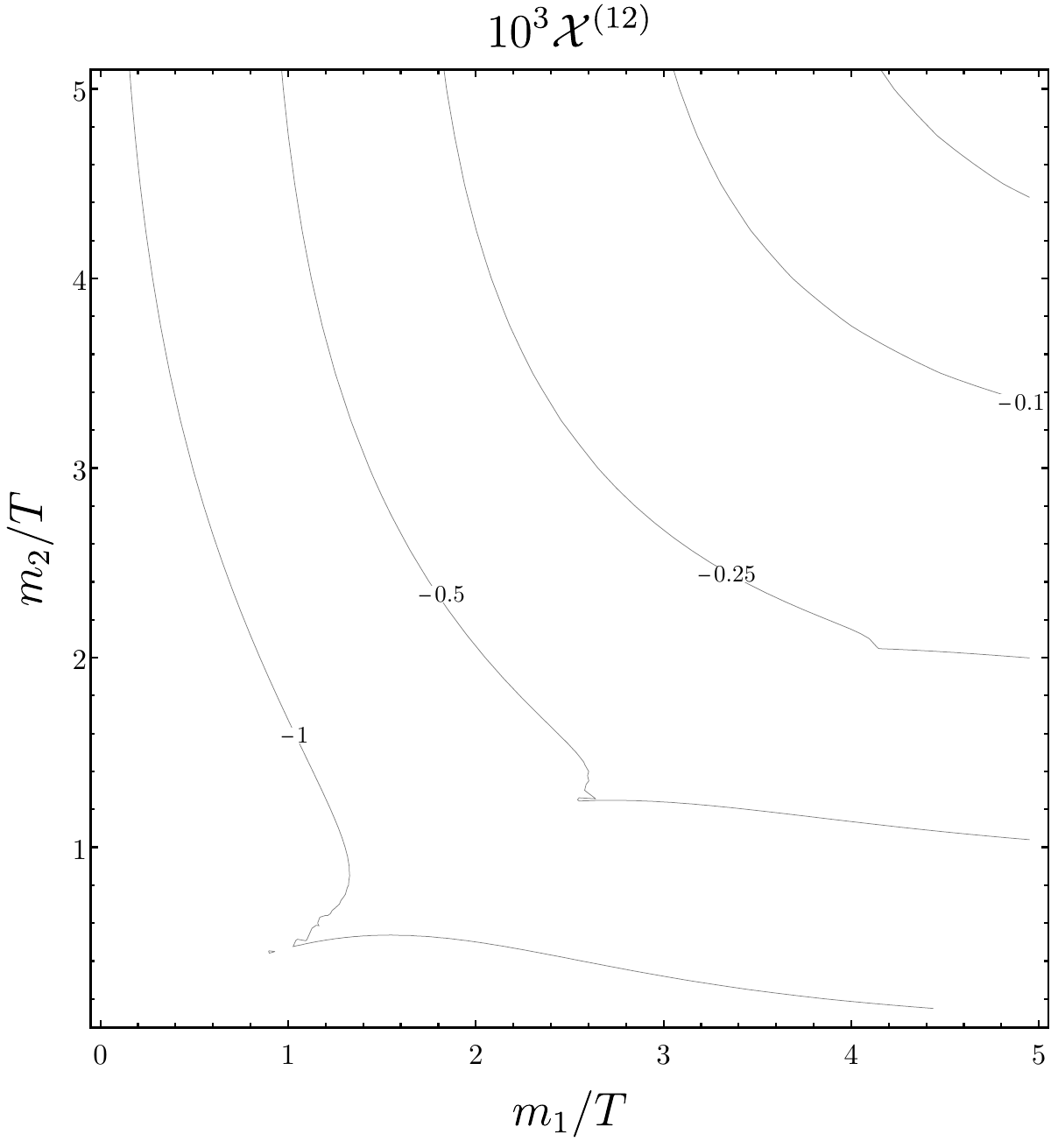} }}%
    &
    {{\includegraphics[width=7.5cm]{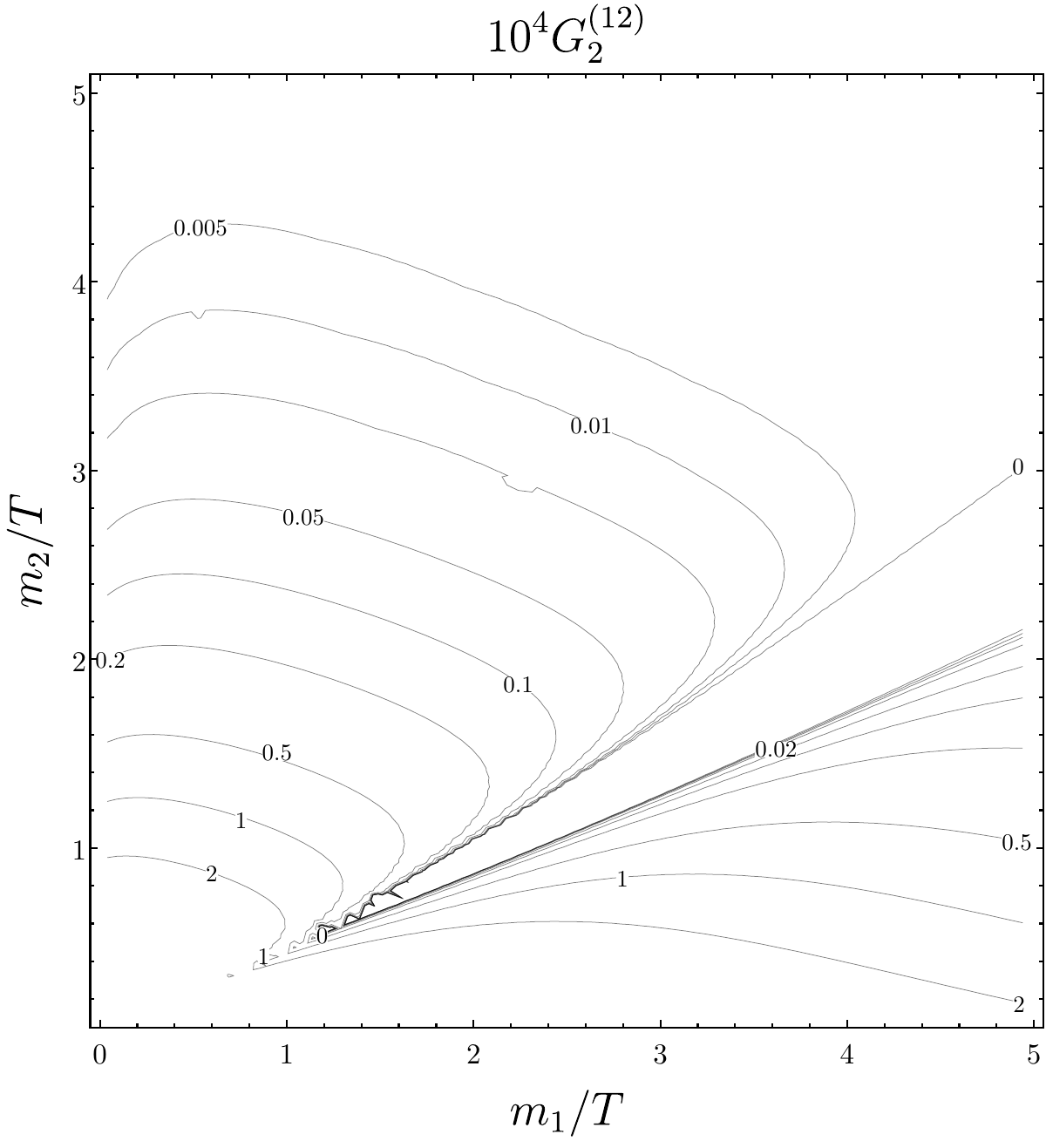} }}
    \end{tabular}
    \caption{We show a plot of the numerical parts of the two scalar sunset, see Eqns.~\eqref{eq:Xsum} and~\eqref{eq:Ysum}. As expected, both vanish when $m_i/T \to \infty$.}
    \label{fig:sunset_mix}
\end{figure}

The generalization to two-scalar theory is straight forward. In our particular example we have two-sunset diagrams. One is with $\phi_1$ propagating in all three lines, and that sunset is equivalent to the one-scalar case. The second sunset is the one with two $\phi_2$ propagators and one $\phi_1$ propagator. This sunset is given by:
\begin{gather}
    G_1+G_2 =\frac{I_1^\mathrm{th.}(m_1^2)}{8\pi^2}\Big(2+\ln\frac{\bar\mu^2}{m_2^2}\Big)
+\frac{I_1^\mathrm{th.}(m_2^2)}{4\pi^2}\Big(2+\ln\frac{\bar\mu^2}{m_1^2}\Big) + \mathcal{X}^{(12)} (m_1^2,m_2^2,T) + G_2^{(12)}
\end{gather}
The $\mathcal{X}_{12}$ and $G_2^{(12)}$ are given as numerical integrals:
\begin{align}
    \mathcal{X}_{12}&=F_1[m_1,m_2;m_2]+2F_1[m_2,m_1;m_2]\label{eq:Xsum}\\
F_1[m_A,m_B;m_C]&=\frac{1}{32\pi^4}\int_0^\infty\!\!dq_1\int_0^\infty\!\!dq_2\;
\frac{n_B^{(A)}(q_1)}{E^{(A)}_{q_1} E^{(B)}_{q_2}}
\left(q_1q_2\log\left|\frac{X_+}{X_-}\right|-2q_1^2\right)
\label{eq:F1def}\\
X_\pm&=\big(E^{(B)}_{q_2}+E^{(C)}_{q_1 \pm q_2}\big)^2-\big(E^{(A)}_{q_1}\big)^2 ,
\label{eq:Xpm}
\end{align}
and:
\begin{align}
G_2^{(12)}&=2F_2[m_1,m_2;m_2]+F_2[m_2,m_2;m_1],
\label{eq:Ysum}\\
F_2[m_A,m_B;m_C]&=\frac{1}{32\pi^4}\int_0^\infty\!\!dq_1\int_0^\infty\!\!dq_2\;
\frac{q_1q_2\;n_B^{(A)}(q_1)\,n_B^{(B)}(q_2)}{E^{(A)}_{q_1} E^{(B)}_{q_2}}
\ln\left|\frac{Y_+}{Y_-}\right|\label{eq:F2def}\\
Y_\pm&=\Big[\big(E^{(A)}_{q_1}+E^{(B)}_{q_2}\big)^2-\big(E^{(C)}_{q_1\pm q_2}\big)^2\Big]
\Big[\big(E^{(C)}_{q_1\pm q_2}\big)^2-\big(E^{(A)}_{q_1}-E^{(B)}_{q_2}\big)^2\Big].
\label{eq:Ypm}
\end{align}
The $\mathcal{X}_{12}$ and $G_2^{(12)}$ are evaluated numerically, as a functions of $m_1^2/T^2$ and $m_2^2/T^2$. We attach the call-up table for these integrals in the \LaTeX folder. In Fig.~\ref{fig:sunset_single} and~\ref{fig:sunset_mix} we show single scalar ($\mathcal{X}_1$ and $G_2^{(1)}$) and two scalar function ($\mathcal{X}_{12}$ and $G_2^{(12)}$).

\section{Proof of $\mathcal{O}(\beta^2)$ Accuracy of HPD for Two Scalar Theory in High-$T$ Limit} \label{a.beta2proof}

In this section, we show that HPD gives the correct potential to $\mathcal{O}(\beta^2)$ for a two-scalar theory where one scalar gets a VEV in the high temperature limit. The thermal integrals used in this section are defined in Appendix~\ref{thermalint_loopints}.

\subsection{Definitions}
The tree level potential is given by:
\begin{equation}
     V_\textrm{tree} =  \frac{1}{2} \nu_1^2 \phi_1^2 + \frac{1}{2} \nu_2^2 \phi_2^2 + \frac{\lambda_{1}}{4!} \phi_1^4 + \frac{\lambda_{2} }{4!} \phi_2^4 + \frac{\lambda_{12}}{4}\phi_1^2\phi_2^2 \ ,
\end{equation}
where we define the field-dependent masses and the cubic couplings
\begin{align}
m_1^2 = \frac{\partial^2 V_0}{\partial \phi_1^2}, \quad \quad
m_2^2 = \frac{\partial^2 V_0}{\partial \phi_2^2}, \quad \quad
c_1 = \frac{\partial^3 V_0}{\partial \phi_1^3} = \lambda_{1} \phi_1, \quad \quad
c_{12} = \frac{\partial^3 V_0}{\partial \phi_1 \phi_2^2} = \lambda_{12} \phi_1 \ .
\end{align}
We assume $\nu_2^2 > 0$ and that only $\phi_1$ acquires a VEV.
Note that the other cubic couplings are zero in this case. 

\subsubsection{Diagrams}
We illustrate $\phi_1$ ($\phi_2$) propagators with solid (dashed) lines and the non-zero (zero) mode loops with $\times$ ($\bullet$). The leading tadpoles are 
\begin{equation}
-\frac{1}{2}\left(
        \includegraphics[valign=c, scale=1]{ProveHPDtwoScalarDiagrams/Tadpoles/TAD_NonzeroScalar1.pdf} + \includegraphics[valign=c, scale=1]{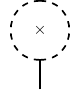} \right)
        =
        \left[ c_1  + c_{12} \right] \frac{T^2}{24} \ . 
\end{equation}   
The subleading diagrams are categorized by factoring out either $c_1T^2$ or $c_{12}T^2$, and then use these parameters 
\begin{equation}
    \alpha_{ij} = \frac{\lambda_{ij}}{48}\frac{T^2}{m_i^2 }, \quad \quad 
    \beta_{ij} = \frac{\lambda_{ij}}{16 \pi}\frac{T}{m_i} , \quad \quad
    \gamma = \frac{\phi_1^2}{T^2} \ . 
    \label{AppBETA:parameters}
\end{equation}
where the field labels $i,j \in \{1,2\}$ do not commute (so $\beta_{12} \neq \beta_{21}$),
$m_{1,2} = \nu_{1,2}^2 + \frac{1}{2}\lambda_{1,2} \phi_{1,2} + \frac{1}{2}\lambda_{12} \phi_{2,1}^2$,
$\lambda_{11} = \lambda_{1}$, $\lambda_{22} = \lambda_{2}$, and $\lambda_{21} = \lambda_{12}$. 

To categorize some of the lollipop and sunset tadpoles in Sections \ref{appBETA:secLollipop} and \ref{appBETA:secSunset}, we need a extra mixed-loop $\beta$-type parameter
\begin{equation}
    \beta_{i}^{(*)} = \frac{\lambda_{12}}{8\pi} \frac{T}{m_1 +m_2} \ ,
    \label{AppBETA:MIXparameters}
\end{equation}

To classify diagrams according to their importance for the finite-temperature calculation, we label each diagram by its order $\mathcal{O}(\alpha^r \beta^s \gamma^t)$, grouping together different types of the above parameters.

\subsubsection{Hybrid Partial Dressing}

The procedure for HPD starts by solving the gap equations
\begin{align} \begin{array}{ccccccc}
    \includegraphics[valign=c,width=1cm]{ProveHPDtwoScalarDiagrams/Gap_equations/DressProg_scalar1.pdf}
    & = & 
    \includegraphics[valign=c,width=1cm]{ProveHPDtwoScalarDiagrams/Gap_equations/Prog_scalar1.pdf}
    & + &
    \raisebox{1.5pt}{\includegraphics[width=1.8cm]{ProveHPDtwoScalarDiagrams/Gap_equations/Quartic_Progscalar1_ResumLoopscalar1.pdf}}
    & + &
    \raisebox{1.5pt}{\includegraphics[width=1.8cm]{ProveHPDtwoScalarDiagrams/Gap_equations/Quartic_Progscalar1_ResumLoopscalar2.pdf}} \\
    M_{1}^{2} &=&  m_{1}^2 &+& \lambda_{1} \left(\frac{T^2}{24} - \frac{T m_1}{8 \pi}\right) &+& \lambda_{12} \left(\frac{T^2}{24} - \frac{T m_2}{8 \pi}\right) \ , \\
    \includegraphics[valign=c,width=1cm]{ProveHPDtwoScalarDiagrams/Gap_equations/DressProg_scalar2.pdf}
    & = & 
    \includegraphics[valign=c,width=1cm]{ProveHPDtwoScalarDiagrams/Gap_equations/Prog_scalar2.pdf}
    & + &
    \raisebox{1.5pt}{\includegraphics[width=1.8cm]{ProveHPDtwoScalarDiagrams/Gap_equations/Quartic_Progscalar2_ResumLoopscalar1.pdf}}
    & + &
    \raisebox{1.5pt}{\includegraphics[width=1.8cm]{ProveHPDtwoScalarDiagrams/Gap_equations/Quartic_Progscalar2_ResumLoopscalar2.pdf}} \\
    M_{2}^{2} &=&  m_{2}^2 &+& \lambda_{12} \left(\frac{T^2}{24} - \frac{T m_1}{8 \pi}\right) &+& \lambda_{2} \left(\frac{T^2}{24} - \frac{T m_2}{8 \pi}\right) \ .
\end{array} \end{align}
Then, we insert the solutions of the gap equation into the tadpole
\begin{equation}
    V'_{\text{HPD}} = \frac{-1}{2} \left( \includegraphics[valign=c,width=1.5cm]{ProveHPDtwoScalarDiagrams/Tadpoles/TAD_RESUM_LoopScalar1.pdf} +
    \includegraphics[valign=c,width=1.5cm]{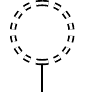} \right) +
    \frac{d}{d\phi_1} \left[ \frac{-1}{12} \includegraphics[valign=c,width=1.5cm]{ProveHPDtwoScalarDiagrams/Tadpoles/VACUUMsunset_RESUM_Scalar111.pdf} 
    - \frac{1}{4} \includegraphics[valign=c,width=1.5cm]{ProveHPDtwoScalarDiagrams/Tadpoles/VACUUMsunset_RESUM_Scalar221.pdf} \right] \ .
\end{equation} 
The first term gives the familiar daisy and superdaisy tadpoles, separating contributions coming from the zero mode  and non-zero mode:
\begin{align}
    V'^\bullet_{\text{DAISY}} &= \frac{-1}{2} \left( \includegraphics[valign=c,width=1.5cm]{ProveHPDtwoScalarDiagrams/Tadpoles/TAD_RESUM_ZeroScalar1.pdf} +
    \includegraphics[valign=c,width=1.5cm]{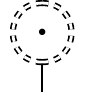} \right) \label{AppBeta2:DaisyZeroTAD} \ ,\\
    V'^\times_{\text{DAISY}} &= \frac{-1}{2} \left( \includegraphics[valign=c,width=1.5cm]{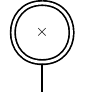} +
        \includegraphics[valign=c,width=1.5cm]{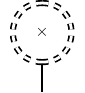} \right) \label{AppBeta2:DaisyNONTAD}  \ .
\end{align}
The total derivative can be expanded to obtain the lollipop and sunset tadpoles:
\begin{align}
    V'_{\text{LOL}} &= 
    -\frac{1}{6} \includegraphics[valign=c,width=1.5cm]{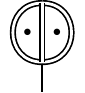}  -
    \frac{1}{2} \includegraphics[valign=c,width=1.5cm]{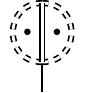}  \label{AppBeta2:lolliTAD}\ , \\ 
    V'_{\text{SUN}} &= 
    - \frac{1}{4} \,\includegraphics[valign=c,width=1.1cm]{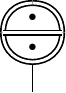} - 
    \frac{1}{2} \,\includegraphics[valign=c,width=1.1cm]{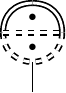} - 
    \frac{1}{4} \,\includegraphics[valign=c,width=1.1cm]{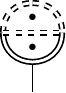}  \label{AppBeta2:SunsetTAD}  \ .
\end{align}

\subsection{Comparison of HPD and Feynman diagrams}

\label{s.HPDvsfeynman}

We now show that the four tadpoles in Eqns.~(\ref{AppBeta2:DaisyZeroTAD} - \ref{AppBeta2:SunsetTAD}) 
recover the complete thermal potential to all  orders in $\alpha, \gamma$, up to and including order $\beta^2$. We obviously can not go explicitly to all orders in $\alpha$. However, by going to order $\beta^2$ for each tadpole type and adding an extra non-zero mode loop, we capture the first term of the non-zero mode daisy chain of that diagram. We can then infer that HPD also covers the rest of the daisy chain.

Accordingly, the daisy and super-daisy contributions are evaluated at 3 loops in the zero‑mode tadpole and at 2 loops in the non‑zero mode tadpole. Additionally, the lollipop and sunset tadpoles are evaluated to 3 loops, as they are of order $\beta^2$ at 2 loops. Note that we present only contributions up to order $\beta^2$, as any $\beta^3$ terms arising even at the loop orders considered are beyond the accuracy we aim for.

This will show that the final HPD tadpole result, correct to order $\beta^2$, is
 \begin{equation}
     \begin{aligned}
     V'_{\text{HPD}} = & \frac{c_1}{2} \left[\frac{T^2}{12} -\frac{TM_1}{4\pi} - \frac{M_1^2 L_R}{16\pi^2}  \right] + \frac{c_{12}}{2} \left[\frac{T^2}{12} -\frac{TM_2}{4\pi}  - \frac{M_2^2 L_R}{16\pi^2} \right] \\
     &+\frac{d}{d \phi_1}\left[
     c_1^2 \frac{T^2}{384 \pi^2}  \left\{ \log\left(\frac{M_1^2}{\Bar{\mu}T}\right) + 1.65 \right\} \right] \\
     &+\frac{d}{d \phi_1}\left[
    c_{12}^2 \frac{T^2}{128 \pi^2}  \left\{\log\left(\frac{(M_1 + 2M_2)^2}{9\Bar{\mu}T}\right) + 1.65 \right\} 
     \right] \ ,
     \end{aligned}
\end{equation}
 \begin{align}
    M_{1}^{2} &=  m_{1}^2+ \frac{\lambda_{1}}{2} \left(\frac{T^2}{12} - \frac{T M_1}{4 \pi} -\frac{M_1^2}{16\pi^2}L_R  \right) 
    +  \frac{\lambda_{12}}{2} \left(\frac{T^2}{12} - \frac{T M_2}{4 \pi}  -\frac{M_2^2}{16\pi^2}L_R   \right) \ , \\
    M_{2}^{2} &=  m_{2}^2 +  \frac{\lambda_{12}}{2} \left(\frac{T^2}{12} - \frac{T M_1}{4 \pi} -\frac{M_1^2}{16\pi^2}L_R  \right) 
    +  \frac{\lambda_{2}}{2} \left(\frac{T^2}{12} - \frac{T M_2}{4 \pi} -\frac{M_2^2}{16\pi^2}L_R  \right) \ .
\end{align}
where  $c_1 = \lambda_{1} \phi_1$ and $c_{12} = \lambda_{12} \phi_1$.
\subsubsection{Daisy and Super-Daisy contributions to 3 loops from zero mode tadpole}

The daisy and super-daisy contributions from the zero mode tadpole (Eq.~\ref{AppBeta2:DaisyZeroTAD}) is given by:
\begin{equation}
    V'^\bullet_{\text{DAISY}} = -\frac{T}{8 \pi} \left( c_1 M_1 + c_{12} M_2 \right) \ .
\end{equation}
We solve the gap equations to second-order in quartic couplings to generate daisy and super-daisy diagrams to 3 loops. The HPD tadpole 
solution to $\mathcal{O}(\beta^2)$ is
\begin{equation}\begin{aligned}
V'^\bullet_{\text{DAISY}} = &-\frac{T}{8 \pi } \left(  c_1  m_1 + c_{12} m_2 \right) \\
&-\frac{T^3}{384 \pi} \left[ \frac{c_1}{m_1} \left( \lambda_{1} + \lambda_{12} \right) + \frac{c_{12}}{m_2} \left( \lambda_{12} + \lambda_{2} \right)  \right]  \\
&+ \frac{T^2}{128 \pi ^2} \left[  c_1 \left(\lambda_{1} + \lambda_{12} \frac{m_2 }{m_1} \right) + c_{12} \left(\lambda_{12} \frac{m_1}{m_2} + \lambda_{2} \right)  \right]  \\
&+\frac{T^5}{18432\pi} \left[ \frac{c_1}{m_1^3} \left(\frac{ \lambda_{1}^2}{2} +  \lambda_{12}\lambda_{1} + \frac{\lambda_{12}^2}{2}  \right) \right] \quad  \\
&+\frac{T^5}{18432\pi} \left[ \frac{c_{12}}{m_2^3 } \left(\frac{\lambda_{12}^2}{2} + \lambda_{12}\lambda_{2} + \frac{\lambda_{2}^2}{2}  \right) \right] \quad  \\
&+ \frac{T^4 \lambda_{12}}{6144 \pi^2 } \left[ 
 \frac{c_1 }{m_1 m_2} \left( \lambda_{12} + \lambda_{2} \right) + \frac{c_{12} }{m_1 m_2} \left( \lambda_{1} + \lambda_{12} \right) \right]  \\
&- \frac{T^4 \lambda_{12}}{6144 \pi^2} \left[ 
c_1  \frac{m_2}{m_1^3} \left(  \lambda_{1} + \lambda_{12} \right) + c_{12}  \frac{m_1}{m_2^3} \left( \lambda_{12} + \lambda_{2} \right) \right]  \ .
\label{eq:AppD_HPD_Zerodaisy}
\end{aligned}\end{equation}

The explicit $\beta^2$-result for the Feynman diagrams to 3 loops is
{\setlength{\belowdisplayskip}{0pt}%
\begin{equation} 
    -\frac12\left(
    \includegraphics[valign=c,scale=1]{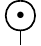} +
    \includegraphics[valign=c,scale=1]{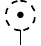} 
    \right)= -\frac{T}{8 \pi } \left(  c_1  m_1 + c_{12} m_2 \right), \qquad \mathcal{O}(\alpha^{-1} \beta) \label{eq:AppD_Diag1_Zerodaisy}
\end{equation}}
{\setlength{\abovedisplayskip}{0pt}%
\setlength{\belowdisplayskip}{0pt}%
\begin{align}
    -\frac14\left(\includegraphics[valign=c,scale=1]{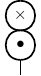} +
    \includegraphics[valign=c,scale=1]{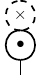} 
    \right) &= -\frac{T^3}{384 \pi} \left[ \frac{c_1}{m_1} \left( \lambda_{1} + \lambda_{12} \right) \right], \qquad  \mathcal{O}( \beta) \\
    -\frac14\left(
    \includegraphics[valign=c,scale=1]{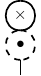} +
    \includegraphics[valign=c,scale=1]{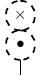} 
    \right) &= -\frac{T^3}{384 \pi} \left[ \frac{c_{12}}{m_2} \left( \lambda_{12} + \lambda_{2} \right)  \right], \qquad  \mathcal{O}( \beta)
\end{align}
\begin{align}
    -\frac14\left(\includegraphics[valign=c,scale=1]{ProveHPDtwoScalarDiagrams/DiagramTables/Chapter2/Chap2_zeroDAISY/Chap2_zeroDAISY_Table-Diag3.1.pdf} +
    \includegraphics[valign=c,scale=1]{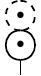} 
    \right) = 
    \frac{T^2}{128 \pi ^2} \left[  c_1 \left( \lambda_{1} + \lambda_{12} \frac{m_2 }{m_1} \right) \right], \qquad \mathcal{O}(\alpha^{-1} \beta^2) \label{eq:fig8_tadpole1}
\end{align}
\begin{align}
    -\frac14\left(
    \includegraphics[valign=c,scale=1]{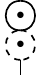} +
    \includegraphics[valign=c,scale=1]{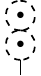} 
    \right) = 
    \frac{T^2}{128 \pi ^2} \left[  c_{12} \left(\lambda_{12} \frac{m_1}{m_2} + \lambda_{2}  \right) \right],  \qquad \mathcal{O}(\alpha^{-1} \beta^2) \label{eq:fig8_tadpole2}
\end{align}
\begin{align}
     -\frac18\left(\includegraphics[valign=c,scale=1]{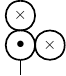} +
    2\includegraphics[valign=c,scale=1]{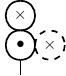} +
    \includegraphics[valign=c,scale=1]{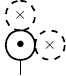}
    \right)= \frac{T^5}{18432\pi} \frac{c_1}{m_1^3} \left(\frac{ \lambda_{1}^2}{2} +  \lambda_{12}\lambda_{1} + \frac{\lambda_{12}^2}{2}  \right) ,  \, \mathcal{O}(\alpha \beta)
\end{align}
\begin{align}
    -\frac18\left(\includegraphics[valign=c,scale=1]{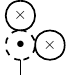} +
    2 \includegraphics[valign=c,scale=1]{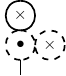} +
    \includegraphics[valign=c,scale=1]{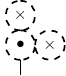}
    \right)= \frac{T^5}{18432\pi} \frac{c_{12}}{m_2^3 } \left(\frac{\lambda_{12}^2}{2} + \lambda_{12}\lambda_{2} + \frac{\lambda_{2}^2}{2}  \right),  \, \mathcal{O}(\alpha \beta)
\end{align}
\begin{align}
     -\frac18\left(\includegraphics[valign=c,scale=1]{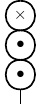} +
    \includegraphics[valign=c,scale=1]{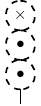}\right) -\frac14\left(
    \includegraphics[valign=c,scale=1]{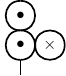} +
    \includegraphics[valign=c,scale=1]{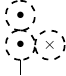}
    \right) = 0,  \qquad \mathcal{O}(\beta^2)
\end{align}
\begin{align}
    -\frac18\left(\includegraphics[valign=c,scale=1]{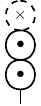} +
    \includegraphics[valign=c,scale=1]{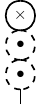} \right)
    -\frac14\left(\includegraphics[valign=c,scale=1]{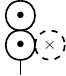} +
    \includegraphics[valign=c,scale=1]{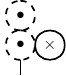}
    \right) =0,  \qquad \mathcal{O}(\beta^2)
\end{align}
\begin{align}
    -\frac18\left(\includegraphics[valign=c,scale=1]{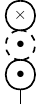} +
    \includegraphics[valign=c,scale=1]{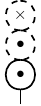}
    \right) &= \frac{T^4 \lambda_{12}}{6144 \pi^2 } \left[ \frac{c_1 }{m_1 m_2} \left( \lambda_{12} + \lambda_{2} \right) \right],  \quad \mathcal{O}(\beta^2) \\
    -\frac18\left(
    \includegraphics[valign=c,scale=1]{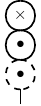} +
    \includegraphics[valign=c,scale=1]{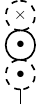}
    \right) &= \frac{T^4 \lambda_{12}}{6144 \pi^2 } \left[ \frac{c_{12} }{m_1 m_2} \left( \lambda_{1} + \lambda_{12} \right) \right],  \quad \mathcal{O}(\beta^2)
\end{align}
\begin{align}
    -\frac14\left(\includegraphics[valign=c,scale=1]{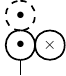} +
    \includegraphics[valign=c,scale=1]{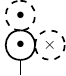} 
    \right) = - \frac{T^4 \lambda_{12}}{6144 \pi^2} \left[ c_1  \frac{m_2}{m_1^3} \left(  \lambda_{1} + \lambda_{12} \right) \right],  \qquad \mathcal{O}(\beta^2)
\end{align}}
{
\setlength{\abovedisplayskip}{0pt}
\begin{align}
    -\frac14\left(
    \includegraphics[valign=c,scale=1]{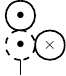} +
    \includegraphics[valign=c,scale=1]{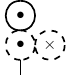}
    \right) = - \frac{T^4 \lambda_{12}}{6144 \pi^2} \left[ c_{12}  \frac{m_1}{m_2^3} \left( \lambda_{12} + \lambda_{2} \right) \right],  \qquad \mathcal{O}(\beta^2)
    \label{eq:AppD_Diaglast_Zerodaisy} \ .
\end{align}}

By comparing each diagrams (\ref{eq:AppD_Diag1_Zerodaisy} - \ref{eq:AppD_Diaglast_Zerodaisy}) to Eqn. \eqref{eq:AppD_HPD_Zerodaisy}, we find that HPD reproduces the correct result to $\mathcal{O}(\beta^2)$.

\subsubsection{Daisy and Super-Daisy contributions to 2 loops from non-zero mode tadpole }

The daisy and super-daisy contributions from the non-zero mode tadpole (Eq.~\ref{AppBeta2:DaisyNONTAD}) is given by:
\begin{equation}
    V'^\times_{\text{DAISY}}  = \frac{T^2}{24} \left( c_1 +  c_{12} \right)
     - \frac{L_R}{32 \pi^2} \left(  M_1^2 c_1 +  M_2^2 c_{12} \right) \ .
\end{equation}
The first term is the leading contribution, which is clearly captured by HPD. We focus on the second term, which can be resummed because of the mass-dependence. We solve the gap equations to first-order in quartic couplings to generate daisy diagrams to 2 loops. The HPD tadpole solution to $\mathcal{O}(\beta^2)$ is 
\begin{equation}\begin{aligned}
V'^{\times, \,\text{log}}_{\text{DAISY}} = &- \frac{L_R}{32 \pi^2} \left(  m_1^2 c_1 +  m_2^2 c_{12} \right) \\
& -\frac{T^2 L_R}{768 \pi^2} \left[ c_1 (\lambda_{1} + \lambda_{12})  + c_{12} (\lambda_{12}   + \lambda_{2}) \right] \ .  
\label{eq:AppD_HPD_NONdaisy}
\end{aligned}\end{equation}
The explicit $\beta^2$-result for the Feynman diagrams to 2 loops is
\begin{align}
    -\frac12\left(
    \includegraphics[valign=c,scale=1]{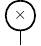} +
    \includegraphics[valign=c,scale=1]{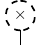} 
    \right) &= - \frac{L_R}{32 \pi^2} \left(  m_1^2 c_1 +  m_2^2 c_{12} \right),  \quad &&\mathcal{O}(\alpha^{-2} \beta^2) \label{eq:AppD_Diag1_NONdaisy}\\
    -\frac14\left(\includegraphics[valign=c,scale=1]{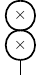} +
    \includegraphics[valign=c,scale=1]{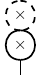}
    \right) &= -\frac{T^2 L_R}{768 \pi^2} \left[ c_1 ( \lambda_{1} + \lambda_{12}) \right],  \quad &&\mathcal{O}(\alpha^{-1} \beta^2) \\
    -\frac14\left(
    \includegraphics[valign=c,scale=1]{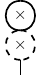} +
    \includegraphics[valign=c,scale=1]{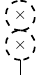} 
    \right) &= -\frac{T^2 L_R}{768 \pi^2} \left[ c_{12} (\lambda_{12}   + \lambda_{2}) \right],  \quad &&\mathcal{O}(\alpha^{-1} \beta^2) 
    \label{eq:AppD_Diaglast_NONdaisy} \ .
\end{align}

By comparing each diagrams (\ref{eq:AppD_Diag1_NONdaisy} - \ref{eq:AppD_Diaglast_NONdaisy}) to Eqn.~\eqref{eq:AppD_HPD_NONdaisy}, we find that HPD reproduces the correct result to $\mathcal{O}(\beta^2)$.

\subsubsection{Lollipop contributions to 3 loops} 
\label{appBETA:secLollipop}
The lollipop contributions (Eq. \ref{AppBeta2:lolliTAD}) is given by:
\begin{equation}
    V'_{\text{LOL}} = 
    \frac{c_1  \lambda_{1}  T^2}{192 \pi^2}  \left[ \log\left(\frac{M_1^2}{\Bar{\mu}T}\right) + 1.65 \right] + 
    \frac{ c_{12} \lambda_{12} T^2}{64 \pi^2}  \left[ \log\left(\frac{(M_1 + 2M_2)^2}{9\Bar{\mu}T}\right) + 1.65 \right] \ . 
\end{equation}
We solve the gap equations to first-order in quartic couplings to generate the lollipop diagrams to 3 loops. The HPD tadpole solution to $\mathcal{O}(\beta^2)$ is 
\begin{equation}\begin{aligned}
    V'_{\text{LOL}} &= 
     \frac{c_1 \lambda_{1} T^2}{192 \pi^2}  \left[ \log\left(\frac{m_1^2}{\Bar{\mu}T}\right) + 1.65 \right] + 
     \frac{c_{12} \lambda_{12} T^2}{64 \pi^2}  \left[ \log\left(\frac{(m_1 + 2m_2)^2}{9\Bar{\mu}T}\right) + 1.65 \right] \\
    &+ \frac{T^4 c_1 \lambda_{1}}{4608 \pi^2 m_1^2}  \left[ \lambda_{1} + \lambda_{12} \right] + \frac{T^4 c_{12} \lambda_{12}}{768 \pi^2 (m_1 + 2m_2)} \left[\frac{\lambda_{1} + \lambda_{12}}{2 m_1} + \frac{\lambda_{12} + \lambda_{2}}{ m_2} \right] \ . 
    \label{eq:AppD_HPD_lol}
\end{aligned}\end{equation}

The explicit $\beta^2$-result for the Feynman diagrams to 3 loops is
{\setlength{\belowdisplayskip}{0pt}%
\begin{align}
    -\frac16
    \includegraphics[valign=c,scale=1]{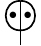} &= c_1 \lambda_{1} \frac{T^2}{192 \pi^2}  \left[ \log\left(\frac{m_1^2}{\Bar{\mu}T}\right) + 1.65 \right],  &&\mathcal{O}(\alpha^{-1} \beta^2) \label{eq:AppD_Diag1_lol}\\
    -\frac12
    \includegraphics[valign=c,scale=1]{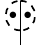} &= 
    c_{12} \lambda_{12} \frac{T^2}{64 \pi^2}  \left[ \log\left(\frac{(m_1 + 2m_2)^2}{9\Bar{\mu}T}\right) + 1.65 \right],  \quad &&\mathcal{O}(\alpha^{-1} \beta^2)
\end{align}}
{
\setlength{\abovedisplayskip}{0pt}%
\begin{align}
    -\frac14\left(
    \includegraphics[valign=c,scale=1]{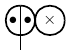} +
    \includegraphics[valign=c,scale=1]{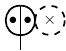}
    \right) &=  \frac{T^4 c_1 \lambda_{1}}{4608 \pi^2 m_1^2}  \left[ \lambda_{1} + \lambda_{12} \right],  \quad &&\mathcal{O}( \beta^2)\\
    -\frac14\left(
    \includegraphics[valign=c,scale=1]{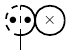} +
    \includegraphics[valign=c,scale=1]{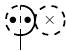} \right)  &= \frac{T^4 c_{12} \lambda_{12}}{768 \pi^2 (m_1 + 2m_2)} \left[\frac{\lambda_{1} + \lambda_{12}}{2 m_1} \right],  \quad &&\mathcal{O}(  \beta^2)\\
     - \frac12\left(
    \includegraphics[valign=c,scale=1]{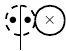} +
    \includegraphics[valign=c,scale=1]{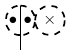}
    \right) &= \frac{T^4 c_{12} \lambda_{12}}{768 \pi^2 (m_1 + 2m_2)} \left[ \frac{\lambda_{12} + \lambda_{2}}{ m_2} \right],  \quad &&\mathcal{O}( \beta^2)
    \label{eq:AppD_Diaglast_lol} \ .
\end{align}}
To determine $\mathcal{O}(\alpha^r \beta^s \gamma^t)$ of each diagram, we use the approximate sunset integrals with the overlapping momenta neglected (\ref{eq:H4highT_NOoverlapping}--\ref{eq:H52highT_NOoverlapping}), which allows us to employ the $\beta$-type parameters defined above. By comparing each diagrams (\ref{eq:AppD_Diag1_lol} - \ref{eq:AppD_Diaglast_lol}) to Eqn.~\eqref{eq:AppD_HPD_lol}, we find that HPD reproduces the correct result to $\mathcal{O}(\beta^2)$.

\subsubsection{Sunset contributions to 3 loops}
\label{appBETA:secSunset}
The sunset contributions (Eq. \ref{AppBeta2:SunsetTAD}) is given by:
\begin{equation}
    V'_{\text{SUN}} = 
    \frac{T^2}{128 \pi^2} \left[ \frac{c_1^3}{3 M_1^2} + \frac{c_1 c_{12}^2 }{M_1 (M_1 + 2M_2)} + \frac{ 2 c_{12}^3}{M_2 (M_1 + 2M_2)} \right] \ .
\end{equation}
We solve the gap equations to first-order in quartic couplings to generate the sunset diagrams to 3 loops. The HPD tadpole solution to $\mathcal{O}(\beta^2)$ is

\begin{equation}\begin{aligned}
    V'_{\text{SUN}} &= 
    \frac{T^2}{128 \pi^2} \left[ \frac{c_1^3}{3 m_1^2} + \frac{c_1 c_{12}^2 }{m_1 (m_1 + 2m_2)} + \frac{ 2 c_{12}^3}{m_2 (m_1 + 2m_2)} \right] \\
    & -\frac{T^4}{9216 \pi^2} \frac{c_1^3}{m_1^4} \left[ \lambda_{1} + \lambda_{12}\right] 
    - \frac{T^4}{3072\pi^2} \frac{c_{12}^2 \left[ c_{12} \lambda_{1} + c_1 \lambda_{2} \right]}{m_1 m_2 (m_1 +2 m_2)^2}  \\
    & -\frac{T^4 c_{12}^2}{3072 \pi^2} 
    \left[  
    \frac{ c_1 \lambda_{1} (m_1 + m_2)  }{ m_1^3(m_1 + 2m _2)^2} + \frac{ c_{12} \lambda_{2} ( m_1 + 4m_2)}{m_2^3 (m_1 + 2m _2)^2}  \right] \\
    & -\frac{T^4 c_{12}^2 \lambda_{12}}{3072 \pi^2} 
    \left[ 
    \frac{ c_1 ( m_1^2 + m_1 m_2 + m_2^2)}{m_1^3 m_2 (m_1 + 2m _2)^2} + 
    \frac{ c_{12} ( m_1^2 + 4 m_1 m_2 + m_2^2) }{ m_1 m_2^3 (m_1 + 2m _2)^2} \right] 
    \label{eq:AppD_HPD_sun} \ .
\end{aligned}\end{equation}
    
The explicit $\beta^2$-result for the Feynman diagrams to 3 loops is
{\setlength{\belowdisplayskip}{0pt}%
\begin{align}
    -\frac14
    \includegraphics[valign=c,scale=1]{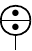} 
    -\frac14
    \includegraphics[valign=c,scale=1]{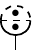} 
    -\frac12
    \includegraphics[valign=c,scale=1]{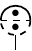}  
    = \frac{T^2}{128 \pi^2} \left[ \frac{c_1^3}{3 m_1^2} + \frac{c_{12}^2 }{(m_1 + 2m_2)} \left(\frac{c_1}{m_1 } + \frac{ 2 c_{12}}{m_2}\right) \right], \quad \mathcal{O}(\beta^2 \gamma) \label{eq:AppD_Diag1_sun}
\end{align}}
{\setlength{\abovedisplayskip}{0pt}%
\setlength{\belowdisplayskip}{0pt}%
\begin{align}
    -\frac14\left(
    \includegraphics[valign=c,scale=1]{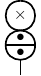} +
    \includegraphics[valign=c,scale=1]{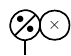} +
    \includegraphics[valign=c,scale=1]{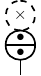}+
    \includegraphics[valign=c,scale=1]{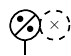}
    \right) = -\frac{T^4}{9216 \pi^2} \frac{c_1^3}{m_1^4} \left[ \lambda_{1} + \lambda_{12}\right], \quad \mathcal{O}(\alpha^1 \beta^2 \gamma) 
\end{align}
\begin{align}
    -\frac14\left(
    \includegraphics[valign=c,scale=1]{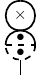} +
    \includegraphics[valign=c,scale=1]{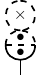}
    \right) &= - \frac{T^4}{3072\pi^2} \frac{c_{12}^2 \left[ c_{12} \lambda_{1} + c_1 \lambda_{2} \right]}{m_1 m_2 (m_1 +2 m_2)^2}, \quad &&\mathcal{O}(\alpha^1 \beta^2 \gamma)\\
    - \frac14\left(
    \includegraphics[valign=c,scale=1]{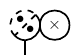} \right) &= -\frac{T^4 c_{12}^2}{3072 \pi^2} 
    \left[  
    \frac{ c_1 \lambda_{1} (m_1 + m_2)  }{ m_1^3(m_1 + 2m _2)^2} \right] , \quad &&\mathcal{O}(\alpha^1 \beta^2 \gamma)\\
    -\frac14
    \includegraphics[valign=c,scale=1]{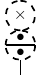}-\frac12
    \includegraphics[valign=c,scale=1]{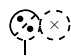} &= -\frac{T^4 c_{12}^2}{3072 \pi^2} 
    \left[   \frac{ c_{12} \lambda_{2} ( m_1 + 4m_2)}{m_2^3 (m_1 + 2m _2)^2}  \right] , \quad &&\mathcal{O}(\alpha^1 \beta^2 \gamma)\\
    -\frac14\left(
    \includegraphics[valign=c,scale=1]{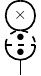} +
    \includegraphics[valign=c,scale=1]{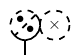}\right) &= -\frac{T^4 c_{12}^2 \lambda_{12}}{3072 \pi^2} 
     \frac{ c_1 ( m_1^2 + m_1 m_2 + m_2^2)}{m_1^3 m_2 (m_1 + 2m _2)^2} , \, &&\mathcal{O}(\alpha^1 \beta^2 \gamma)
\end{align}}
{\setlength{\abovedisplayskip}{0pt}%
\begin{align}
    -\frac14 \left(
    \includegraphics[valign=c,scale=1]{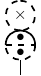}+
    \includegraphics[valign=c,scale=1]{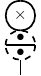} \right)  -\frac12
    \includegraphics[valign=c,scale=1]{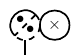} &= -\frac{T^4 c_{12}^3 \lambda_{12}}{3072 \pi^2} 
     \frac{ m_1^2 + 4 m_1 m_2 + m_2^2}{ m_1 m_2^3 (m_1 + 2m _2)^2} , \, &&\mathcal{O}(\alpha^1 \beta^2 \gamma)
     \label{eq:AppD_Diaglast_sun} \ . 
\end{align}}

To determine $\mathcal{O}(\alpha^r \beta^s \gamma^t)$ of each diagram, we use the approximate sunset integrals with the overlapping momenta neglected (\ref{eq:H4highT_NOoverlapping}--\ref{eq:H52highT_NOoverlapping}), which allows us to employ the $\beta$-type parameters defined above. By comparing each diagram (\ref{eq:AppD_Diag1_sun} - \ref{eq:AppD_Diaglast_sun}) to Eqn.~\eqref{eq:AppD_HPD_sun}, we find that HPD reproduces the correct result to $\mathcal{O}(\beta^2)$.

\subsection{Comparison of HPD to NLO DR}

\label{s.HPDvsDR}

Another way to prove that the two-field HPD calculation is accurate to $\mathcal{O}(\beta^2)$ is to directly prove that the NLO DR and HPD calculations capture exactly the same diagrams and give the same analytical result in the high-temperature limit.  The NLO DR potential for two-scalar is given by:
\begin{align}
V_{3 d}= & \frac{m_{1,3 d}^2 \phi_{3 d}^2}{2}+\frac{\lambda_{1,3 d} \phi_{3 d}^4}{24}-\frac{\left(m_{1,3 d}^2+\frac{\lambda_{1,3 d} \phi_{3 d}^2}{2}\right)^{3 / 2}}{12 \pi}-\frac{\left(m_{2,3 d}^2+\frac{\lambda_{12,3 d} \phi_{3 d}^2}{2}\right)^{3 / 2}}{12 \pi}\nonumber\\&+\frac{\lambda_{1,3 d}\left(m_{1,3 d}^2+\frac{\lambda_{1,3 d} \phi_{3 d}^2}{2}\right)}{128 \pi^2} 
  +\frac{\lambda_{12,3 d} \sqrt{m_{1,3 s}^2+\frac{\lambda_{1,3 d} \phi_{3 d}^2}{2}} \sqrt{m_{2,3 s}^2+\frac{\lambda_{12,3 d} \phi_{3 d}^2}{2}}}{64 \pi^2}\nonumber \\ &+\frac{\lambda_{2,3 d}\left(m_{2,3 s}^2+\frac{\lambda_{12,3 d} \phi_{3 d}^2}{2}\right)}{128 \pi^2} +V_\textrm{sun} \ .
\end{align}

Here, $V_\textrm{sun}$ is a 3D equivalent of the sunset correction term we add in HPD, and as such automatically equivalent. Now, let us compare 4D tadpole from 3D EFT potential with HPD tadpole. First notice $V_{4d} = T V_{3d}$. We will do our comparison to order $\beta^2$. To make the comparison easier, let us first define some notation:
\begin{gather}
    M_{1,3d}^2 = m_{1,3d}^2 + \frac{\lambda_1 \phi^2}{2} \simeq m_1^2 + \frac{\lambda_1+ \lambda_{12}}{24}T^2 \nonumber \ , \\
    M_{2,3d}^2 = m_{2,3d}^2 + \frac{\lambda_{12} \phi^2}{2} \simeq m_2^2 + \frac{\lambda_2+ \lambda_{12}}{24}T^2 \label{e.DR_m} \ .
\end{gather}
Now, let us take a derivative:
\begin{align}
V'_{4d} = m_{1,3d}^2 \phi &+ \frac{\lambda_1 \phi^3}{6} -\frac{\lambda _1 T \phi  M_{1,3 d}}{8 \pi }-\frac{\lambda _{12} T \phi  M_{2,3 d}}{8 \pi }
\nonumber
\\&+\frac{\lambda _{12}^2 T^2 \phi  M_{1,3 d}}{128 \pi ^2 M_{2,3 d}}+\frac{\lambda _1 \lambda _{12} T^2 \phi  M_{2,3 d}}{128 \pi ^2 M_{1,3 d}}
+\frac{\lambda _{12} \lambda_2 T^2 \phi }{128 \pi ^2}+\frac{\lambda _1^2 T^2 \phi }{128 \pi ^2} \ . 
\end{align}
Now, if we use equation~\ref{e.DR_m}, and expand in terms of couplings, we obtain:
\begin{align}
    V_{4d}'&=\phi  m_{1,3 d}^2+\frac{\lambda _1 \phi ^3}{6}\nonumber\\
    &+T^5 \left(\frac{\lambda _1^3 \phi }{36864 \pi  m_1^3}+\frac{\lambda _{12} \lambda _1^2 \phi }{18432 \pi  m_1^3}+\frac{\lambda _{12}^2 \lambda _1 \phi }{36864 \pi  m_1^3}+\frac{\lambda _{12}^3 \phi }{36864 \pi  m_2^3}+\frac{\lambda _2 \lambda _{12}^2 \phi }{18432 \pi  m_2^3}+\frac{\lambda _2^2 \lambda _{12} \phi }{36864 \pi  m_2^3}\right)\nonumber\\
    &+T^4 \left(\frac{\lambda _{12}^3 \phi }{6144 \pi ^2 m_1 m_2}-\frac{\lambda _{12}^3 m_1 \phi }{6144 \pi ^2 m_2^3}+\frac{\lambda _1 \lambda _{12}^2 \phi }{3072 \pi ^2 m_1 m_2}-\frac{\lambda _1 \lambda _{12}^2 m_2 \phi }{6144 \pi ^2 m_1^3}\right. \nonumber \\& \quad \quad \quad \quad \quad \quad \quad \quad \quad \quad\quad\quad\quad\quad\quad\quad\quad\quad\quad\left.-\frac{\lambda _2 \lambda _{12}^2 m_1 \phi }{6144 \pi ^2 m_2^3}+\frac{\lambda _1 \lambda _2 \lambda _{12} \phi }{6144 \pi ^2 m_1 m_2}-\frac{\lambda _1^2 \lambda _{12} m_2 \phi }{6144 \pi ^2 m_1^3}\right)\nonumber\\
    &+T^3 \left(-\frac{\lambda _1^2 \phi }{384 \pi  m_1}-\frac{\lambda _{12} \lambda _1 \phi }{384 \pi  m_1}-\frac{\lambda _{12}^2 \phi }{384 \pi  m_2}-\frac{\lambda _2 \lambda _{12} \phi }{384 \pi  m_2}\right)\\
    &+T^2 \left(\frac{\lambda _1^2 \phi }{128 \pi ^2}+\frac{\lambda _2 \lambda _{12} \phi }{128 \pi ^2}+\frac{\lambda _{12} \lambda _1 m_2 \phi }{128 \pi ^2 m_1}+\frac{\lambda _{12}^2 m_1 \phi }{128 \pi ^2 m_2}\right)\nonumber\\
    &+T \left(-\frac{\lambda _1 m_1 \phi }{8 \pi }-\frac{\lambda _{12} m_2 \phi }{8 \pi }\right)\nonumber \ ,
\end{align}
which agrees term by term up to order $\beta^2$ with ~\ref{eq:AppD_HPD_Zerodaisy}. Let us check the purely non-zero Matsubara modes (given in ~\ref{eq:AppD_Diag1_NONdaisy} and \ref{eq:AppD_Diaglast_NONdaisy}). The first terms simply come from 4D running of $m_1^2$ and $\lambda_1$.
\begin{align}
    V_{4d}' \supset m_{1,3d}^2 \phi + \frac{1}{6}\lambda_1 \phi^3 &\supset \frac{\lambda_1 m_1^2 + \lambda_{12} m_2^2 }{32\pi^2} \phi L_R+ \frac{ (\lambda_1^2 + \lambda_{12}^2)}{64\pi^2} L_R \nonumber \\ 
    &= \frac{L_R}{32\pi^2} (\lambda_1 \phi m_1^2(\phi) + \lambda_{12} \phi m_2^2(\phi)) \ , 
\end{align}
matching exactly the \ref{eq:AppD_Diag1_NONdaisy}. The terms in \ref{eq:AppD_Diaglast_NONdaisy} come from matching relations:
\begin{gather}
    m_{1,3d}^2 \supset -\frac{T^2 L_R}{768 \pi^2} (\lambda_1^2 + \lambda_1 \lambda_{12} + \lambda_2\lambda_{12} + \lambda_{12}^2) \ .
\end{gather}
Plugging these in, we get:
\begin{gather}
    V_{4d}' \supset  m_{1,3d}^2 \phi \supset -\frac{T^2 \phi L_R}{768 \pi^2} (\lambda_1^2 + \lambda_1 \lambda_{12} + \lambda_2\lambda_{12} + \lambda_{12}^2) \ ,
\end{gather}
again, exactly matching \ref{eq:AppD_Diaglast_NONdaisy}. To complete the proof consider sunset topologies. In both methods, these are added by hand (generated via $V_\textrm{sun}$) which exactly agree between the two.

\section{Comment on Optimized Partial Dressing}
\label{a.OPD}
In~\cite{Curtin:2022ovx}, it was analytically shown that a two-scalar partial dressing effective potential can be RG-improved to yield a residual scale variation of $\mathcal{O}(\lambda^3)$, equivalent to NLO DR.
However, the numerical studies of the effective potential in that analysis, which focused on the Optimized Partial Dressing procedure,  did not actually implement RG improvement. 
The reason can be understood as follows. First, OPD was being studied as a way of obtaining reliable predictions beyond the high-temperature regime. It was, at the time, not clear whether the high-temperature proof of vanishing scale variation at order $\lambda$ would generalize to all temperatures. In part, this was because the OPD procedure mixed a high-temperature expansion of the gap equation with a thermal potential that did not rely on the high-temperature expansion, which obscured both how to generalize the proof of small scale variation beyond high temperature, and whether the RGI procedure introduced any double-countings between the leading-log and thermal resummations of the specific OPD procedure. As a result, this also left unanswered how to correctly choose the renormalization scale within the RGI thermal potential. Apart from fixing the incorrect sunset terms in the OPD procedure (see Section~\ref{s.twofieldpartialdressing}), our HPD procedure never relies on the high-temperature approximation and is easy to numerically implement owing to the simplified form of the truncated HPD gap equation, circumvented all these difficulties and making it easy to implement RG improvement and prove that $\beta^2$ accuracy is maintained in RGHPD.

\section{Computing Gravitational Wave Observables of Phase Transitions}
\label{s.gravwave}
We now briefly review how we compute the stochastic gravitational wave spectrum of a strong first order phase transition in the early universe from the effective potential. Our simple gravitational wave calculation is sufficient to demonstrate the relative impact of different effective potential calculations, but is very much not state-of-the-art. For example, using $V_\mathrm{eff}$ in the classical bounce action neglects finite-temperature derivative and wave function corrections~\cite{Ai:2020sru,Strumia:1998nf,Ekstedt:2021kyx,Ekstedt:2022tqk,Kierkla:2025qyz}. 

For our two scalar model, we will assume that SM particle content is coupled strongly enough to the two scalars so that kinetic equilibrium is maintained, but weakly enough so that it does not affect gravitational waves cosmology. We assume that the main contribution to the gravitational wave amplitude is acoustic, and we calculate it using sound shell model \cite{Hindmarsh:2019phv,Hindmarsh:2013xza,Hindmarsh:2015qta,Hindmarsh:2017gnf}. In the sound shell model, the spectrum is completely determined by the temperature of the transition, the fluid velocity, the mean bubble separation, the life time of the sound waves and the fraction of energy released that becomes converted to sound waves. All of these quantities can, to a good approximation, be calculated using four macroscopic quantities: bubble wall velocity, the transition temperature, the trace anomaly normalized by critical density and inverse lifetime of the transition. We assume that bubble wall velocity is relativistic ($v_w \approx 1$). In a full, non-equilibrium calculation, one would need to compute the wall velocity using contemporary techniques (e.g. \cite{Ai:2025bjw,Ekstedt:2024fyq}), however this is outside of the scope of this paper. The trace anomaly normalized by the critical density is given by:
\begin{gather}
    \alpha = \frac{\Delta V -T \frac{d\Delta V}{dT}}{\rho_c} \ ,
\end{gather}
and the inverse lifetime of transition is given by:
\begin{gather}
    \frac{\beta}{H_*} =  T \frac{d \left(S_E/T\right)}{dT} \ .
\end{gather}
We compute the Euclidian  action $S_E$ from the effective potential using the Mathematica package FindBounce \cite{Guada:2020xnz}. 
The sound-shell model maps these four parameters to two observables: peak frequency
\begin{gather}
    f_{\mathrm{SW}}=1.9 \times 10^{-5} \frac{1}{v_w}\left(\frac{\beta}{H_n}\right)\left(\frac{T_n}{100 \mathrm{GeV}}\right)\left(\frac{g_*}{100}\right)^{1 / 6} \mathrm{~Hz} \ ,
\end{gather}
and peak amplitude \cite{Guo:2020grp,Hindmarsh:2013xza,Hindmarsh:2015qta,Hindmarsh:2017gnf,Hindmarsh:2019phv}:
\begin{gather}
    h^2 \Omega_{\mathrm{GW}}=8.5 \times 10^{-6}\left(\frac{100}{g_*}\right)^{1 / 3} \kappa^2\left(\frac{H_*}{\beta}\right) v_w \Upsilon\left(\bar{U}_{f, \max }, R_*\right) \ ,
\end{gather}
where the $g_*$ counts the number of relativistic degrees of freedom. The efficiency factor, $\kappa$, can be approximated for relativistic wall velocities as:
\begin{gather}
\kappa \simeq \frac{\alpha}{0.73+0.083 \sqrt{\alpha}+\alpha} \ ,
\end{gather}
and the suppression factor from the finite lifetime of the source is:
\begin{gather}
    \Upsilon=1-\frac{1}{\sqrt{1-2 \tau_{\mathrm{sw}} H_s}} \ ,
\end{gather}
where $\tau_{\mathrm{sw}}=R_* / \bar{U}_f$ with the fluid velocity and the mean bubble separation having the form $U_f^2 \sim \frac{3}{4} \kappa \alpha$ and $R_*=(8 \pi)^{1 / 3} v_w / \beta$, respectively. Once we have peak amplitude and peak frequency, in sound-shell model, the spectral shape is approximately given as:
\begin{gather}
    S_{\mathrm{SW}}(f)=\left(\frac{f}{f_{\mathrm{SW}}}\right)^3\left[\frac{7}{4+3\left(f / f_{\mathrm{SW}}\right)^2}\right]^{7 / 2} \ .
\end{gather}
Finally, the percolation temperature of the bubbles can be found solving the equations:
\begin{gather}
    \frac{S_E(T)}{T} = 131 - \log(A/T^4)- 4 \log (T/100 GeV) - 4 \log \left(\frac{\beta(T)/H}{100}\right) +3 \log (v_w) \ ,
\end{gather}
where $A$ is a constant, usually chosen as $\log\left(A/T^4\right)\simeq 14$.

\bibliography{biblio}
\end{document}